\documentclass[a4paper,usenames,dvipsnames,11pt]{article}
\pdfoutput=1

\usepackage{jheppub}
\usepackage{slashed}
\usepackage{mathrsfs,booktabs,multirow,tabularx}
\usepackage{stmaryrd}
\usepackage{xspace}
\usepackage{fancyvrb}
\usepackage[makeroom]{cancel}
\usepackage{amsmath}    
\usepackage{amssymb}    %
\usepackage{graphicx}   
\usepackage{verbatim}   
\usepackage{lscape}
\usepackage{subfig}
\usepackage{listings}

\def\beq{\begin{equation}}
\def\beqn{\begin{eqnarray}}
\def\eeq{\end{equation}}
\def\eeqn{\end{eqnarray}}

\def\xidistr#1{\left(\frac{1}{\xi}\right)_{\!#1}}
\def\pdistr#1#2{\left(\frac{1}{#1}\right)_{\!#2}}
\def\lxidistr#1{\left(\frac{\log\xi}{\xi}\right)_{\!#1}}
\def\lpdistr#1#2{\left(\frac{\log#1}{#1}\right)_{\!#2}}
\def\lppdistr#1#2{\left(\frac{\log\left(#1\right)}{#1}\right)_{\!#2}}

\def\EWeq#1{eq.~({\bf I}.#1)}

\def\PDF#1#2{\Gamma_{\!#1/#2}}
\def\Fragm#1#2{D_{\!#1/#2}}

\newcommand\sss{\scriptscriptstyle}
\newcommand\mydot{\!\cdot\!}
\newcommand\ep{\epsilon}
\newcommand\half{\frac{1}{2}}

\newcommand\aem{\alpha}
\newcommand\aemotpi{\frac{\aem}{2\pi}}
\newcommand\gs{g_{\sss S}}

\newcommand{\bq}{\bar{q}}

\newcommand{\epem}{e^+e^-}
\newcommand{\lp}{e^+}
\newcommand{\lm}{e^-}
\newcommand{\lpm}{e^{\pm}}
\newcommand{\Zp}{Z_+}
\newcommand{\Zm}{Z_-}
\newcommand{\Zpm}{Z_{\pm}}
\newcommand{\zp}{z_+}
\newcommand{\zm}{z_-}

\newcommand{\yp}{y_+}
\newcommand{\ym}{y_-}
\newcommand{\ub}{\bar{u}}
\newcommand{\uub}{u\ub}
\newcommand{\plp}{p_{\lp}}
\newcommand{\plm}{p_{\lm}}
\newcommand{\plpm}{p_{\lpm}}
\newcommand{\pb}{\bar{p}}
\newcommand{\pblp}{\bar{p}_{\lp}}
\newcommand{\pblm}{\bar{p}_{\lm}}
\newcommand{\pblpm}{\bar{p}_{\lpm}}
\newcommand{\pu}{p_u}
\newcommand{\pub}{p_{\ub}}
\newcommand{\puub}{p_{u\ub}}
\newcommand{\pga}{p_\gamma}
\newcommand{\hp}{\hat{p}}
\newcommand{\klp}{k_{\lp}}
\newcommand{\klm}{k_{\lm}}
\newcommand{\klpm}{k_{\lpm}}
\newcommand{\ku}{k_u}
\newcommand{\kub}{k_{\ub}}
\newcommand{\kuub}{k_{u\ub}}
\newcommand{\kga}{k_\gamma}
\newcommand{\hk}{\hat{k}}
\newcommand{\ord}{{\cal O}}

\newcommand\CF{C_{\sss F}}
\newcommand\TR{T_{\sss R}}

\newcommand\hsig{\hat{\sigma}}
\newcommand\bsig{\bar{\sigma}}
\newcommand\hbsig{\hat{\bsig}}
\newcommand\bbsig{\bar{\bsig}}

\newcommand\ee{e_e}
\newcommand\emu{e_\mu}
\newcommand\qu{q_u}
\newcommand\boost{{\mathbb B}}
\newcommand\iboost{{\mathbb B}^{-1}}

\newcommand\shat{\hat{s}}
\newcommand\ycm{y_{\rm\sss CM}}
\newcommand\vet{\vec{e}}
\newcommand\amp{{\cal A}}
\newcommand\ampsq{{\cal M}}
\newcommand\ampsqnt{\ampsq^{(n,0)}}
\newcommand\ampsqtwt{\ampsq^{(2,0)}}

\newcommand\ampsqtht{\ampsq^{(3,0)}}

\newcommand\ampsqtwl{\ampsq^{(2,1)}}
\newcommand\bampsq{\overline{\cal M}}
\newcommand\bampsqtwt{\bampsq^{(2,0)}}
\newcommand\bampsqtht{\bampsq^{(3,0)}}
\newcommand\bampsqtwl{\bampsq^{(2,1)}}
\newcommand\bampsqnpot{\bampsq^{(n+1,0)}}
\newcommand\rampsq{{\mathbb M}}

\newcommand\rampsqtwt{\rampsq^{(2,0)}}

\newcommand\rampsqtht{\rampsq^{(3,0)}}

\newcommand\rbampsq{\overline{\mathbb M}}
\newcommand\rbampsqtwt{\rbampsq^{(2,0)}}
\newcommand\rbampsqtht{\rbampsq^{(3,0)}}

\newcommand\Tr{{\rm Tr}}
\newcommand\QES{Q_{\rm ES}}
\newcommand\QESt{Q_{\rm ES}^2}

\newcommand\ordmos{{\ord\left(\!\frac{m^2}{s}\!\right)}}
\newcommand\MSb{\overline{\rm MS}}

\newcommand\bm{\overline{m}}
\newcommand\emtem{e\mu\to e\mu}
\newcommand\emtemg{e\mu\to e\mu\gamma}
\newcommand\emtembgb{e\mu\to e\mu(\gamma)}
\newcommand\gmtgm{\gamma\mu\to\gamma\mu}
\newcommand\mtos{\frac{m^2}{s}}
\newcommand\mtosl{m^2/s}
\newcommand\Sfun{{\cal S}}
\newcommand\Sfunij{\Sfun_{ij}}

\title{Initial conditions for electron and photon structure
and fragmentation functions}

\author[a]{S. Frixione,}
\affiliation[a]{INFN, Sezione di Genova, Via Dodecaneso 33, I-16146, 
Genoa, Italy}

\emailAdd{Stefano.Frixione@cern.ch}

\abstract{In the computation of short-distance cross sections
initiated by electrons and photons one can adopt the so-called
structure-function approach, in which these particles play formally
the same roles as hadrons do in QCD factorisation theorems, and must
thus be associated with PDFs (equivalently known as structure 
functions in this context). At variance with their QCD counterparts,
such PDFs are entirely calculable in QED. In this paper we present the 
results, at the next-to-leading order in the QED coupling constant $\aem$, 
for the initial conditions of the unpolarised electron and photon PDFs, 
which are a necessary ingredient for their eventual collinear evolution 
at the next-to-leading logarithmic accuracy. We also compute the analogous
final-state quantities, namely the initial conditions for fragmentation 
functions into electrons and photons.
}


\begin{document}
\maketitle
\flushbottom

\section{Introduction\label{sec:intro}}
It is not unreasonable to assume that the medium- to long-term
future of high-energy physics will involve an $\epem$ collider.
From the theoretical point of view, whether such a collider will
be a linear or a circular one is relatively unimportant, since the
primary goal will be that of computing production short-distance
cross sections, which are independent of the collider type. What
matters is that these cross sections are initiated by electrons
and positrons (and, if beam dynamics play a significant role, by
photons as well). We hasten to stress that although electrons/positrons 
and photons might feature in the initial states of hard reactions
occurring at hadron colliders, the corresponding cross sections
are fundamentally different with respect to their $\epem$-collider
counterparts. This is because in hadronic collisions leptons and
photons emerge from the low-scale collinear and soft dynamics of
the incoming hadrons; their momenta are not fixed, but distributed
according to their respective Parton Distribution Functions (PDFs henceforth).
Conversely, at an $\epem$ collider, the incoming $e^+$, $e^-$, and $\gamma$ 
must be regarded as having definite\footnote{Strictly speaking, this is not 
true because of beamstrahlung effects (which are also the source of photons 
to start with). However, such effects have a different origin, and a different
outcome, w.r.t.~the collinear dynamics which is our primary concern here.} 
and known momenta. Therefore, and at least when working in QED, physical 
distributions at $\epem$ colliders can be obtained by using solely
ingredients that can be computed from first principles, i.e.~matrix
elements and phase spaces.

The previous statement, if true, is however mostly academic when
the computations are performed strictly in a perturbative series
truncated at any given order in the QED coupling constant $\aem$,
because of the presence in the matrix elements of terms such as
\mbox{$\log^k E/m$}, with $m$ the electron mass and $E$ a scale of the 
order of the hardness of the collision (e.g.~the collider c.m.~energy).
These terms are numerically large and compensate the suppression due to 
the coupling constant, thus preventing the perturbative series from being 
well behaved. Fortunately, a large class of them is also process-independent,
and can therefore be accounted for in a universal manner, thanks to
the fact that their physical origin is well understood, and stem from
collinear emissions off an electron/positron line: the finite electron
mass screens potential collinear divergences, but leaves logarithmic
leftovers. This is reminiscent of the situation in QCD, and one way
to address the problem is indeed QCD-inspired: the structure-function 
approach~\cite{Kuraev:1985hb,Ellis:1986jba}, whereby one collects all of 
the logarithmic terms in some universal factors (the PDFs, a.k.a.~structure
functions. Here, for brevity we shall mainly refer to them by the former 
name), which are then resummed by means of the DGLAP evolution 
equations~\cite{Gribov:1972ri,Lipatov:1974qm,Altarelli:1977zs,
Dokshitzer:1977sg}. The key difference w.r.t.~QCD is that in QED one 
can calculate perturbatively not only the evolution kernels, which are 
the standard Altarelli-Parisi ones~\cite{Altarelli:1977zs}, but also the 
initial conditions for such an evolution\footnote{See sect.~\ref{sec:PDFcoll}
for the analogy between these initial conditions and related QCD objects.}.

The electron structure functions that are used nowadays
result from the solution of the evolution equations where both
the kernels and the initial conditions are leading-order (LO)
accurate~\cite{Skrzypek:1990qs,Skrzypek:1992vk,Cacciari:1992pz} -- 
at this order, the latter are in fact trivial, and equal to a Dirac 
delta. However, the accuracy targets of any future $\epem$ collider must 
be matched by equally accurate theoretical computations, which thus demand
to increase the precision of both the matrix element and the structure
function predictions. The aim of this paper is to address the latter 
aspect, with the goal of calculating the initial conditions at the
next-to-leading order (NLO) in QED; these are a prerequisite for an 
evolution accurate at the next-to-leading logarithmic (NLL) 
level~\cite{BCCFS}. When working at the NLO, the electron-photon 
mixing cannot formally be neglected any longer as is done at the LO; this
applies to photon-initiated processes as well. Therefore, results will
be presented for the initial conditions of all of the four possible
combinations of electron and photons emerging from the collinear dynamics
of either electron or photon evolution.

From the technical point of view, electron and photon structure functions
are quite analogous to their final-state counterparts, namely to the
{\em fragmentation} functions of either an electron or a photon
emerging from the multiple collinear branchings of either an electron 
or a photon, which in turn are (some of) the outgoing particles of
a hard scattering. Thus, we shall apply the methods employed in
the case of the structure functions to compute the initial conditions
for these fragmentation functions as well. Needless to say, owing to
their being associated with final-state properties, these results will
be equally relevant to lepton as to hadron colliders.

This paper is organised as follows: in sect.~\ref{sec:gen} we introduce
the factorisation formulae which constitute the core of the structure
function method. We shall employ two different procedures to compute
the structure-function initial conditions, which we sketch in 
sect.~\ref{sec:PDF}; the corresponding calculations are reported
in sects.~\ref{sec:PDFxsec} and~\ref{sec:PDFcoll}, respectively. Initial 
conditions for fragmentation functions are presented in sect.~\ref{sec:FF}. 
We summarise our results and draw our conclusions in sect.~\ref{sec:concl}. 
Some technical material is collected in the appendices.

\section{Cross sections and notation\label{sec:gen}}
At an $\epem$ collider a generic differential cross section 
$d\Sigma_{\epem}$ relevant to the process
\beq
e^+(P_{\lp})\,+e^-(P_{\lm})\;\longrightarrow\; X
\label{eeX}
\eeq
is written as follows (throughout this paper, we sum over all polarisation
states):
\beq
d\Sigma_{\epem}(P_{\lp},P_{\lm})=\sum_{kl}\int d\yp d\ym\,
{\cal B}_{kl}(\yp,\ym)\,d\sigma_{kl}(\yp P_{\lp},\ym P_{\lm})\,,
\label{beamstr}
\eeq
with
\beq
k\;\in\;\{\lp,\gamma\}\,,\;\;\;\;\;\;\;\;
l\;\in\;\{\lm,\gamma\}\,.
\label{klrange}
\eeq
In eq.~(\ref{eeX}) $X$ denotes a set of final-state particles, 
which does not play any role in what follows, and it is
thus ignored notation-wise.
The functions ${\cal B}_{kl}$ parametrise the beam dynamics, 
in particular beamstrahlung effects. Their forms are typically
extracted from fits to Monte Carlo simulations, and are strictly
dependent on the collider one considers; they also will play no role
in this paper. Thus, eq.~(\ref{beamstr}) is relevant here only 
insofar that it shows how the measured cross section is the incoherent
sum of four terms, associated with short-distance cross sections
$d\sigma_{kl}$ whose initial states are $\epem$, $e^+\gamma$,
$\gamma e^-$, and $\gamma\gamma$ pairs\footnote{We thus assume 
beamstrahlung conversions $e^\pm\to e^\mp$ to be negligible.
Likewise, short-distance processes initiated by $\mu$'s or $\tau$'s
are ignored.\label{ft:one}.}. As far as the latter cross sections are
concerned, in the so-called structure-function approach one writes
them in a way analogous to that of factorisation theorems in QCD, namely
(see e.g.~appendix A.1 of ref.~\cite{Beenakker:1996kt}):
\beqn
d\sigma_{kl}(\pb_k,\pb_l)&=&\sum_{ij}\int d\zp d\zm\,
\PDF{i}{k}(\zp,\mu^2,m^2)\,\PDF{j}{l}(\zm,\mu^2,m^2)
\nonumber\\*&&\phantom{\sum_{ij}\int}\times
d\hsig_{ij}(\zp\pb_k,\zm\pb_l,\mu^2,m^2)\,,
\label{factth0}
\eeqn
with $d\hsig_{ij}$ a suitably defined quantity, which we shall
specify later. By $m$ we have denoted the electron mass, $\mu$ is an 
arbitrary mass scale $\mu\gg m$, the indices $i$ and $j$ are such that
\beq
i\,,j\;\in\;\{e^+,e^-,\gamma\}\,,
\label{ijrange}
\eeq
and
\beq
\pb_q=m_q^2\,,\;\;\;\;\;\;
s=(\pb_k+\pb_l)^2\,,
\label{sdef}
\eeq
with either $m_q=m$ (when $q=e^\pm$) or $m_q=0$ (when $q=\gamma$). Two
observations are in order. Firstly, the specific values assumed by the
indices $i$ and $j$ are possibly subsets of those in eq.~(\ref{ijrange}),
that depend on both the incoming particles $k$ and $l$, and the perturbative
order one is working at. Regardless of this, and consistently with what
has been done for beamstrahlung effects, $\mu$ and $\tau$ contributions
are ignored (lifting this limitation is straightforward, but it complicates
the notation unnecessarily). Secondly, strictly speaking there is a 
kinematical inconsistency when eq.~(\ref{factth0}) is used in
eq.~(\ref{beamstr}), owing to eq.~(\ref{sdef}) and $P_{\lpm}^2=m^2$. 
This is however irrelevant, since we shall show that factorisation
formulae such as eq.~(\ref{factth0}) are best employed with massless
incoming momenta (as is done in QCD). We shall often adopt a shorthand
notation for eq.~(\ref{factth0}), which we write as follows:
\beq
d\sigma_{kl}=\sum_{ij}\PDF{i}{k}\star\PDF{j}{l}\star
d\hsig_{ij}\,.
\label{factth}
\eeq
The idea behind eq.~(\ref{factth0}) is that the cross section
for the process of eq.~(\ref{eeX}) will depend on ratios such as:
\beq
\frac{m^2}{Q^2}\,,\;\;\;\;\;\;\;\;Q^2\sim s\,,
\label{mos}
\eeq
which might spoil the perturbative ``convergence'' since $m^2\ll s$. 
One then collects the {\em universal} dependence on the ratio~(\ref{mos})
in the Initial-State Radiation (ISR) structure functions $\PDF{i}{k}$ (which, 
owing to the strict analogy of eq.~(\ref{factth0}) with the QCD case, we shall 
call PDFs henceforth), and resums large logarithmic terms (at least, 
those of collinear origin) by means of the DGLAP evolution equations:
\beq
\frac{\partial \PDF{i}{k}}{\partial\log\mu^2}=
\frac{\aem}{2\pi}P_{ij}\otimes\PDF{j}{k}\,.
\label{APeq2}
\eeq
At any given perturbative order $\aem^u$, the universal dependence of the 
cross section on $m^2/Q^2$ can be identified with logarithms of this ratio,
times the cross section of order $\aem^{u-1}$ with all such logarithms
already removed. This definition is always arbitrary to a certain extent,
which is why one introduces the scale $\mu$; note, however, that
$\mu^2\sim s$. Furthermore, from now onwards we shall understand the 
rightmost relationship in eq.~(\ref{mos}), and simply write $m^2/s$ 
to denote a generic ``small'' ratio.

While this is all fully analogous to its QCD counterpart, the key difference 
is that the PDFs $\PDF{i}{k}$ are perturbatively calculable in QED. Indeed, 
they are available in a closed form to leading logarithmic (LL) accuracy, 
which account also for soft-photon effects~\cite{Skrzypek:1990qs,
Skrzypek:1992vk,Cacciari:1992pz}.
By expanding such a closed form in a series of $\aem$, and by replacing
the result thus obtained in eq.~(\ref{factth0}), one can solve for the 
subtracted cross sections $d\hsig_{ij}$. This subtraction is mandatory,
lest one double-count the universal terms already included in the PDFs. 
When working at the NLO, one uses the following expansions:
\beqn
\PDF{i}{k}&=&\PDF{i}{k}^{[0]}+\aemotpi\,\PDF{i}{k}^{[1]}+\ord(\aem^2)\,,
\label{Gexp}
\\
d\sigma_{kl}&=&d\sigma_{kl}^{[0]}+\aemotpi\,d\sigma_{kl}^{[1]}+\ord(\aem^2)\,,
\label{sMexp}
\\
d\hsig_{ij}&=&d\hsig_{ij}^{[0]}+\aemotpi\,d\hsig_{ij}^{[1]}+\ord(\aem^2)\,.
\label{shexp}
\eeqn
By replacing these expressions into eq.~(\ref{factth}) and by solving
order by order in $\aem$, one obtains:
\beq
d\sigma_{kl}^{[0]}=\sum_{ij}\PDF{i}{k}^{[0]}\star
\PDF{j}{l}^{[0]}\star d\hsig_{ij}^{[0]}
\label{eq0}
\eeq
at $\ord(\aem^0)$ (relative to the power of $\aem$ implicit in
the Born contribution), and:
\beq
d\sigma_{kl}^{[1]}=\sum_{ij}\left(
\PDF{i}{k}^{[1]}\star\PDF{j}{l}^{[0]}\star d\hsig_{ij}^{[0]}+
\PDF{i}{k}^{[0]}\star\PDF{j}{l}^{[1]}\star d\hsig_{ij}^{[0]}+
\PDF{i}{k}^{[0]}\star\PDF{j}{l}^{[0]}\star d\hsig_{ij}^{[1]}
\right)
\label{eq1}
\eeq
at ${\cal O}(\aem)$. Whatever the form of $\PDF{i}{k}$, it must 
be such that the physically-obvious zeroth-order condition
\beq
\PDF{i}{k}^{[0]}(z,\mu^2,m^2)=\delta_{ik}\,\delta(1-z)
\label{G0sol}
\eeq
is fulfilled. Therefore, from eq.~(\ref{eq0}):
\beq
d\hsig_{kl}^{[0]}=d\sigma_{kl}^{[0]}\,.
\label{eq0sol}
\eeq
In fact, one can arrive at eq.~(\ref{eq0sol}) without employing
eq.~(\ref{G0sol}), by means of physical considerations: at the leading
order, there are no logarithmic terms, and any mass dependence (bar for
that in the flux, upon which we shall comment later) is specific to the 
process considered, and thus non-universal. Therefore, the subtracted
cross section must be identical to the cross section that emerges
directly from matrix element and phase space computations. 
As far as the ${\cal O}(\aem)$ term is concerned, from eqs.~(\ref{eq1}) 
and~(\ref{G0sol}) we obtain:
\beq
d\hsig_{kl}^{[1]}=d\sigma_{kl}^{[1]}-\sum_i\left(
\PDF{i}{k}^{[1]}\star d\hsig_{il}^{[0]}+
\PDF{i}{l}^{[1]}\star d\hsig_{ki}^{[0]}
\right)\,.
\label{eq1sol}
\eeq
In order to clarify a physics argument that stems from eqs.~(\ref{eq0sol})
and~(\ref{eq1sol}), let us introduce the following naming conventions:
\begin{itemize}
\item $d\Sigma_{\epem}$: collider-level cross section.
\item $d\sigma_{kl}$: particle-level cross section.
\item $d\hsig_{ij}$: (subtracted) parton-level cross section.
\end{itemize}
Thus, in eqs.~(\ref{beamstr}) and~(\ref{factth0}), $k$ and $l$ are
particle indices, while $i$ and $j$ are parton indices. This might
be confusing, since all of these indices assume the same values
(see eqs.~(\ref{klrange}) and~(\ref{ijrange})). The possible confusion 
ultimately originates from the fact that an $\lpm$ or $\gamma$
plays a double role in the context of the structure-function approach.
Namely, it can be one of the objects that emerge from the beamstrahlung
process, and is thus one of the incoming {\em particles} in the cross section
on the r.h.s. of eq.~(\ref{beamstr}) or the l.h.s. of eq.~(\ref{factth0}). 
But it can also be an object that emerges from the PDF evolution (i.e.~from 
ISR), and is thus one of the incoming {\em partons} in the cross section on 
the r.h.s.~of eq.~(\ref{factth0}). 

This duality is not without practical consequences. In particular,
eq.~(\ref{eq0sol}) and~(\ref{eq1sol}) must be interpreted as definitions
of the subtracted partonic cross sections {\em only} when the parton
indices coincide with the particle indices. While this is a trivial
statement as far as eq.~(\ref{eq0sol}) is concerned (since there is
actually no subtraction in that equation), it is not for eq.~(\ref{eq1sol}).
Indeed, strictly speaking eq.~(\ref{eq1sol}) should be supplemented by:
\beq
d\hsig_{ij}^{[1]}=d\sigma_{ij}^{[1]}\,,\;\;\;\;\;\;
i\ne k~~{\rm or/and}~~j\ne l\,,
\label{eq1solex}
\eeq
for any given particle indices $k$ and $l$. Thus, a partonic cross
section will have, or will not have, to be subtracted depending on
the particle cross section it contributes to. For examples, 
$d\hsig_{\gamma\lm}^{[1]}$ will be derived from $d\sigma_{\gamma\lm}^{[1]}$
according to eq.~(\ref{eq1sol}) if contributing to the particle cross
section initiated by a $\gamma\lm$ pair, and according to eq.~(\ref{eq1solex})
if contributing to a particle cross section initiated by any pair not
equal to $\gamma\lm$.

The second part of the previous statement, however, stems from a strict
interpretation of perturbative results. The difference between adopting
eq.~(\ref{eq1sol}) and~(\ref{eq1solex}) is beyond NLO (in particular,
it is of NNLO if either $i\ne k$ or $j\ne l$, and of NNNLO if both
$i\ne k$ and $j\ne l$), and therefore either choice is acceptable
from a phenomenological viewpoint.

\section{Initial conditions for PDFs: generalities\label{sec:PDF}}
As was mentioned in sect.~\ref{sec:intro}, the key difference between
QCD and QED factorisation formulae is that the PDFs that enter the latter
are fully calculable in perturbation theory. Given the fact that evolution
equations are determined perturbatively also in QCD, this is equivalent
to saying that in QED one can compute the initial conditions for the PDFs
(i.e.~their values at a given scale $\mu=\mu_0$) in a perturbative manner.
This is what we seek to do here.

There are several ways in which the PDF initial conditions can be 
determined, and we shall consider two of them here. 
\begin{itemize}
\item An approach based on explicit short-distance cross section 
computations for specific (but arbitrary) processes. 
\item An approach that exploits universal factorisation properties in
the collinear limit. 
\end{itemize}
We point out that analogous procedures have been
employed in QCD in computations relevant to heavy-quark fragmentation
functions (see ref.~\cite{Mele:1990cw} and ref.~\cite{Cacciari:2001cw},
respectively). We start with the former method, and will return to the 
latter one in sect.~\ref{sec:PDFcoll}, thereby showing that the two lead 
to identical results.

We begin by observing that by letting $m\to 0$ in the subtracted 
parton-level cross sections one drops terms suppressed by powers of 
$m^2/s$, and thus one obtains finite quantities\footnote{This might seem
not to be true if one has final-state electrons, and defines mass-sensitive
observables (e.g.~associated with bare electrons). However, these can be 
dealt with by introducing lepton fragmentation functions, so that the 
previous statement on short-distance cross sections is in fact correct 
in these cases as well.}. 
One makes the assumption, justified by the smallness of the electron mass, 
that the dropped terms are numerically negligible. Hence, we shall use
the following rule:
\begin{itemize}
\item {\em R.1: Henceforth, all short-distance partonic cross sections 
$d\hsig_{ij}$ are understood to be computed with massless electrons.}
\end{itemize}
With such a rule, eq.~(\ref{factth0}) holds up to power-suppressed terms.
It is thus useful to define the quantity:
\beq
d\bsig_{kl}=d\sigma_{kl}+
\ord\left(\!\left(\frac{m^2}{s}\right)^p\,\right)\,,\;\;\;\;\;\;\;\;
p\ge 1\,.
\label{leadingM}
\eeq
In other words, $d\bsig_{\epem}$ is obtained by computing
$d\sigma_{\epem}$ (with massive electrons), by Taylor-expanding
the result, and by keeping only the terms\footnote{For an early (two-loop)
example where a similar strategy is employed, see
refs.~\cite{Penin:2005kf,Penin:2005eh}.} which are either proportional
to a logarithm (possibly to some power) of $m^2/s$, or are independent 
of $m$. By introducing the momenta:
\beq
p_q=\lim_{m_q\to 0}\pb_q
\;\;\;\;\Longrightarrow\;\;\;\;
p_q^2=0\;\;\;\;\;\;(q=k,l)\,,
\label{pdef}
\eeq
which is trivial when $q=\gamma$, one then obtains from eq.~(\ref{factth0}) 
and rule {\em R.1}:
\beqn
d\bsig_{kl}(\pb_k,\pb_l)&=&\sum_{ij}\int d\zp d\zm\,
\PDF{i}{k}(\zp,\mu^2,m^2)\,\PDF{j}{l}(\zm,\mu^2,m^2)
\nonumber\\*&&\phantom{\sum_{ij}\int}\times
d\hsig_{ij}(\zp p_k,\zm p_l,\mu^2)\,.
\label{master0}
\eeqn
This equation is by construction a simplified version of the original
factorisation formula. However, it can also be seen as the {\em definition}
of $\PDF{i}{k}$ at a given $\mu$. Thus, eq.~(\ref{master0}) can be
solved for $\PDF{i}{k}$ after having computed $d\bsig_{kl}$
and $d\hsig_{ij}$; the solutions obtained in this way are then 
interpreted as initial conditions for PDF evolution at $\mu=\mu_0\sim m$.

We stress that this procedure, while it leads again to eqs.~(\ref{eq0})
and~(\ref{eq1}) at $\ord(\aem)$, has the opposite logic w.r.t.~that employed
in sect.~\ref{sec:gen}. Namely, rather than using a known result for
the PDFs in order to define the subtraction terms for the parton-level 
cross sections, one determines the PDFs by computing both the particle
cross section (and by keeping its leading behaviour according to
eq.~(\ref{leadingM})) and the partonic cross section. For the latter
to be sensible, a suitable zero-mass subtraction scheme must be introduced
(e.g.~$\MSb$). Therefore, as is customary in all cases beyond the LL, 
the PDFs will be specific to that scheme. Note that, by proceeding in 
this way, eq.~(\ref{G0sol}) must be the zeroth-order solution of 
eq.~(\ref{master0}). The fact that it does constitutes a first 
consistency check of the procedure advocated here.

\section{Initial conditions for PDFs through cross section 
computations\label{sec:PDFxsec}}
All of the NLO cross sections will be written by employing the FKS
subtraction~\cite{Frixione:1995ms,Frixione:1997np}. It should be clear,
however, that the final results for the PDFs will be independent of
the specific IR-subtraction method chosen in the intermediate steps
of the calculations. A recent paper where the FKS method is written so as 
to encompass both QCD and QED subtractions is ref.~\cite{Frederix:2018nkq};
in the following, we shall denote equation~(x.y) of that paper by \EWeq{x.y}.

\subsection{Determination of $\PDF{\lpm}{\lpm}$\label{sec:Gee}}
In order to perform a definite computation, we shall consider the process(es):
\beq
\epem\;\longrightarrow\;\uub(\gamma)\,,
\label{eeuu}
\eeq
with the final-state photon only present in the real-emission contributions.
The $u$ quark is taken to be massless, and we set $N_c=1$. The matrix elements
for the processes of eq.~(\ref{eeuu}) factorise the following coupling
combinations:
\beqn
&&\aem^2\,\ee^2\,\qu^2\,,
\\
&&\aem^3\,\ee^2\,\qu^2
\left[\ee^2(\ldots)+\ee\qu(\ldots)+\qu^2(\ldots)\right]\,,
\eeqn
at the LO and NLO respectively. We have denoted by $\ee=1$ and $\qu=2/3$ 
the electric charges of the positron and of the $u$ quark in units of the
positron charge. We shall use the following rule:
\begin{itemize}
\item {\em R.2: At the NLO, only contributions proportional to
$\aem^3\,\ee^4\,\qu^2$ will be kept.}
\end{itemize}

\subsubsection{Kinematics\label{sec:kin}}
Owing to the definition of $s$ in eq.~(\ref{sdef}), in the
particle c.m.~frame we have:
\beq
\pblpm=\frac{\sqrt{s}}{2}(1,0,0,\pm\beta)\,,
\;\;\;\;\;\;\;\;
\beta=\sqrt{1-\frac{4m^2}{s}}\,.
\label{hadcm}
\eeq
The corresponding massless-electron momenta introduced in eq.~(\ref{pdef})
thus read:
\beq
\plpm=\frac{\sqrt{s}}{2}(1,0,0,\pm 1)
\;\;\;\;\Longrightarrow\;\;\;\;
s=(\plp+\plm)^2\,.
\label{smass}
\eeq
Note that other forms are possible (eq.~(\ref{pdef}) can be imposed in
a different frame): what is done here has the advantage that the invariant
$s$ plays the same role in the massive and massless kinematics. The 
final result must be independent of arbitrary choices of this kind.
In the particle c.m.~frame, the processes of eq.~(\ref{eeuu}) are
assigned the following kinematic configurations:
\beqn
\lp(\pblp)+\lm(\pblm)&\longrightarrow&u(\pu)+\ub(\pub)+\gamma(\pga)\,,
\label{kinhad1}
\\
\lp(\plp)+\lm(\plm)&\longrightarrow&u(\pu)+\ub(\pub)+\gamma(\pga)\,,
\label{kinhad2}
\\
\pblp+\pblm=\plp+\plm&=&\pu+\pub+\pga\,.
\label{kinhad3}
\eeqn
In the partonic c.m.~frame we use instead:
\beqn
\lp(\klp)+\lm(\klm)&\longrightarrow&u(\ku)+\ub(\kub)+\gamma(\kga)\,,
\\
\klp+\klm&=&\ku+\kub+\kga\,.
\eeqn
By construction:
\beq
\klp=\boost(\zp\plp)\,,\;\;\;\;\;\;\;\;
\klm=\boost(\zm\plm)\,,
\eeq
with $\boost$ the longitudinal boost from the particle to the parton
c.m.~frame. We also define:
\beq
\shat=(\klp+\klm)^2=\zp\zm s\,,
\label{shat}
\eeq
whence:
\beq
\klpm=\frac{\sqrt{\shat}}{2}(1,0,0,\pm 1)\,.
\eeq
By using these equations, it is immediate to see that the rapidity
of $\boost$ is:
\beq
\ycm=\half\log\frac{\zp}{\zm}\,.
\eeq
Finally, by following the FKS procedure we parametrise the momentum of 
the outgoing photon as follows:
\beq
\kga=\frac{\sqrt{\shat}}{2}\,\xi\left(1,\vet\,\sqrt{1-y^2},y\right)\,.
\label{kgadef}
\eeq
In full analogy with that, we parametrise the photon momentum in the 
particle c.m.~frame (see eq.~(\ref{hadcm})) as follows:
\beq
\pga=\frac{\sqrt{s}}{2}\,\xi\left(1,\vet\,\sqrt{1-y^2},y\right)\,.
\label{pgadef}
\eeq
We stress that the variables $\xi$ and $y$ that appear in eq.~(\ref{kgadef})
are {\em not} the same as those that appear in eq.~(\ref{pgadef}). Since
the computations in the two frames are performed separately, no confusion
is possible, and by using the same symbols the notation is simplified.

\subsubsection{Observables\label{sec:obs}}
We can solve eq.~(\ref{master0}) for the PDFs only after having specified
how to integrate over the phase space. A complete integration obviously
would not do, because it would turn the procedure into a complicated 
inverse problem. Since we have two PDFs, we need to consider at least 
doubly-differential cross sections in order to have a purely algebraic
problem. We start by introducing the following quantities:
\beqn
T&=&\frac{1}{s}\left(\iboost(\kuub)\right)^2\,,
\label{Tdef}
\\
Y&=&\half\log
\frac{\left(\iboost(\kuub)\right)^0+\left(\iboost(\kuub)\right)^3}
{\left(\iboost(\kuub)\right)^0-\left(\iboost(\kuub)\right)^3}\,,
\label{Ydef}
\eeqn
having defined the momentum of the outgoing $u\ub$ pair:
\beq
\kuub=\ku+\kub\,.
\eeq
By using the explicit momentum parametrisations given in sect.~\ref{sec:kin}
this leads to:
\beqn
T&=&\zp\zm(1-\xi)\,,
\label{Tres}
\\
Y&=&\ycm+\half\log\frac{2-\xi(1+y)}{2-\xi(1-y)}\,.
\label{Yres}
\eeqn
The results relevant to a $2\to 2$ (i.e.~Born and virtual) kinematics
can be obtained in the $\xi\to 0$ limit from eqs.~(\ref{Tres})
and~(\ref{Yres}), namely:
\beqn
T&=&\zp\zm\,,
\label{T0res}
\\
Y&=&\ycm\,.
\label{Y0res}
\eeqn
We finally introduce the observables that will define our doubly-differential
cross section:
\beqn
\Zp&=&\sqrt{T}\exp\left(Y\right)\,,
\\
\Zm&=&\sqrt{T}\exp\left(-Y\right)\,.
\eeqn
Eqs.~(\ref{Tres}) and~(\ref{Yres}) then lead to (with the $\xi$ and $y$ 
variables of eq.~(\ref{kgadef})):
\beqn
\Zp&=&\zp\,\sqrt{(1-\xi)\,\frac{2-\xi(1+y)}{2-\xi(1-y)}}\,,
\label{Zpres}
\\
\Zm&=&\zm\,\sqrt{(1-\xi)\,\frac{2-\xi(1-y)}{2-\xi(1+y)}}\,,
\label{Zmres}
\eeqn
whose ($2\to 2$)-level counterparts are:
\beqn
\Zp&=&\zp\,,
\label{Zp0res}
\\
\Zm&=&\zm\,.
\label{Zm0res}
\eeqn
The observables $\Zp$ and $\Zm$ thus obtained are constructed by means of
parton-level momenta, and are therefore meant to be used on the r.h.s.~of
the factorisation formula. As far as the l.h.s.~of such a formula is 
concerned, the analogous definitions can be readily given by observing 
that $\iboost(\kuub)$ is the momentum of the $u\ub$ pair in the particle
c.m.~frame. Therefore, by directly using the momenta defined in that frame 
(see eqs.~(\ref{kinhad1})--(\ref{kinhad3})), one has:
\beqn
T&=&\frac{\puub^2}{s}\,,
\label{Thdef}
\\
Y&=&\half\log\frac{\puub^0+\puub^3}{\puub^0-\puub^3}\,,
\label{Yhdef}
\eeqn
where
\beq
\puub=\pu+\pub\,.
\eeq
It is immediate to see that the r.h.s.'s of eqs.~(\ref{Thdef}) 
and~(\ref{Yhdef}) can be formally read from the r.h.s.'s of
eqs.~(\ref{Tres}) and~(\ref{Yres}) with $\zp\to 1$ and $\zm\to 1$ there.
In turn, this implies that, in terms of particle-level variables,
the results for $\Zp$ and $\Zm$ will read (with the $\xi$ and $y$ 
variables of eq.~(\ref{pgadef})):
\beqn
\Zp&=&\sqrt{(1-\xi)\,\frac{2-\xi(1+y)}{2-\xi(1-y)}}\,,
\label{Zphres}
\\
\Zm&=&\sqrt{(1-\xi)\,\frac{2-\xi(1-y)}{2-\xi(1+y)}}\,,
\label{Zmhres}
\eeqn
at the real-emission level, and:
\beqn
\Zp&=&1\,,
\label{Zp0hres}
\\
\Zm&=&1\,.
\label{Zm0hres}
\eeqn
at the ($2\to 2$)-level.

These results allow one to compute the doubly-differential cross sections 
in the following way. Let $d\varsigma$ denote a contribution to either 
the particle-level or parton-level cross section, and let $d\phi$ be the
corresponding phase space, parametrised in terms of a set of variables
we generically denote by $\Omega$; in the case of parton-level contributions,
$\Omega$ includes $\zp$ and $\zm$ as well. We have:
\beq
\frac{d\varsigma}{d\Zp d\Zm}=\int\! d\phi\,
\frac{d\varsigma}{d\phi}\,\delta\!\left(\Zp-\Zp(\Omega)\right)\,
\delta\!\left(\Zm-\Zm(\Omega)\right)\,,
\label{dsdzdz}
\eeq
where the integration is performed over the whole of the phase space,
and $\Zpm(\Omega)$ are the functional forms, relevant to the given
cross section contribution $d\varsigma$, given in the r.h.s.'s of
eqs.~(\ref{Zpres}) and~(\ref{Zmres}), or eqs.~(\ref{Zp0res}) 
and~(\ref{Zm0res}), or eqs.~(\ref{Zphres}) and~(\ref{Zmhres}),
or eqs.~(\ref{Zp0hres}) and~(\ref{Zm0hres}).

\subsubsection{Matrix elements\label{sec:MEs}}
We denote by $\ampsqtwt$, $\ampsqtht$, and $\ampsqtwl$ the $2\to 2$ tree-level,
$2\to 3$ tree-level, and $2\to 2$ one-loop matrix elements, respectively; 
according to the FKS conventions, they include the flux and average factors, 
and hence must be only multiplied by the phase space in order to obtain the
corresponding cross section contributions. This notation is relevant
to massless-electron matrix elements; their massive-electron counterparts,
will be denoted by $\bampsqtwt$, $\bampsqtht$, and $\bampsqtwl$, respectively.

Note that the barred notation has been introduced in eq.~(\ref{leadingM})
to denote {\em cross sections} where only terms that are not power-suppressed 
are retained. One cannot discard power-suppressed terms directly in matrix
elements, because logarithmic terms only become apparent after phase-space
integration (in spite of the fact that all of the final-state particles in
the process we are interested in are massless). So there is an abuse of
notation here, which is however harmless since fully-massive cross
section computations will not be relevant in what follows.

It is easy to see that the matrix elements we need to consider in
our computations (see rule {\em R.2} in sect.~\ref{sec:Gee}) have
the following forms:
\beqn
\ampsqtwt&=&e^4\ee^2\qu^2\,L_{(2,0)}^{\mu\nu}\,H^{\mu\nu}(\ku,\kub)\,,
\label{ME1}
\\
\bampsqtwt&=&e^4\ee^2\qu^2\,\overline{L}_{(2,0)}^{\mu\nu}
H^{\mu\nu}(\pu,\pub)\,,
\label{ME2}
\\
\ampsqtht&=&e^6\ee^4\qu^2\,L_{(3,0)}^{\mu\nu}H^{\mu\nu}(\ku,\kub)\,,
\label{ME3}
\\
\bampsqtht&=&e^6\ee^4\qu^2\,\overline{L}_{(3,0)}^{\mu\nu}
H^{\mu\nu}(\pu,\pub)\,,
\label{ME4}
\\
\ampsqtwl&=&e^6\ee^4\qu^2\,L_{(2,1)}^{\mu\nu}H^{\mu\nu}(\ku,\kub)\,,
\label{ME5}
\\
\bampsqtwl&=&e^6\ee^4\qu^2\,\overline{L}_{(2,1)}^{\mu\nu}
H^{\mu\nu}(\pu,\pub)\,,
\label{ME6}
\eeqn
with
\beq
H^{\mu\nu}(q_1,q_2)=\frac{1}{Q^2}\,
\Tr\Big[\slashed{q}_1\gamma^\mu
\slashed{q}_2\gamma^\nu\Big]\,,
\;\;\;\;\;\;\;\;
Q=q_1+q_2\,.
\label{Hdef}
\eeq
In other words, regardless of whether one considers tree-level or one-loop
matrix elements, the $u\ub$ pair will always emerge from an $s$-channel
photon splitting. We shall call the quantity introduced in eq.~(\ref{Hdef})
the hadronic tensor.

The explicit expressions of the leptonic parts $L_{(k,l)}^{\mu\nu}$
and $\overline{L}_{(k,l)}^{\mu\nu}$ are of no interest here, and will 
not be reported. The only crucial piece of information, to be used in 
the following, is that they depend upon the momenta of the $u$ and $\ub$ 
quarks only through their sums: $\ku+\kub$ and $\pu+\pub$ at the parton 
and particle level, respectively.

\subsubsection{Phase spaces\label{sec:phis}}
Owing to the forms of both the matrix elements and of the observables
defined in sect.~\ref{sec:obs}, it is convenient to use factorised
phase-space expressions. Let us now write these in the case of the
parton kinematics in order to be definite; they will apply to the 
particle-level case as well, simply by means of a change of notation 
and $\shat\to s$ where relevant. We have:
\beq
d\phi_2=\frac{d\kuub^2}{2\pi}\,d\phi_1(\klp+\klm;\kuub)\,
d\phi_2(\kuub;\ku,\kub)\,,
\label{phi2}
\eeq
for a $2\to 2$ process, and
\beq
d\phi_3=\frac{d\kuub^2}{2\pi}\,d\phi_2(\klp+\klm;\kuub,\kga)\,
d\phi_2(\kuub;\ku,\kub)\,,
\label{phi3}
\eeq
for a $2\to 3$ process. The facts that the two rightmost two-body phase spaces
in eqs.~(\ref{phi2}) and~(\ref{phi3}) are identical to each other, that
the hadronic tensor factorises in all of the matrix elements in
eqs.~(\ref{ME1})--(\ref{ME6}), and that the observables $\Zp$ and $\Zm$
are constructed with the sum of the $u$ and $\ub$ quark momenta imply that,
at the level of cross sections, the relevant quantity will always be
the {\em integrated} hadronic tensor, namely:
\beq
h^{\mu\nu}(Q)=\int\!d\phi_2(Q;q_1,q_2)\,H^{\mu\nu}(q_1,q_2)\,.
\label{Hint}
\eeq
The tensor $h^{\mu\nu}(Q)$ can only be a linear combination of
$g^{\mu\nu}$ and $Q^\mu Q^\nu$. Furthermore, from eq.~(\ref{Hdef})
we see that $Q^\mu H^{\mu\nu}=Q^\nu H^{\mu\nu}=0$. Therefore:
\beq
h^{\mu\nu}(Q)=\left(-g^{\mu\nu}+\frac{Q^\mu Q^\nu}{Q^2}\right)h(Q^2)\,.
\label{Hint2}
\eeq
The function $h(Q^2)$ can be determined by replacing the r.h.s.~of
eq.~(\ref{Hint2}) into the l.h.s.~of eq.~(\ref{Hint}), and by contracting
both sides of the equation thus obtained with $g^{\mu\nu}$. In $d$
dimensions, the result reads as follows:
\beq
h(Q^2)=-\frac{1}{d-1}\int\!d\phi_2(Q;q_1,q_2)\,H^{\mu\nu}(q_1,q_2)\,
g^{\mu\nu}\,.
\eeq
With the explicit expression of the hadronic tensor of eq.~(\ref{Hdef}),
one obtains:
\beq
H^{\mu\nu}(q_1,q_2)\,g^{\mu\nu}=-4(1-\ep)\,,
\eeq
having set $d=4-2\ep$. With the $d$-dimensional two-body phase space:
\beq
d\phi_2(Q;q_1,q_2)=\frac{2^{-4(1-\ep)}}{\pi^{1-\ep}}
\frac{(Q^2)^{-\ep}}{\Gamma(1-\ep)}\left(1-\cos^2\theta\right)^{-\ep}
d\cos\theta\,,
\eeq
we have:
\beq
h(Q^2)=\frac{(1-\ep)(16\pi)^\ep (Q^2)^{-\ep}}
{8\sqrt{\pi}\Gamma(5/2-\ep)}\,,
\label{hQresd}
\eeq
or, in four dimensions:
\beq
h(Q^2)=\frac{1}{6\pi}\,.
\label{hQres4}
\eeq
The integration over the $u\ub$ phase space of the matrix elements of
eqs.~(\ref{ME1})--(\ref{ME6}) therefore amounts to formally replacing
$H^{\mu\nu}$ there with $h^{\mu\nu}$, the latter being given in closed
form in eqs.~(\ref{Hint2}) and~(\ref{hQresd}) (or eq.~(\ref{hQres4})),
and to multiplying by the leftover measures in eqs.~(\ref{phi2}) 
and~(\ref{phi3}). The crucial point is that this is possible also when 
considering the doubly-differential cross sections of eq.~(\ref{dsdzdz}),
because as already pointed out both the $\Zpm$ observables and the
leptonic tensors in eqs.~(\ref{ME1})--(\ref{ME6}) only depend on the
momentum of the $u\ub$ pair. More explicitly, as far as the matrix 
elements are concerned, the quantities we shall deal with are:
\beqn
\rampsq^{(k,l)}&=&\int\!d\phi_2(\kuub;\ku,\kub)\,\ampsq^{(k,l)}
\nonumber
\\&=&
e^{4+2(k-2)+2l}\ee^{2+2(k-2)+2l}\qu^2\,L_{(k,l)}^{\mu\nu}\,
h^{\mu\nu}(\ku,\kub)\,,
\label{iMdef}
\eeqn
and analogously for their massive counterparts $\rbampsq^{(k,l)}$.
For what concerns the leftover measures, they turn out to be 
(by using the momentum parametrisations introduced in 
sect.~\ref{sec:kin}):
\beqn
&&d\mu_2\equiv\frac{d\kuub^2}{2\pi}\,d\phi_1(\klp+\klm;\kuub)=1\,,
\label{mu2res}
\\
&&d\mu_3\equiv\frac{d\kuub^2}{2\pi}\,d\phi_2(\klp+\klm;\kuub,\kga)=
\frac{V_\ep\shat}{8(2\pi)^3}\,\xi^{1-2\ep} d\xi\,(1-y^2)^{-\ep} dy\,,
\phantom{aaa}
\label{mu3res}
\eeqn
with $V_\ep$ a volume factor (equal to $2\pi$ when $\ep=0$), whose explicit
form will not be needed in the following. The integration over the
transverse degrees of freedom $\vet$ of eq.~(\ref{kgadef}) has already
been carried out in eq.~(\ref{mu3res}), since there is no dependence on
them in the matrix elements. Both eqs.~(\ref{mu2res}) and~(\ref{mu3res})
are in $d$ dimensions. The former is needed in the computation of the
virtual matrix elements. Conversely, the {\em only} reason for having 
the latter for $d\ne 4$ is a peculiarity related to the computation of
the doubly-differential cross section we are interested in (see
appendix~\ref{plvsdel})  -- as far as the subtraction mechanism 
is concerned, we remind the reader that the FKS formalism allows 
one to work directly in four dimensions. 

In summary, we have established the identities:
\beqn
\int_{\phi_2(u,\ub)}\!\!\!\!\!\!d\phi_k\,\,\ampsq^{(k,l)}&=&
d\mu_k\,\,\rampsq^{(k,l)}\,,
\label{id1}
\\
\int_{\phi_2(u,\ub)}\!\!\!\!\!\!d\phi_k\,\,\bampsq^{(k,l)}&=&
d\mu_k\,\,\rbampsq^{(k,l)}\,.
\label{id2}
\eeqn
As a check, we have explicitly computed both sides of these equations in the 
Born case (i.e.~for \mbox{$(k,l)=(2,0)$}), and found that the two results 
coincide in $d$ dimensions, as expected. In the real-emission case, the
analytical computation of the l.h.s.~of eqs.~(\ref{id1}) and~(\ref{id2})
would be extremely challenging -- this is the reason why we shall instead
exploit the r.h.s.~of those equations.

The above implies that eq.~(\ref{dsdzdz}) can be re-written as follows:
\beq
\frac{d\varsigma}{d\Zp d\Zm}=\int\! d\mu_k\,
\frac{d\varsigma}{d\mu_k}\,\delta\!\left(\Zp-\Zp(\Omega)\right)\,
\delta\!\left(\Zm-\Zm(\Omega)\right)\,,
\label{dsdzdz2}
\eeq
with $d\mu_k$ the leftover measure relevant to the contribution $d\varsigma$.

\subsubsection{Short-distance cross sections\label{sec:res}}
Our master equation for the determination of the PDFs is eq.~(\ref{master0}).
By using the perturbative expansions of eqs.~(\ref{Gexp})--(\ref{shexp})
(with $d\bsig$ in place of $d\sigma$), and by considering the 
doubly-differential cross section advocated in eq.~(\ref{dsdzdz2}),
we obtain what follows. At the LO (we omit here to write the explicit 
dependence on $m$ and $\mu$, in order to shorten the notation):
\beq
\int d\zp d\zm \PDF{\lp}{\lp}^{[0]}(\zp)\PDF{\lm}{\lm}^{[0]}(\zm)\,
\frac{d\hsig_{\epem}^{[0]}(\zp\plp,\zm\plm)}{d\Zp d\Zm}=
\frac{d\bsig_{\epem}^{[0]}(\pblp,\pblm)}{d\Zp d\Zm}\,,
\label{masterLO}
\eeq
while at the NLO:
\beqn
&&\!\!\!\!\!\!\!\!
\int d\zp d\zm \left[
\PDF{\lp}{\lp}^{[1]}(\zp)\PDF{\lm}{\lm}^{[0]}(\zm)+
\PDF{\lp}{\lp}^{[0]}(\zp)\PDF{\lm}{\lm}^{[1]}(\zm)\right]
\label{masterNLO}
\\*&&\phantom{aaaaaaaaaaa}\times
\frac{d\hsig_{\epem}^{[0]}(\zp\plp,\zm\plm)}{d\Zp d\Zm}=
\nonumber\\*&&\;
\frac{d\bsig_{\epem}^{[1]}(\pblp,\pblm)}{d\Zp d\Zm}-\int d\zp d\zm
\PDF{\lp}{\lp}^{[0]}(\zp)\PDF{\lm}{\lm}^{[0]}(\zm)
\frac{d\hsig_{\epem}^{[1]}(\zp\plp,\zm\plm)}{d\Zp d\Zm}\,.
\nonumber
\eeqn
Note that the $\ord(\aem)$ terms $\PDF{\gamma}{\lpm}^{[1]}$ terms,
present in eq.~(\ref{master0}), do not contribute to eq.~(\ref{masterNLO})
since there are no $\ord(\aem^2)$ $\gamma\lpm\to u\ub$ processes.
The idea is that of an iterative solution. One first determines
$\PDF{\lpm}{\lpm}^{[0]}$ from eq.~(\ref{masterLO}), then plugs this
solution into eq.~(\ref{masterNLO}) to obtain $\PDF{\lpm}{\lpm}^{[1]}$.
A further simplification is due to the fact that, at least in QED:
\beq
\PDF{\lp}{\lp}^{[i]}=\PDF{\lm}{\lm}^{[i]}\equiv\PDF{e}{e}^{[i]}\,,
\;\;\;\;\;\;\;\;\forall\,i\,.
\eeq
As far as the short distance cross sections are concerned, 
the massless-electron case is written as follows with the
standard FKS notation:
\beq
\aemotpi\,d\hsig_{\epem}^{[1]}=d\hsig_{\epem}^{(3)}+d\hbsig_{\epem}^{(3)}+
d\hsig_{\epem}^{(2)}\,,
\label{hsig1def}
\eeq
the three terms on the r.h.s.~of this equation being the real-emission,
degenerate $(n+1)$-body\footnote{The bar that appears in this term is 
inherited from the FKS notation, and it 
has of course nothing to do with the one introduced in 
eq.~(\ref{leadingM}). As we shall point out later, the corresponding 
contribution in the massive computation vanishes.}, and $n$-body 
contributions respectively. The pattern of photon radiation in the
matrix elements we are considering is such that the introduction of
the ${\cal S}$ functions is not necessary, and the parametrisation
of eq.~(\ref{kgadef}) is sufficient to deal with the collinearity 
of $\gamma$ to both $\lp$ and $\lm$. In turn, this implies that
one can write:
\beq
d\hsig_{\epem}^{(3)}=\half\xidistr{+}
\left[\pdistr{1-y}{+}+\pdistr{1+y}{+}\right]
\Big(\xi^2(1-y^2)\rampsqtht\Big)\frac{d\mu_3}{\xi}\,.
\label{hsreal}
\eeq
The degenerate $(n+1)$-body cross section can be read directly
from \EWeq{3.22}. Finally, the $n$-body cross section is:
\beq
d\hsig_{\epem}^{(2)}=d\hsig_{\epem}^{(C,2)}+d\hsig_{\epem}^{(S,2)}+
d\hsig_{\epem}^{(V,2)}\,.
\label{h2bdy}
\eeq
The quantities on the r.h.s.~of this equation are the collinear,
soft, and finite virtual contributions respectively, whose forms can
be found in \EWeq{3.26}, \EWeq{3.28}, and \EWeq{3.29}.

The case where the electron is massive has never been explicitly 
considered in FKS before. However,
it is easy to convince oneself that a simplified version of the usual
formulae applies to this case as well. By observing that the electron
mass screens against collinear singularities (which will be ``replaced''
by \mbox{$\log m^2/s$} terms) we quickly arrive at:
\beq
\aemotpi\,d\bsig_{\epem}^{[1]}=d\bsig_{\epem}^{(3)}+d\bsig_{\epem}^{(2)}\,.
\label{bsig1def}
\eeq
Note the absence of the degenerate $(n+1)$-body cross section: this
has a pure collinear origin, and thus must vanish in the case of
massive emitters. For the same reason, the real-emission contribution
can be read from eq.~(\ref{hsreal}), by observing that the $y$-plus
distributions there behave as ordinary functions, and therefore:
\beq
\pdistr{1-y}{+}+\pdistr{1+y}{+}\;\;\longrightarrow\;\;\;
\frac{2}{1-y^2}\,\,.
\eeq
Thus:
\beq
d\bsig_{\epem}^{(3)}=\xidistr{+}
\Big(\xi^2\rbampsqtht\Big)\frac{d\mu_3}{\xi}\,.
\label{bsreal}
\eeq
Finally, the analogue of eq.~(\ref{h2bdy}) is:
\beq
d\bsig_{\epem}^{(2)}=d\bsig_{\epem}^{(S,2)}+d\bsig_{\epem}^{(V,2)}\,,
\label{b2bdy}
\eeq
where again one remarks the absence of the term of collinear origin.

As a concluding observation, we point out that we have set the standard FKS
free parameters as follows: $\xi_c=1$ and $\delta_I=2$. In the present
context, they are not particularly helpful, and keeping their analytical
dependence might significantly complicate the computations (obviously, 
the final results for the PDFs would be independent of them).

\vskip 0.3truecm
\noindent
$\blacklozenge$ {\bf Born}

\noindent
In four dimensions:
\beq
\rbampsqtwt(\pblp,\pblm)=\lim_{m\to 0}B(s,m^2)=B(s,0)\,,
\eeq
where
\beq
B(s,m^2)=\frac{\pi\alpha^2\ee^2\qu^2}{2s\beta}\,
\frac{8}{3}\left(1+\frac{2m^2}{s}\right)\,,
\label{Bfunres}
\eeq
and $\beta$ given in eq.~(\ref{hadcm}) -- hence, $\beta=1$ when $m=0$.
Therefore:
\beq
\rampsqtwt(\zp\plp,\zm\plm)=B(\shat,0)=\frac{1}{\zp\zm}\,B(s,0)\,.
\eeq
With these results, and by using eqs.~(\ref{dsdzdz2}), (\ref{Zp0hres}),
and~(\ref{Zm0hres}), the r.h.s.~of eq.~(\ref{masterLO}) reads as follows:
\beq
\frac{d\bsig_{\epem}^{[0]}(\pblp,\pblm)}{d\Zp d\Zm}=
B(s,0)\,\delta(\Zp-1)\,\delta(\Zm-1)\,.
\eeq
As far as the cross section on the l.h.s.~of eq.~(\ref{masterLO}) is
concerned, by employing eqs.~(\ref{Zp0res}) and~(\ref{Zm0res}) one obtains:
\beq
\frac{d\hsig_{\epem}^{[0]}(\zp\plp,\zm\plm)}{d\Zp d\Zm}=
\frac{1}{\zp\zm}\,B(s,0)\,\delta(\Zp-\zp)\,\delta(\Zm-\zm)\,.
\label{Bornres}
\eeq
Thus, eq.~(\ref{masterLO}) becomes:
\beq
\PDF{e}{e}^{[0]}(\Zp)\PDF{e}{e}^{[0]}(\Zm)\,\frac{1}{\Zp\Zm}\,B(s,0)=
B(s,0)\,\delta(\Zp-1)\,\delta(\Zm-1)\,,
\eeq
whence:
\beq
\PDF{e}{e}^{[0]}(z)=\delta(1-z)\,,
\label{G0sol2}
\eeq
which is the expected result (see eq.~(\ref{G0sol})).

In keeping with the iterative procedure advocated before, by employing
eq.~(\ref{G0sol2}) one can immediately simplify eq.~(\ref{masterNLO}),
which becomes:
\beqn
&&\!\!\!\!\!\!\!\!
\int d\zp \PDF{e}{e}^{[1]}(\zp)
\frac{d\hsig_{\epem}^{[0]}(\zp\plp,\plm)}{d\Zp d\Zm}+
\int d\zm \PDF{e}{e}^{[1]}(\zm)
\frac{d\hsig_{\epem}^{[0]}(\plp,\zm\plm)}{d\Zp d\Zm}
\nonumber\\*&&\phantom{aaaaaaa}
=\frac{d\bsig_{\epem}^{[1]}(\pblp,\pblm)}{d\Zp d\Zm}-
\frac{d\hsig_{\epem}^{[1]}(\plp,\plm)}{d\Zp d\Zm}\,.
\label{masterNLO2}
\eeqn
An implication of this result is that the $\ord(\aem)$ parton-level
contributions, which enter the second term of the r.h.s.~of 
eq.~(\ref{masterNLO2}), have to be computed with $\zp=\zm=1$.
This implies that $s$ will be used rather than $\shat$ (see 
eq.~(\ref{shat})) and, importantly, that eqs.~(\ref{Zpres}) 
and~(\ref{Zmres}) will coincide with eqs.~(\ref{Zphres}) 
and~(\ref{Zmhres}) (because the $\xi$ and $y$ variables of 
eq.~(\ref{kgadef}) will be the same as those of eq.~(\ref{kgadef})); 
likewise, eqs.~(\ref{Zp0res}) and~(\ref{Zm0res}) will coincide with 
eqs.~(\ref{Zp0hres}) and~(\ref{Zm0hres}). Eq.~(\ref{masterNLO2})
can be further simplified by using the Born result of eq.~(\ref{Bornres}):
\beqn
&&\!\!\!\!\!\!\!\!
\left(\frac{\PDF{e}{e}^{[1]}(\Zp)}{\Zp}\,\delta(\Zm-1)+
\frac{\PDF{e}{e}^{[1]}(\Zm)}{\Zm}\,\delta(\Zp-1)\right)B(s,0)
\nonumber\\*&&\phantom{aaaaaaa}
=\frac{d\bsig_{\epem}^{[1]}(\pblp,\pblm)}{d\Zp d\Zm}-
\frac{d\hsig_{\epem}^{[1]}(\plp,\plm)}{d\Zp d\Zm}\,.
\label{masterNLO3}
\eeqn
Equation~(\ref{masterNLO3}) has a remarkably simple dependence
on the quantity it must be solved for, $\PDF{e}{e}^{[1]}$. We shall
show in what follows that its r.h.s.~also factorises $B(s,0)$, which
is a necessary condition for the PDF initial conditions to be
process independent.

We conclude this section with a remark on the electron-mass dependence.
We see in eq.~(\ref{Bfunres}) that the mass enters in $\beta$ and in the
second term in the round brackets. While the latter is process-specific,
the former is not, being due to the flux. That renders it universal,
in spite of being of power-suppressed type when $\beta$ is Taylor expanded.
As such, it could be included in the massless computations, e.g.~by simply
assigning the massive flux to those cross sections. As far as the 
determination of the initial conditions for the PDFs are concerned,
this would not lead to any changes in the final results, and therefore
such an approach will not be pursued here.

\vskip 0.3truecm
\noindent
$\blacklozenge$ {\bf Soft and collinear $n$-body contributions}

\noindent
We consider here the non-virtual contributions to the cross sections 
of eqs.~(\ref{h2bdy}) and~(\ref{b2bdy}). The term of collinear origin
is only present in the case of massless electrons. By using \EWeq{3.27}
and \EWeq{A.16}, we obtain:
\beq
\frac{d\hsig_{\epem}^{(C,2)}}{d\mu_2}=-\aemotpi\,3\ee^2\,
\log\frac{\mu^2}{\QESt}\,B(s,0)\,,
\label{hsigC}
\eeq
with $\mu$ the factorisation scale, and $\QES$ the Ellis-Sexton scale
(an arbitrary mass scale used in FKS as a reference). For massless
electrons, the soft cross section receives a single contribution,
due to the eikonal connecting the two incoming particles. From \EWeq{3.28}:
\beq
d\hsig_{\epem}^{(S,2)}=\aemotpi\,{\cal E}_{12}^{(0,0)}\,
\ampsq_{12}^{(2,0)}\,d\phi_2\,.
\label{Shsig}
\eeq
By using \EWeq{3.12} and eq.~(A.6) of ref.~\cite{Frederix:2009yq} to obtain 
the expressions of the charge-linked Born and of the ${\cal E}_{12}^{(0,0)}$ 
eikonal factor, we obtain:
\beq
\frac{d\hsig_{\epem}^{(S,2)}}{d\mu_2}=\aemotpi\,2\ee^2
\left(\half\log^2\frac{\QESt}{s}-\frac{\pi^2}{6}\right)
B(s,0)\,.
\label{hsigS}
\eeq
For what concerns the massive-electron cross section, one must also
take into account the self-eikonals of the two incoming lines.
The analogue of eq.~(\ref{Shsig}) thus reads:
\beq
d\bsig_{\epem}^{(S,2)}=\aemotpi\left(
{\cal E}_{12}^{(m,m)}\,\bampsq_{12}^{(2,0)}
+{\cal E}_{11}^{(0,m)}\,\bampsq_{11}^{(2,0)}
+{\cal E}_{22}^{(0,m)}\,\bampsq_{22}^{(2,0)}
\right)d\phi_2\,.
\eeq
Using again the results of refs.~\cite{Frederix:2018nkq,Frederix:2009yq} 
for the definitions of the charge-linked Borns and of the eikonal factors,
we arrive at:
\beqn
\frac{d\bsig_{\epem}^{(S,2)}}{d\mu_2}&=&-\aemotpi\,2\ee^2
\Bigg(\frac{\pi^3}{3}+\half\log^2\frac{m^2}{s}-
\log\frac{m^2}{s}\log\frac{\QESt}{s}
\nonumber\\*&&\phantom{-\aemotpi\,2\ee^2}
+\log\frac{m^2}{s}-\log\frac{\QESt}{s}\Bigg)
B(s,0)\,.
\label{bsigS}
\eeqn
Finally, thanks to eqs.~(\ref{dsdzdz2}) and~(\ref{mu2res}), and bearing
in mind that all $2\to 2$ contributions are kinematically identical to
the Born one, the results of eqs.~(\ref{hsigC}), (\ref{hsigS}), 
and~(\ref{bsigS}) can be turned into the doubly-differential cross
section contributions we are interested in simply by multiplying
them by
\beq
\delta(\Zp-1)\,\delta(\Zm-1)\,.
\label{dZdZ}
\eeq

\vskip 0.3truecm
\noindent
$\blacklozenge$ {\bf Degenerate $(n+1)$-body contribution}

\noindent
This contribution is absent in the massive-electron case, while for
massless electrons it reads as follows (see \EWeq{3.22}):
\beq
d\hbsig_{\epem}^{(3)}=d\hbsig_{\lp}^{(3)}+d\hbsig_{\lm}^{(3)}\,.
\label{eedeg}
\eeq
Bearing in mind that eq.~(\ref{masterNLO2}) dictates that this
cross section be computed with $\zp=\zm=1$, and hence with $s$ 
rather than with $\shat$, one has:
\beqn
d\hbsig_{\lp}^{(3)}(\plp,\plm)&=&\aemotpi\,{\cal K}_{ee}(1-\xi)\,
d\hsig_{\epem}^{[0]}((1-\xi)\plp,\plm)\,d\xi\,,
\label{hbsigpl}
\\
d\hbsig_{\lm}^{(3)}(\plp,\plm)&=&\aemotpi\,{\cal K}_{ee}(1-\xi)\,
d\hsig_{\epem}^{[0]}(\plp,(1-\xi)\plm)\,d\xi\,,
\label{hbsigmn}
\eeqn
where, from \EWeq{3.21}:
\beqn
{\cal K}_{ee}(1-\xi)&=&\xi P_{ee}^{<}(1-\xi)
\left[\xidistr{+}\log\frac{s}{\mu^2}+2\lxidistr{+}\right]
\nonumber
\\
&-&\xi P_{ee}^{\prime<}(1-\xi)\xidistr{+}-\ee^2 K_{ee}(1-\xi)\,.
\label{Kdef}
\eeqn
The explicit expressions of the terms originating from the 
Altarelli-Parisi kernels can be found e.g.~in \EWeq{A.1}:
\beqn
P_{ee}^{<}(z)&=&\ee^2\,\frac{1+z^2}{1-z}\,,
\label{Pee}
\\
P_{ee}^{\prime<}(z)&=&-\ee^2\,(1-z)\,.
\label{Peep}
\eeqn
Conversely, the term $K_{ee}(z)$ is arbitrary, and is defined only after 
having selected a given subtraction scheme. In $\MSb$, $K_{ee}(z)\equiv 0$;
in a generic scheme, such a term is a distribution.
Finally, by using the results obtained before in the case of the Born
contribution, we have:
\beqn
&&d\hsig_{\epem}^{[0]}((1-\xi)\plp,\plm)=
d\hsig_{\epem}^{[0]}(\plp,(1-\xi)\plm)=
\nonumber\\*&&\phantom{aaaaaaaa}
B((1-\xi)s,0)=\frac{1}{1-\xi}\,B(s,0)\,.
\eeqn

In order to turn eq.~(\ref{eedeg}) into the doubly-differential
distribution we need, we have to multiply by the $\delta$'s that 
enforce the definitions of $\Zpm$, according to eq.~(\ref{dsdzdz2})
(note that $d\mu_2=1$, see eq.~(\ref{mu2res})). The relevant expressions
are those of eqs.~(\ref{Zphres}) and~(\ref{Zmhres}), computed in the
collinear configurations $y=1$ (for $d\hbsig_{\lp}^{(3)}$) and
$y=-1$ (for $d\hbsig_{\lm}^{(3)}$), which thus read:
\beqn
&&\delta\!\left(\Zp-(1-\xi)\right)\,\delta(\Zm-1)\,,
\label{ddpl}
\\
&&\delta(\Zp-1)\,\delta\!\left(\Zm-(1-\xi)\right)\,,
\label{ddmn}
\eeqn
respectively. Owing to the symmetry between the two contributions,
let us consider only $d\hbsig_{\lp}^{(3)}$ for the time being -- the
results for $d\hbsig_{\lm}^{(3)}$ will then follow straightforwardly.
It is tempting to use the first of the $\delta$'s in eq.~(\ref{ddpl})
to get rid of the $\xi$ integration in eq.~(\ref{hbsigpl}). Although
we shall show that this ultimately leads to the correct result, it
is a dangerous procedure (being formally incorrect), because ${\cal K}_{ee}$
is not a regular function, but a distribution. This renders the formal
use of the $\delta$, as e.g.~in the following expression
\beq
\xidistr{+}\delta\!\left(\Zp-(1-\xi)\right)d\xi=
\pdistr{1-\Zp}{+}
\label{xipdelZ}
\eeq
an ill-defined mathematical procedure. The interested reader can
find in appendix~\ref{plvsdel} the proof that eq.~(\ref{xipdelZ}),
and its analogues stemming from eq.~(\ref{Kdef}), is indeed an identity.
Here, we limit ourselves to reporting the final result, which as was
anticipated is the following:
\beq
\frac{d\hbsig_{\lp}^{(3)}(\plp,\plm)}{d\Zp d\Zm}=\aemotpi\,
\frac{1}{\Zp}\,{\cal K}_{ee}(\Zp)\,\delta(\Zm-1)\,B(s,0)\,,
\label{hbsigplres}
\eeq
where by ${\cal K}_{ee}(\Zp)$ we understand the formal replacement of 
\mbox{$1-\xi$} by $\Zp$ on the r.h.s.~of eq.~(\ref{Kdef}) (thus obtaining 
distributions such as that on the r.h.s.~of eq.~(\ref{xipdelZ})).
Likewise:
\beq
\frac{d\hbsig_{\lm}^{(3)}(\plp,\plm)}{d\Zp d\Zm}=\aemotpi\,
\frac{1}{\Zm}\,{\cal K}_{ee}(\Zm)\,\delta(\Zp-1)\,B(s,0)\,.
\label{hbsigmnres}
\eeq

\vskip 0.3truecm
\noindent
$\blacklozenge$ {\bf Real $(n+1)$-body contributions}

\noindent
In the massless-electron case, the relevant cross section stems from 
eqs.~(\ref{hsreal}) and~(\ref{dsdzdz2}):
\beqn
\frac{d\hsig_{\epem}^{(3)}}{d\Zp d\Zm}&=&\half\xidistr{+}
\left[\pdistr{1-y}{+}+\pdistr{1+y}{+}\right]
\Big(\xi^2(1-y^2)\rampsqtht\Big)
\nonumber\\&\times&
\delta\!\left(\Zp-\Zp(\xi,y)\right)\,
\delta\!\left(\Zm-\Zm(\xi,y)\right)
\frac{d\mu_3}{\xi}\,.
\label{hsrealdZZ}
\eeqn
Conversely, in the massive-electron case we need to consider 
eqs.~(\ref{bsreal}) and~(\ref{dsdzdz2}):
\beq
\frac{d\bsig_{\epem}^{(3)}}{d\Zp d\Zm}=\xidistr{+}
\Big(\xi^2\rbampsqtht\Big)\,
\delta\!\left(\Zp-\Zp(\xi,y)\right)\,
\delta\!\left(\Zm-\Zm(\xi,y)\right)\,
\frac{d\mu_3}{\xi}\,.
\label{bsrealdZZ}
\eeq
The direct computation of the matrix elements that enter  
eq.~(\ref{bsrealdZZ}) leads to the following result:
\beq
\xi^2\rbampsqtht(\xi,y)=\bm_0^{(3,0)}(\xi,y)+
\frac{m^2}{s}\,\bm_2^{(3,0)}(\xi,y)+
\frac{m^4}{s^2}\,\bm_4^{(3,0)}(\xi,y)\,,
\eeq
where:
\beqn
\bm_0^{(3,0)}&=&
\left[\frac{1-y}{(1-\beta y)^2}+
\frac{1+y}{(1+\beta y)^2}\right]C_{00}
\nonumber\\*&+&
\left[\frac{(1-\beta)^2}{(1-\beta y)^2}+
\frac{(1-\beta)^2}{(1+\beta y)^2}\right]C_{01}\,,
\label{bm30}
\\
\bm_2^{(3,0)}&=&
\left[\frac{1}{(1-\beta y)^2}+
\frac{1}{(1+\beta y)^2}\right]C_{21}\,,
\label{bm32}
\\
\bm_4^{(3,0)}&=&
\left[\frac{1}{(1-\beta y)^2}+
\frac{1}{(1+\beta y)^2}\right]C_{41}\,.
\label{bm34}
\eeqn
We remind the reader that $\beta$ has been defined in eq.~(\ref{hadcm}),
which implies:
\beq
\frac{m^2}{s}=\frac{1}{4}\left(1-\beta^2\right)\,.
\label{mosvsbe}
\eeq
The coefficients $C_{ij}$ are functions of $\xi$, $y$, and $\beta$, whose
crucial property is that of being regular for any values of $\xi$ and $y$,
for any given $\beta$ (including, importantly, $\beta=1$, which corresponds 
to the massless limit). Their explicit forms are not relevant here, except
for that of $C_{00}$, that reads as follows:
\beq
C_{00}=\frac{1}{2\beta}\,m^{(3,0)}\,,
\label{P00expr}
\eeq
where $m^{(3,0)}$ is the reduced matrix element that enter 
eq.~(\ref{hsrealdZZ}):
\beq
m^{(3,0)}(\xi,y)=\xi^2(1-y^2)\rampsqtht(\xi,y)\,,
\eeq
which again is a regular function, finite for any values of $\xi$ and $y$.

The direct computation of eq.~(\ref{hsrealdZZ}) is easily doable, while
that of eq.~(\ref{bsrealdZZ}) is much more demanding. Fortunately, neither
is actually necessary (nevertheless, the former has been carried out for 
checking purposes; its result is reported in appendix~\ref{plvsdel}).
This is because, as is shown by our master formula for determining the 
PDFs, eq.~(\ref{masterNLO3}), only the difference between the massive-
and massless-electron real-emission cross sections is needed. Such a
difference can be computed by starting from the following observation.
Let $\rho$ be a small positive dimensionless parameter:
\beq
0<\rho\ll 1\,.
\label{rocond}
\eeq
One can prove the following identities among distributions:
\beqn
\frac{1\pm y}{\left(1\pm\frac{y}{1+\rho}\right)^2}&=&
-\left(\log\frac{\rho}{2}+1\right)\delta(1\pm y)
+\pdistr{1\pm y}{+}+\ord(\rho)\,,
\label{ybe1}
\\
\frac{1}{\left(1\pm\frac{y}{1+\rho}\right)^2}&=&
\left(\frac{1}{\rho}+\frac{3}{2}\right)\delta(1\pm y)
+\left(\log\frac{\rho}{2}+1\right)\delta^\prime(1\pm y)
\nonumber\\*&+&
\sum_{j=2}^\infty \frac{(-1)^j\,2^{j-1}}{(j-1)\,j!}\,\delta^{(j)}(1-y)
+\ord(\rho)\,.\phantom{aaaa}
\label{ybe2}
\eeqn
These can be used in eqs.~(\ref{bm30})--(\ref{bm34}), by setting:
\beq
\rho=\frac{1-\beta}{\beta}\equiv 2\,\frac{m^2}{s}
+\ord\left(\!\frac{m^4}{s^2}\!\right)\,.
\label{rovsbe}
\eeq
Note that the rightmost side of eq.~(\ref{rovsbe}) is indeed consistent
with eq.~(\ref{rocond}) in the region $m^2\ll s$ of interest to us.
By doing so, one observes the following facts:
\begin{enumerate}
\item Thanks to eq.~(\ref{P00expr}), the contributions emerging from
the plus distribution in eq.~(\ref{ybe1}) become identical (bar for an
irrelevant $1/\beta$ prefactor) to those of the massless-electron case,
and account for all of them.
\item Because of eq.~(\ref{mosvsbe}), only the $1/\rho$ term of
eq.~(\ref{ybe2}) gives a non-zero contribution to $\bm_2^{(3,0)}$.
\item For the same reason, the contribution due to $\bm_4^{(3,0)}$
can be discarded.
\end{enumerate}
The bottom line is that the difference we are interested in is
proportional to $\delta(1\pm y)$ factors, and thus becomes 
essentially straightforward to compute. The final result is:
\beq
\frac{d\bsig_{\epem}^{(3)}}{d\Zp d\Zm}=
\ee^2\,\aemotpi\left(\bar{l}_0\log\frac{m^2}{s}+
\Delta\bar{f}_0\right)B(s,0)+\frac{d\hsig_{\epem}^{(3)}}{d\Zp d\Zm}\,,
\label{bsrealdZZfin}
\eeq
where:
\beqn
\bar{l}_0&=&-\frac{1+\Zp^2}{\Zp}\pdistr{1-\Zp}{+}\delta(\Zm-1)
+\Big[\Zp\,\longleftrightarrow\,\Zm\Big]\,,
\label{bl0res}
\\
\Delta\bar{f}_0&=&-2\pdistr{1-\Zp}{+}\delta(\Zm-1)
+\Big[\Zp\,\longleftrightarrow\,\Zm\Big]\,.
\label{bf0res}
\eeqn

\newpage
\vskip 0.3truecm
\noindent
$\blacklozenge$ {\bf Virtual contributions}

\noindent
Because of rule {\em R.2} in sect.~\ref{sec:Gee}, the only contributions
to $L_{(2,1)}^{\mu\nu}$ and $\overline{L}_{(2,1)}^{\mu\nu}$ in
eqs.~(\ref{ME5}) and~(\ref{ME6}) are due to a 
triangle graph associated with the $\epem\gamma$ vertex, and to a fermion 
bubble in the photon propagator. According to the standard FKS conventions,
one works in the CDR scheme, and thus all momenta are in $d$ dimensions;
as a consequence of that, the integrated hadronic tensor in eq.~(\ref{iMdef}) 
is the one in $d$ dimensions, whose scalar part is given in eq.~(\ref{hQresd}).

The computations are straightforward, and have been carried out with
both Macsyma and Mathematica, including analytical integral reduction.
The results thus obtained have then been renormalised in $\MSb$, by decoupling 
the massive electron where relevant. The renormalised matrix elements
still contain IR poles and, in order to extract the finite part in the CDR 
scheme as demanded by the FKS procedure, we employ \EWeq{3.29}--\EWeq{3.32}. 
Apart from obtaining in this way the sought finite parts, we also check 
that the IR singularity structures agree with what is expected from the
general FKS formulae. Note that, in CDR, the Born and charge-linked Born
matrix elements that constitute part of the pole residues must be
evaluated in $d$ dimensions. The results read as follows:
\beqn
\frac{d\hsig_{\epem}^{(V,2)}}{d\Zp d\Zm}&=&\ee^2\,\aemotpi\Bigg[
-\frac{92}{9}+\pi^2-\frac{4}{3}\log\frac{\mu^2}{s}
-3\log\frac{\QESt}{s}
\nonumber\\*&&\phantom{\ee^2\aemotpi\Bigg[}
-\log^2\frac{\QESt}{s}
\Bigg]B(s,0)\,\delta(\Zp-1)\,\delta(\Zm-1)\,,
\label{hsigVZZ}
\\
\frac{d\bsig_{\epem}^{(V,2)}}{d\Zp d\Zm}&=&\ee^2\,\aemotpi\Bigg[
-\frac{56}{9}+\frac{4}{3}\pi^2+\log^2\frac{m^2}{s}
-\frac{1}{3}\left(7+6\log\frac{\QESt}{s}\right)\log\frac{m^2}{s}
\nonumber\\*&&\phantom{\ee^2\aemotpi\Bigg[}
-2\log\frac{\QESt}{s}
\Bigg]B(s,0)\,\delta(\Zp-1)\,\delta(\Zm-1)\,.\phantom{aa}
\label{bsigVZZ}
\eeqn
One finally needs to note that, owing to the decoupling of the massive
lepton in the renormalisation prescription adopted here, the two $\aem$'s
in eqs.~(\ref{hsigVZZ}) and~(\ref{bsigVZZ}) are not the same. This issue
can be addressed by a simple change of scheme, using the QED analogue
of e.g.~eq.~(3.5) of ref.~\cite{Cacciari:1998it} or eq.~(2.1) of
ref.~\cite{Cacciari:2001td}. Taking the formal QCD$\to$QED translation
into account, that for the aforementioned equations entails
\beq
T_{\sss\rm F}\;\longrightarrow\;\ee^2\,,
\label{TF}
\eeq
we have:
\beq
\aem_{\rm dec}=\aem_{\MSb}-\frac{1}{3\pi}\,\ee^2\aem^2\log\frac{\mu^2}{m^2}\,.
\label{avsa}
\eeq
Eq.~(\ref{avsa}) results in a relative $\ord(\aem)$ term when applied
at the Born level, while it induces terms beyond NLO accuracy when applied
at the NLO level. Only the former must therefore be taken into account, 
which implies adding the following contribution:
\beq
\frac{d\bsig_{\epem}^{({\rm scheme})}}{d\Zp d\Zm}=
-\frac{4}{3}\,\ee^2\,\aemotpi\,\log\frac{\mu^2}{m^2}\,
B(s,0)\,\delta(\Zp-1)\,\delta(\Zm-1)
\label{bsigsch}
\eeq
to the massive-electron cross section.

\subsubsection{Final result\label{sec:PDFres}}
We can now put together all of the results obtained thus far, and
obtain the PDFs. We point out that the $\ord(\aem^0)$ contribution 
has already been determined, and is given in eq.~(\ref{G0sol2}). The 
$\ord(\aem)$ contribution is obtained by solving eq.~(\ref{masterNLO3})
for $\PDF{e}{e}^{[1]}$. We have explicitly computed all of the terms that 
enter the r.h.s.~of eq.~(\ref{masterNLO3}); for the reader's convenience,
we summarise them here. In the massless-electron case these are given 
in eqs.~(\ref{hsigC}), (\ref{hsigS}) (both of these multiplied by the
$\delta$'s of eq.~(\ref{dZdZ}) in order to obtain the doubly-differential
cross section), (\ref{hbsigplres}), (\ref{hbsigmnres}), (\ref{hsrealdZZfin}),
and~(\ref{hsigVZZ}). With massive electrons, we need to use
eqs.~(\ref{bsigS}) (multiplied by the $\delta$'s of eq.~(\ref{dZdZ})),
(\ref{bsrealdZZfin}), (\ref{bsigVZZ}), and~(\ref{bsigsch}). Some trivial
algebra leads to the following result:
\beqn
&&\frac{d\bsig_{\epem}^{[1]}(\pblp,\pblm)}{d\Zp d\Zm}-
\frac{d\hsig_{\epem}^{[1]}(\plp,\plm)}{d\Zp d\Zm}=\ee^2\Bigg[
\left(2+\frac{3}{2}\log\frac{\mu^2}{m^2}\right)\delta(\Zp-1)
\nonumber\\*&&\phantom{aaaaaaa}
+\frac{1}{\Zp}\Bigg((1+\Zp^2)\pdistr{1-\Zp}{+}\log\frac{\mu^2}{m^2}
+K_{ee}(\Zp)
\nonumber\\*&&\phantom{aaaaaaa\frac{1}{\Zp}\Bigg(}
-(1+\Zp^2)\left[2\lppdistr{1-\Zp}{+}+\pdistr{1-\Zp}{+}\right]
\Bigg)\Bigg]
\nonumber\\*&&\phantom{aaa}
\times B(s,0)\,\delta(\Zm-1)
+\Big[\Zp\,\longleftrightarrow\,\Zm\Big]\,,
\eeqn
whose structure is a perfect match of the l.h.s.~of eq.~(\ref{masterNLO3}). 
The solution for $\PDF{e}{e}^{[1]}$ can be given in a compact form by
exploiting the identities:
\beqn
\frac{1+z^2}{(1-z)_+}&=&\left(\frac{1+z^2}{1-z}\right)_+
-\frac{3}{2}\,\delta(1-z)\,,
\label{pluscan}
\\
(1+z^2)\lppdistr{1-z}{+}&=&
\left(\frac{1+z^2}{1-z}\log(1-z)\right)_+
+\frac{7}{4}\,\delta(1-z)\,,\phantom{aaaa}
\label{Lpluscan}
\eeqn
which lead to:
\beq
\PDF{e}{e}^{[1]}(z,\mu^2)=\ee^2\left[\frac{1+z^2}{1-z}\left(
\log\frac{\mu^2}{m^2}-2\log(1-z)-1\right)\right]_+ +\ee^2 K_{ee}(z)\,.
\label{G1sol2}
\eeq
According to the discussion given in sects.~\ref{sec:gen} and~\ref{sec:PDF},
eqs.~(\ref{G0sol2}) and~(\ref{G1sol2}) constitute, for any choice 
$\mu=\mu_0\sim m$, the initial
conditions for the the evolution equation~(\ref{APeq2}) of the electron
PDFs. On the other hand, they could also be seen as a way to obtain the 
leading terms of an NLO massive-electron cross section, whose origin can
be traced back to ISR off an electron/positron leg, by performing only 
short-distance massless-electron cross section computations, according
to eq.~(\ref{master0}). The latter interpretation requires $\mu\sim s$, 
and thus explicitly exposes large logarithms of the electron mass. We
remark that similar considerations apply to photon PDFs and to 
fragmentation functions.

\subsection{Determination of $\PDF{\gamma}{\lpm}$\label{sec:Gge}}
We start by observing that:
\beqn
\PDF{\gamma}{\lp}^{[i]}=\PDF{\gamma}{\lm}^{[i]}
&\equiv&\PDF{\gamma}{e}^{[i]}\,,
\;\;\;\;\;\;\;\;\forall\,i\,,
\\
\PDF{\gamma}{e}^{[0]}(z)&=&0\,,
\label{Gg0sol}
\eeqn
and therefore that the only unknown quantity is $\PDF{\gamma}{e}^{[1]}$,
which is what we want to compute. In order to do that, we shall consider 
the process(es):
\beq
e^-\mu^-\;\longrightarrow\;e^-\mu^-(\gamma)\,,
\label{emem}
\eeq
which factorise the following coupling-constant combinations:
\beqn
&&\aem^2\,\ee^2\,\emu^2\,,
\\
&&\aem^3\,\ee^2\,\emu^2
\left[\ee^2(\ldots)+\ee\emu(\ldots)+\emu^2(\ldots)\right]\,,
\eeqn
at the LO and NLO respectively, with $\emu$ the electric charge of the
muon\footnote{Henceforth, we shall omit the charge indices for the process
of eq.~(\ref{emem}) in order to simplify the notation; note that the 
computation would be identical had we considered positrons and
positively-charged muons.} in units of the positron charge. We shall 
perform the calculation by imposing the following rules:
\begin{itemize}
\item {\em R.3: At the NLO, only contributions proportional to
$\aem^3\,\ee^2\,\emu^4$ will be kept.}
\item {\em R.4: The muon is treated as a massless particle.}
\end{itemize}
The first consequence of rule {\em R.4} is that, also for the particle-level
cross sections, we shall regard the muon as a bare particle, whose collinear
singularities are subtracted in an arbitrary scheme (e.g.~$\MSb$). This
is not equivalent to saying that those cross sections are physical, since
they are not. However, it is irrelevant: owing to collinear factorisation, 
any contribution due to branchings off the muon leg will drop out from
$\PDF{\gamma}{e}^{[1]}$. By taking this fact and rule {\em R.3}
into account, from the factorisation formulae we obtain:
\beq
d\bsig_{\emtembgb}^{[1]}=d\hsig_{\emtembgb}^{[1]}+
\PDF{\gamma}{e}^{[1]}\star d\hsig_{\gmtgm}^{[0]}\,,
\label{eq2}
\eeq
which we shall solve for $\PDF{\gamma}{e}^{[1]}$. Therefore, 
thanks to the simplifying conditions induced by rules {\em R.3} and 
{\em R.4}, eq.~(\ref{eq2}) has the same form as eq.~(\ref{masterNLO3}).
For both of these equations, the quantity ultimately relevant to the 
computation is the difference 
between the particle and the partonic cross sections, evaluated with the 
same four-momentum configurations, and which do not require any further
integration over initial-state degrees of freedom. It is remarkable
that eq.~(\ref{eq2}), at variance with eq.~(\ref{masterNLO3}), can 
be employed without introducing the $\Zpm$ observables.

\subsubsection{Kinematics\label{sec:gkin}}
The massive- and massless-electron kinematics configurations
are denoted as follows in the particle c.m.~frame:
\beqn
e(\pb_1)+\mu(p_2)&\longrightarrow&e(\pb_3)+\mu(p_4)+\gamma(p_5)\,,
\label{gkinhad1}
\\
e(p_1)+\mu(p_2)&\longrightarrow&e(p_3)+\mu(p_4)+\gamma(p_5)\,,
\label{gkinhad2}
\eeqn
Because of eq.~(\ref{eq2}), we shall not need to parametrise explicitly
the kinematics in the partonic c.m.~frame. In keeping with eqs.~(\ref{sdef})
and~(\ref{smass}), we set:
\beq
s=(\pb_1+p_2)^2=(p_1+p_2)^2\,,
\eeq
whence:
\beqn
\pb_1&=&\frac{\sqrt{s}}{2}\left(1+\mtosl,0,0,1-\mtosl\right)\,,
\label{pb1}
\\
p_2&=&\frac{\sqrt{s}}{2}\!\left(1-\mtosl\right)(1,0,0,-1)\,,
\label{p2}
\eeqn
for the kinematic configuration of eq.~(\ref{gkinhad1}), and:
\beqn
p_1&=&\frac{\sqrt{s}}{2}(1,0,0,1)\,,
\\
p_2&=&\frac{\sqrt{s}}{2}(1,0,0,-1)\,,
\eeqn
for the configuration of eq.~(\ref{gkinhad2}). In principle, all of the
final-state partons may play the role of the FKS parton; in practice,
however, we shall show that only the case where such a parton coincides
with the outgoing electron is of interest to us. Clearly, when 
the electron mass is different from zero, the usual momentum
parametrisation must be generalised, which we do in the following way:
\beq
\pb_3=\frac{\sqrt{s}}{2}\,\xi\left(1,\vet\,\sqrt{1-y^2}\,\beta_3,
y\,\beta_3\right)\,,
\;\;\;\;\;\;\;\;
\beta_3=\sqrt{1-\frac{4m^2}{s\,\xi^2}}\,.
\label{pb3def}
\eeq
The massless-electron limit of eq.~(\ref{pb3def}) coincides with the usual 
parametrisation of the momentum of the FKS parton:
\beq
p_3=\frac{\sqrt{s}}{2}\,\xi\left(1,\vet\,\sqrt{1-y^2},y\right)\,.
\label{p3def}
\eeq

\subsubsection{Short-distance cross sections\label{sec:xsecs}}
The $\ord(\aem^3)$ cross sections that appear in eq.~(\ref{eq2})
are written as in eq.~(\ref{hsig1def}):
\beqn
\aemotpi\,d\hsig_{\emtembgb}^{[1]}=d\hsig_{\emtemg}^{(3)}+
d\hbsig_{\emtembgb}^{(3)}+d\hsig_{\emtem}^{(2)}\,,
\label{hsig2def}
\\
\aemotpi\,d\bsig_{\emtembgb}^{[1]}=d\bsig_{\emtemg}^{(3)}+
d\bbsig_{\emtembgb}^{(3)}+d\bsig_{\emtem}^{(2)}\,.
\label{bsig2def}
\eeqn
The massive-electron cross section of eq.~(\ref{bsig2def}) differs from
that of eq.~(\ref{bsig1def}) owing to the presence of the second term
on the r.h.s.~of the former equation. Such a term is non-null here 
owing to the masslessness of the incoming muon.
We point out that, although the $\emtem$ Born cross section does not
appear in eq.~(\ref{eq2}), it will still factor out of several of the
quantities on the r.h.s.'s of eqs.~(\ref{hsig2def}) and~(\ref{bsig2def}).
We shall not need its explicit expression, and we limit ourselves
to remarking that, thanks to eq.~(\ref{eq0sol}), one has:
\beq
d\hsig_{\emtem}^{[0]}=d\bsig_{\emtem}^{[0]}\,.
\label{born2}
\eeq

\vskip 0.3truecm
\noindent
$\blacklozenge$ {\bf $2\to 2$ contributions}

\begin{figure}[htb]
\begin{center}
  \includegraphics[trim=0 30 100 400,clip,scale=0.4]{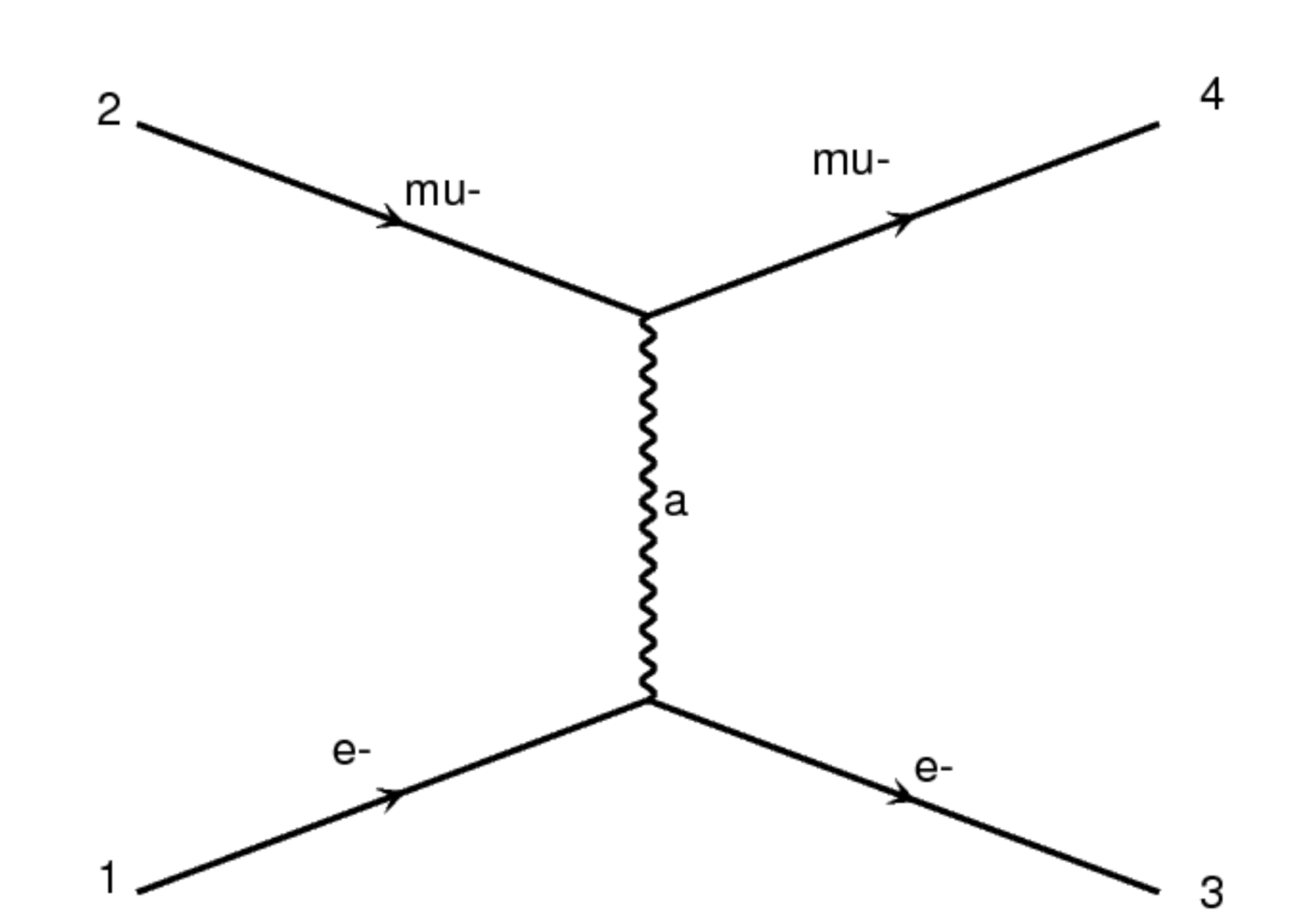}
\end{center}
\caption{\label{fig:emem} 
Born diagram relevant to the process $\emtem$.
}
\end{figure}

\noindent
The $2\to 2$ cross sections on the r.h.s.'s of eqs.~(\ref{hsig2def}) 
and~(\ref{bsig2def}) have the general form already used in 
eq.~(\ref{h2bdy}):
\beqn
d\hsig_{\emtem}^{(2)}&=&d\hsig_{\emtem}^{(C,2)}+d\hsig_{\emtem}^{(S,2)}+
d\hsig_{\emtem}^{(V,2)}\,,
\label{h2bdy2}
\\
d\bsig_{\emtem}^{(2)}&=&d\bsig_{\emtem}^{(C,2)}+d\bsig_{\emtem}^{(S,2)}+
d\bsig_{\emtem}^{(V,2)}\,.
\label{b2bdy2}
\eeqn
The three contributions on the r.h.s.'s of these equations are obtained
by ``dressing'' the Born diagram of fig.~\ref{fig:emem} with collinear
factors, eikonal factors, and one-loop corrections, respectively.
As was the case for the degenerate $(n+1)$-body contribution to 
eq.~(\ref{bsig2def}), the first term on the r.h.s.~of eq.~(\ref{b2bdy2}) 
is non-null because the muon is massless.
From \EWeq{3.27} one obtains\footnote{The explicit forms of the charge
factors $C(a)$, $\gamma(a)$, and $\gamma^\prime(a)$ can be found
e.g.~in ref.~\cite{Frederix:2018nkq}.}:
\beq
d\hsig_{\emtem}^{(C,2)}=\aemotpi\left[
-\gamma(\mu)\log\frac{\mu^2}{\QESt}
+\gamma^\prime(\mu)
-\gamma(\mu)\log\frac{s}{\QESt}
\right]d\hsig_{\emtem}^{[0]}\,.
\label{hsigC2}
\eeq
Note that terms proportional to $\gamma(e)$ and $\gamma^\prime(e)$
are absent, since their contributions violate rule {\em R.3}. By
taking eq.~(\ref{born2}) into account, we also have:
\beqn
d\bsig_{\emtem}^{(C,2)}&=&d\hsig_{\emtem}^{(C,2)}
+\frac{\aem}{\pi}\Bigg[-\gamma(\mu)\log\frac{2E_4}{\sqrt{s}}
+C(\mu)\Bigg(\log\frac{2E_2}{\sqrt{s}}\log\frac{2E_2\sqrt{s}}{\QESt}
\nonumber\\&&\phantom{aaaaaaaaa}
+\log\frac{s}{\QESt}\log\frac{2E_4}{\sqrt{s}}
+\log^2\frac{2E_4}{\sqrt{s}}\Bigg)\Bigg]d\hsig_{\emtem}^{[0]}\,.
\label{bsigC}
\eeqn
A few remarks are in order here. Firstly, terms proportional to $\gamma(e)$ 
and $\gamma^\prime(e)$ would not appear on the r.h.s.~of eq.~(\ref{bsigC}) 
even without rule {\em R.3}, because the electron is massive. Secondly, 
the logarithmic terms are non zero since (see eq.~(\ref{p2})):
\beq
E_2=E_4=\frac{\sqrt{s}}{2}\!\left(1-\mtos\right)\,.
\label{E2E4}
\eeq
Thirdly, while the terms that feature $E_4$ can indeed be read directly
from \EWeq{3.27}, the one that features $E_2$ cannot. This is because
in ref.~\cite{Frederix:2018nkq} it was assumed, as is customary in
FKS, that both of the incoming particles are massless, and therefore that
their energies are equal to $\sqrt{s}/2$. It can be easily shown
that one can relax the latter condition without 
having to change anything in the subtraction procedure, at the price 
of introducing an extra term, which is the one we are discussing here. 
By using eq.~(\ref{E2E4}) and by expanding in $m^2$ we finally arrive at:
\beq
d\bsig_{\emtem}^{(C,2)}=d\hsig_{\emtem}^{(C,2)}+\ordmos\,.
\label{CmC}
\eeq
We now turn to the soft cross sections. Eq.~({\bf I}.3.28) and
rule {\em R.3} imply that $d\hsig_{\emtem}^{(S,2)}$ receives a single
contribution, from the $(24)$ eikonal. The same is true for
$d\bsig_{\emtem}^{(S,2)}$. By taking into account eq.~(\ref{born2}) 
and the fact that charged-link Borns are proportional to the Born,
we obtain:
\beq
d\bsig_{\emtem}^{(S,2)}=d\hsig_{\emtem}^{(S,2)}+\ordmos\,.
\label{SmS}
\eeq
Finally, owing to rule {\em R.3} the sole virtual diagrams one must
consider are those obtained by dressing the graph of fig.~\ref{fig:emem} 
with a triangle on the $\mu\mu\gamma$ vertex,
and with a muon bubble on the $t$-channel propagator. Therefore,
any electron-mass dependence may only arise when contracting the 
electron-line tensor with the one-loop subamplitudes (which are 
mass independent). Thus, no terms proportional to $\log m^2/s$ 
can possibly arise in $d\bsig_{\emtem}^{(V,2)}$, which implies:
\beq
d\bsig_{\emtem}^{(V,2)}=d\hsig_{\emtem}^{(V,2)}+\ordmos\,.
\label{VmV}
\eeq
From eqs.~(\ref{CmC})--(\ref{VmV}) we thus see that none of the $2\to 2$ 
cross sections contribute to eq.~(\ref{eq2}).

\vskip 0.3truecm
\noindent
$\blacklozenge$ {\bf Degenerate $(n+1)$-body contribution}

\noindent
In the massless-electron case, there is one contribution per incoming
leg (one further contribution associated with each leg is discarded 
because of rule {\em R.3}). Explicitly:
\beq
d\hbsig_{\emtembgb}^{(3)}=\aemotpi\Big(
{\cal K}_{\gamma e}\star d\hsig_{\gmtgm}^{[0]}+
{\cal K}_{\mu\mu}\star d\hsig_{\emtem}^{[0]}\Big)\,.
\label{emdeg1}
\eeq
The $\star$ operator in the two terms on the r.h.s.~of this equation
stands for the convolution which is explicitly written in eqs.~(\ref{hbsigpl})
and~(\ref{hbsigmn}), respectively. The factors ${\cal K}_{\gamma e}$
and ${\cal K}_{\mu\mu}$ can be obtained from ${\cal K}_{ee}$ of
eq.~(\ref{Kdef}) by replacing the Altarelli-Parisi kernels and the 
scheme-defining function with those relevant to the branchings
$e\to\gamma e$ and $\mu\to\mu\gamma$, respectively. We have:
\beqn
P_{\mu\mu}^{<}(z)&=&\frac{\emu^2}{\ee^2}\,P_{ee}^{<}(z)\,,
\\
P_{\mu\mu}^{\prime<}(z)&=&\frac{\emu^2}{\ee^2}\,P_{ee}^{\prime<}(z)\,,
\eeqn
and, from \EWeq{A.2}:
\beqn
P_{\gamma e}^{<}(z)&=&P_{ee}^{<}(1-z)=\ee^2\,\frac{1+(1-z)^2}{z}\,,
\\
P_{\gamma e}^{\prime<}(z)&=&P_{ee}^{\prime<}(1-z)=-\ee^2\,z\,.
\eeqn
The massive-electron analogue of eq.~(\ref{emdeg1}) reads instead:
\beq
d\bbsig_{\emtembgb}^{(3)}=\aemotpi\,
\bar{{\cal K}}_{\mu\mu}\star d\hsig_{\emtem}^{[0]}\,,
\label{emdeg2}
\eeq
the only contribution being due to the incoming massless-muon leg,
and with:
\beq
\bar{{\cal K}}_{\mu\mu}={\cal K}_{\mu\mu}+
2\log\frac{2E_2}{\sqrt{s}}\,\xi P_{\mu\mu}^{<}(1-\xi)\xidistr{+}\,.
\label{bKdef}
\eeq
The second term on the r.h.s.~of eq.~(\ref{bKdef}) is due to $E_2$
not being equal to $\sqrt{s}/2$ (see eq.~(\ref{E2E4})), as its $2\to 2$
analogues on the r.h.s.~of eq.~(\ref{bsigC}). Exactly as in the $2\to 2$
case, it thus amounts to a power-suppressed term, and therefore:
\beq
d\bbsig_{\emtembgb}^{(3)}-d\hbsig_{\emtembgb}^{(3)}=
-\aemotpi\,{\cal K}_{\gamma e}\star d\hsig_{\gmtgm}^{[0]}
+\ordmos\,.
\label{degdiff}
\eeq
We finally point out that the plus distributions
contained in ${\cal K}_{\gamma e}$ can be replaced with ordinary functions.
In fact, the subtraction at $\xi=0$ corresponds to a soft electron and 
therefore is not associated with any singularity. Indeed, from the explicit
expression of ${\cal K}_{\gamma e}$ we see that the residue at that
point vanishes.

\vskip 0.3truecm
\noindent
$\blacklozenge$ {\bf Real contributions}

\begin{figure}[thb]
  \begin{center}
  \includegraphics[trim=0 40 0 280,clip,scale=0.65]{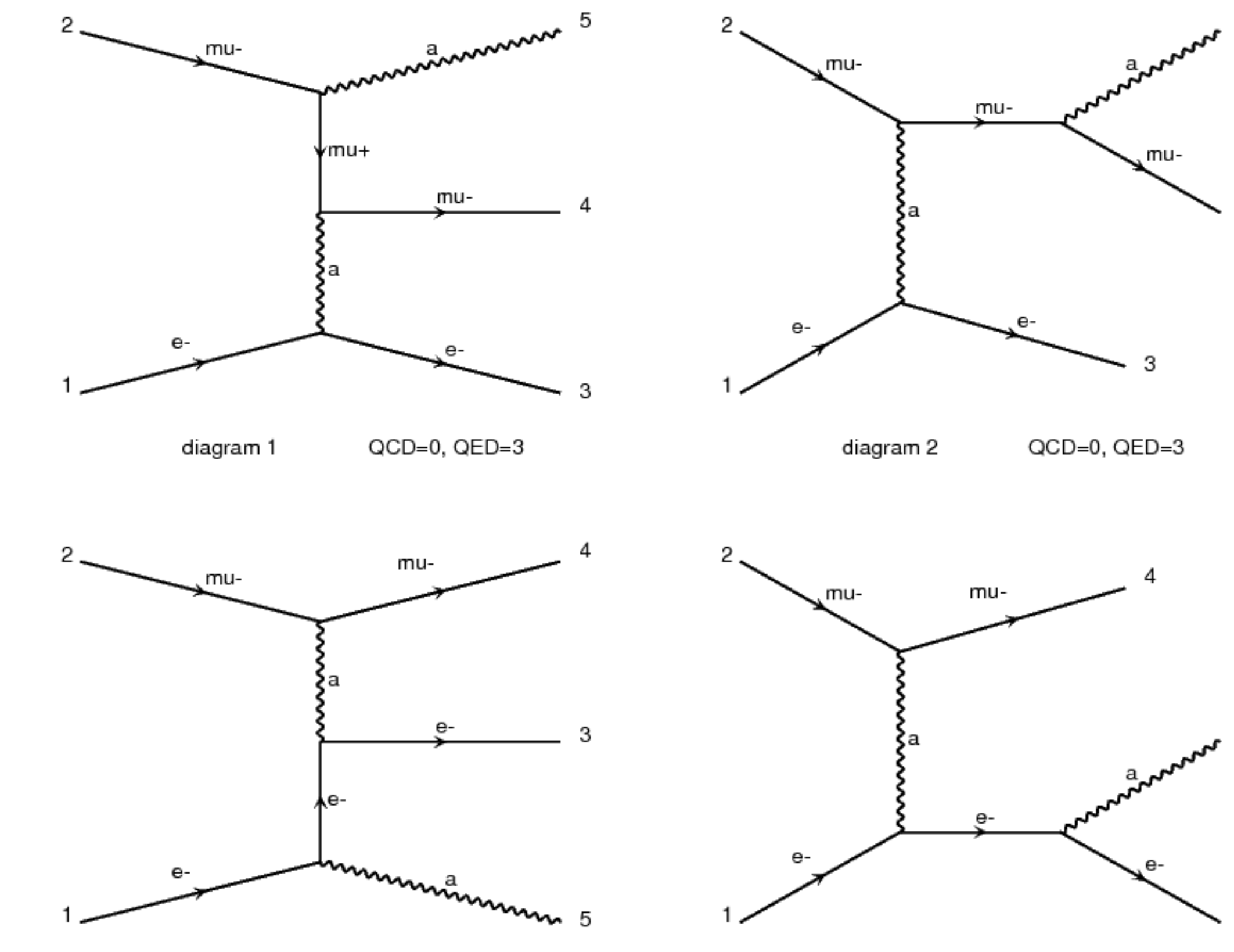}
\caption{\label{fig:ememg} 
Real-emission diagrams relevant to the process $\emtemg$.
}
  \end{center}
\end{figure}

\begin{figure}[thb]
  \begin{center}
  \includegraphics[trim=0 20 300 530,clip,scale=0.5]{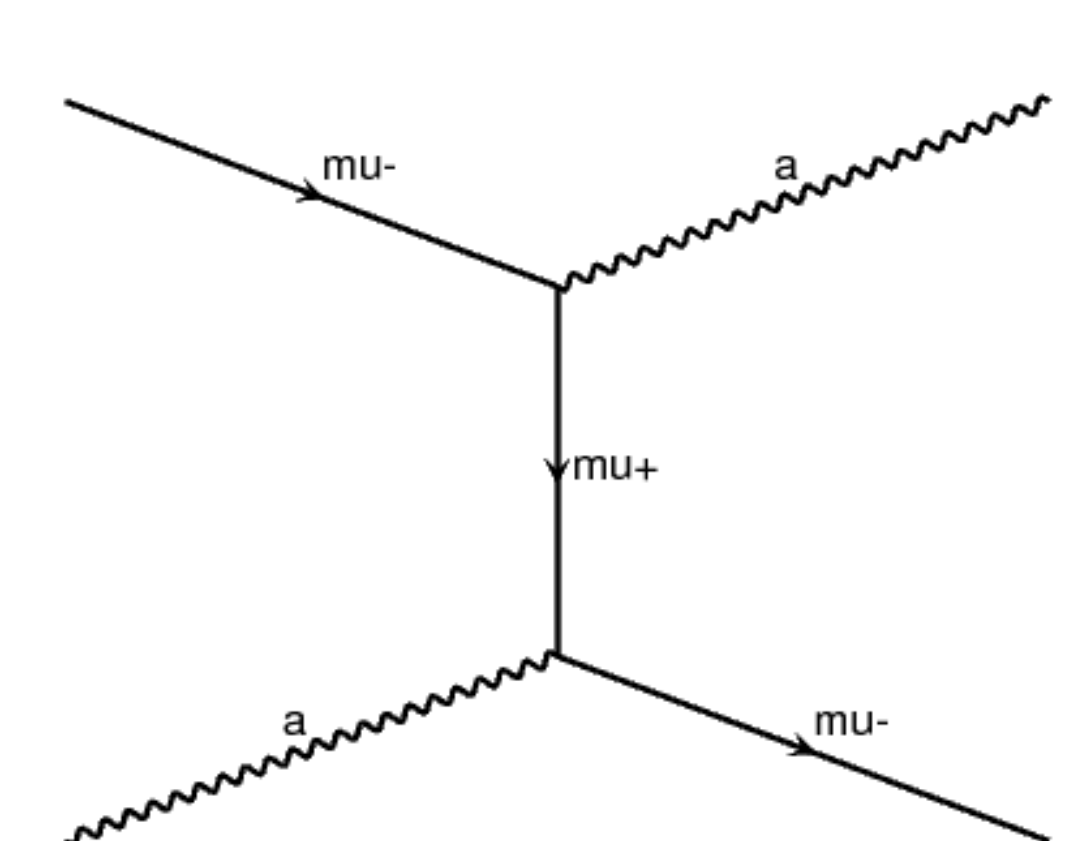}
$\phantom{aaaa}$
  \includegraphics[trim=0 20 300 530,clip,scale=0.5]{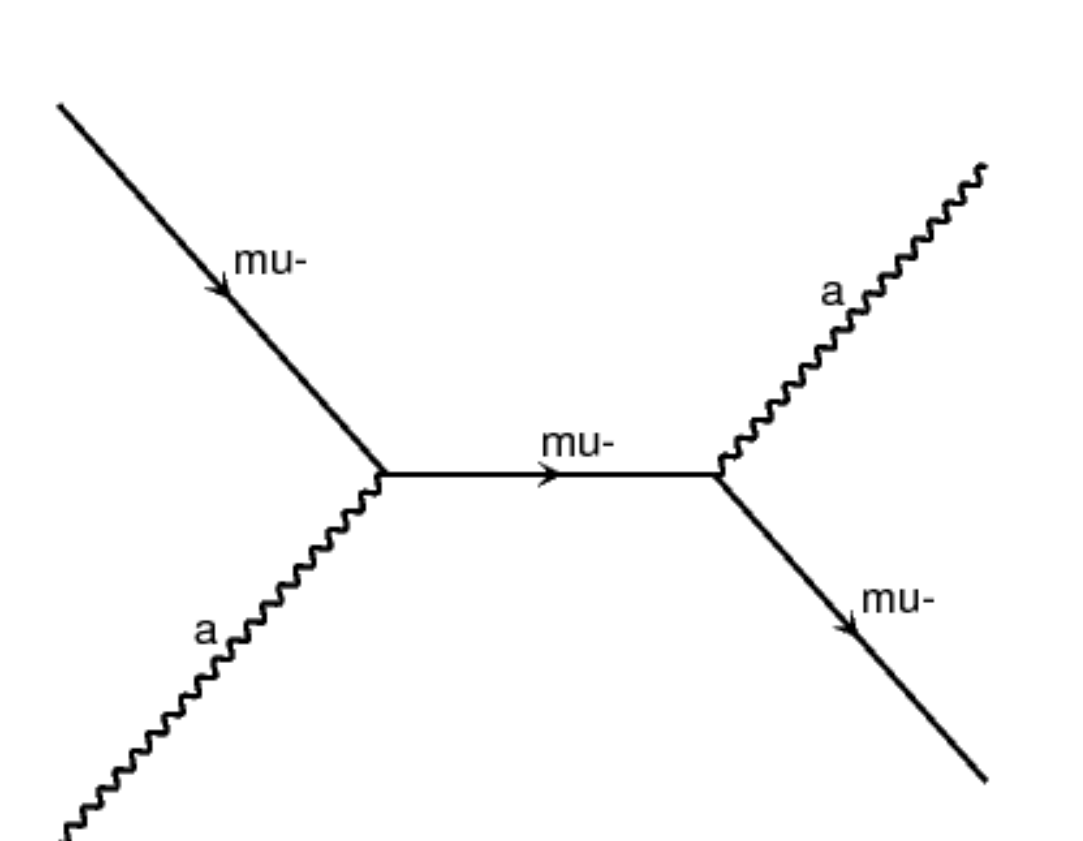}
\caption{\label{fig:gmgm} 
Diagrams relevant to the process $\gmtgm$. These are also the
sub-graphs obtained from diagrams \#1 and \#2 of fig~\ref{fig:ememg} 
by stripping them of the $e\gamma e$ current.
}
  \end{center}
\end{figure}

\noindent
The diagrams that contribute to the real-emission cross sections
are depicted in fig.~\ref{fig:ememg}. The amplitudes associated with 
diagrams \#1 and \#2 (top row) are proportional to $\ee\emu^2$, while those 
associated with diagrams \#3 and \#4 (bottom row) are proportional to 
$\ee^2\emu$. Therefore, owing to rule {\em R.3}, we can consider only 
the former two.
The first implication of this is that the set of the FKS pairs
(see e.g.~ref.~\cite{Frederix:2009yq}) is given by:
\beq
{\cal P}_{\rm FKS}=\Big\{(5,2),(5,4),(3,1)\Big\}\,.
\eeq
The sector $(5,2)$ captures soft-photon and $\mu\!\parallel\!\gamma$
initial-state collinear configurations. Likewise, the sector $(5,4)$ 
is associated with soft-photon and $\mu\!\parallel\!\gamma$
final-state collinear configurations. Finally, the sector $(3,1)$
singles out the $e\!\parallel\! e$ initial-state collinear configurations.
Therefore, it is only the latter sector in the massive-electron case that 
may induce $\log m^2/s$ terms, or constant terms not present in
the massless-electron case, since such terms emerge exclusively from 
quasi-collinear kinematics where potential singularities are screened 
by the electron mass. This implies that:
\beqn
d\bsig_{\emtemg,52}^{(3)}-d\hsig_{\emtemg,52}^{(3)}&=&\ordmos\,,
\label{bsig352}
\\
d\bsig_{\emtemg,54}^{(3)}-d\hsig_{\emtemg,54}^{(3)}&=&\ordmos\,,
\label{bsig354}
\eeqn
having denoted by $d\hsig_{\emtemg,ij}^{(3)}$ and $d\bsig_{\emtemg,ij}^{(3)}$
the contribution to $d\hsig_{\emtemg}^{(3)}$ and $d\bsig_{\emtemg}^{(3)}$,
respectively, proportional to the $\Sfunij$ function relevant to the
$(i,j)$ FKS sector\footnote{We have implicitly assumed that the $\Sfun$
functions for the massive- and massless-electron cross sections are
identical. On top of being allowed by the freedom in the definition of
the $\Sfun$ functions, it can be shown that this is actually the most 
convenient thing to do.}. 
One is thus left with computing the contributions due to the $(3,1)$ sector. 
In keeping with what was done in sect.~\ref{sec:res}, we start by considering 
the massive-electron case, whose cross section we write as follows:
\beq
d\bsig_{\emtemg,31}^{(3)}=\Sfun_{31}\bampsqtht d\bar{\phi}_3\,.
\label{bsig331}
\eeq
Note that since the electron is massive this FKS sector does not
require any subtraction. The matrix element in eq.~(\ref{bsig331})
reads as follows:
\beq
\bampsqtht=\frac{1}{4}\frac{1}{2s}\,e^2\ee^2\,\frac{1}{q^4}\,
\overline{E}^{\rho\sigma}W^{\rho\sigma}\,,
\label{mat331}
\eeq
having denoted by
\beq
q=\pb_1-\pb_3
\eeq
the momentum flowing in the $t$-channel photon of diagrams \#1 and \#2 
of fig.~\ref{fig:ememg}. The tensor associated with the $e\gamma e$ 
current is:
\beq
\overline{E}^{\rho\sigma}=
\Tr\Big[(\slashed{\pb}_3-m)\gamma^\rho
(\slashed{\pb}_1-m)\gamma^\sigma\Big]=
2q^2g^{\rho\sigma}+
4\pb_1^\rho\pb_3^\sigma+
4\pb_3^\rho\pb_1^\sigma\,.
\eeq
The contribution of diagrams \#1 and \#2 stripped of the $e\gamma e$
current has been denoted by $W^{\rho\sigma}$, and it reads as follows:
\beq
W^{\rho\sigma}=\sum_{\rm pol}\Big(\amp_1^{\rho\mu}+\amp_2^{\rho\mu}\Big)
\Big(\amp_1^{\sigma\nu}+\amp_2^{\sigma\nu}\Big)^*
\left(-g^{\mu\nu}\right)\,,
\label{Wdef}
\eeq
where $\amp_i^{\rho\mu}$ denotes the amplitude associated with the
diagrams of fig.~\ref{fig:gmgm}, with the following momenta and
photon-index assignments (see eq.~(\ref{gkinhad2})):
\beq
\gamma^{\rho}(q)+\mu(p_2)\;\longrightarrow\;
\gamma^{\mu}(p_5)+\mu(p_4)\,.
\label{gmgmmom}
\eeq
Note that since $q^2\ne 0$, these amplitudes are off-shell.
In eq.~(\ref{Wdef}), the sum runs over the muon polarization, and
the factor $-g^{\mu\nu}$ is the result of summing over the 
outgoing-photon polarization.

For what concerns the three-body phase space that appears in
eq.~(\ref{bsig331}), we decompose it as follows:
\beq
d\bar{\phi}_3=d\tilde{\phi}_2\,d\phi_1(3)\,,
\eeq
where
\beq
d\tilde{\phi}_2=(2\pi)^4\delta\left(q+p_2-p_4-p_5\right)
\frac{d^3 p_4}{(2\pi)^3 2p_4^0}\frac{d^3 p_5}{(2\pi)^3 2p_5^0}\,,
\label{tphi2}
\eeq
and, by using eq.~(\ref{pb3def}):
\beq
d\phi_1(3)=\frac{d^3 \pb_3}{(2\pi)^3 2\pb_3^0}=
\frac{1}{2(2\pi)^3}\,\frac{s\,\beta_3^3}{4}\,\xi d\xi\,dy\,d\varphi\,,
\eeq
with $\varphi$ an azimuthal angle that parametrises the transverse degrees 
of freedom $\vet$.

We now understand the integration of the r.h.s.~of eq.~(\ref{bsig331})
over $d\tilde{\phi}_2$, having fixed the variables of $d\phi_1(3)$. Thus,
we regard the cross sections that enter eq.~(\ref{eq2}) as inclusive 
quantities in $d\tilde{\phi}_2$ which, as was anticipated, still allows 
us to solve that equation for $\PDF{\gamma}{e}^{[1]}$. In practice, we 
do not need to perform such an integration explicitly. We only exploit it 
in the following way: at the inclusive level, the tensor $W^{\rho\sigma}$ 
may only depend on $q$ and $p_2$. Furthermore, this tensor is transverse
in $q$. Therefore, we can write it in the following form:
\beq
W^{\rho\sigma}=
T_1^{\rho\sigma}W_1+T_2^{\rho\sigma}W_2\,,
\label{Wdec}
\eeq
where
\beqn
T_1^{\rho\sigma}&=&-g^{\rho\sigma}+\frac{q^\rho q^\sigma}{q^2}\,,
\label{T1def}
\\
T_2^{\rho\sigma}&=&-\frac{q^2}{(p_2\mydot q)^2}\,
\left(p_2^\rho-\frac{p_2\mydot q}{q^2}q^\rho\right)
\left(p_2^\sigma-\frac{p_2\mydot q}{q^2}q^\sigma\right)\,,
\label{T2def}
\eeqn
and the scalar functions $W_i$ depend only on $q^2$ and $p_2\mydot q$.
Given $W^{\rho\sigma}$ (e.g.~as computed from amplitudes like those on 
the r.h.s.~of eq.~(\ref{Wdef})), these functions can be projected out 
as follows:
\beqn
W_1&=&-\half\left(g^{\rho\sigma}+\frac{q^2}{(p_2\mydot q)^2}\,
p_2^\rho p_2^\sigma\right)W^{\rho\sigma}\,,
\label{W1proj}
\\
W_2&=&-\half\left(g^{\rho\sigma}+3\,\frac{q^2}{(p_2\mydot q)^2}\,
p_2^\rho p_2^\sigma\right)W^{\rho\sigma}\,.
\label{W2proj}
\eeqn
By using the decomposition of eq.~(\ref{Wdec}) in eq.~(\ref{mat331}),
we see that two relevant quantities are:
\beqn
\overline{E}^{\rho\sigma}T_1^{\rho\sigma}&=&
-4\left(q^2+2m^2\right)\,,
\label{ET1}
\\
\overline{E}^{\rho\sigma}T_2^{\rho\sigma}&=&
-8q^2\frac{F}{(1-F)^2}\,,
\label{ET2}
\eeqn
where
\beq
F=\half\,\xi\left(1+\beta_3 y\right)\,.
\eeq
Therefore:
\beq
\bampsqtht=-\frac{1}{4}\frac{1}{2s}\,e^2\ee^2\,\frac{1}{q^4}
\left[4\left(q^2+2m^2\right)W_1
+8q^2\frac{F}{(1-F)^2}\,W_2\right]\,.
\label{mat331exp}
\eeq
Now observe that:
\beqn
p_2\mydot q&=&\half\,(s-m^2)\,(1-F)\,,
\\
q^2&=&-\frac{s}{2}\left[\xi\left(1+\mtos\right)-\frac{4m^2}{s}\right]
\left(1-\frac{y}{1+\rho}\right)\,,
\label{q2y}
\eeqn
where
\beq
\rho=\frac{\xi(1-\beta_3)-(4-\xi(1+\beta_3))\mtosl}
{\xi\beta_3(1-\mtosl)}=
2\,\frac{(1-\xi)^2}{\xi^2}\,\mtos
+\ord\left(\frac{m^4}{s^2}\right)\,.
\eeq
In the collinear ($y\to 1$) and massless-electron limit, we have $q^2\to 0$ 
as expected, and \mbox{$p_2\mydot q\to s(1-\xi)/2$}. Therefore, by regarding 
the r.h.s.~of eq.~(\ref{mat331exp}) as an expansion in $q^2$, the dominant
terms in the collinear limit will be proportional to $1/q^4$ and $1/q^2$.
However, one must not exclude the possibility that either (or both) of
the residue(s) of these terms is (are) equal to zero. In order to check
this, we must study the behaviour of the $W_i$ functions in the $q^2\to 0$ 
limit. By means of an explicit computation we obtain what follows:
\beq
W_1=W_2=\half B_{\gmtgm}+\ord(q^2)\,,
\label{W1W2res}
\eeq
where $B_{\gmtgm}$ is the matrix element squared and summed over all
polarisations of the process $\gamma\mu\to\gamma\mu$. The momenta
are assigned as in eq.~(\ref{gmgmmom}) with $q^2=0$, which implies
\mbox{$q=(1-\xi)p_1$}. In other words:
\beq
\ampsqtwt_{\gmtgm}\big((1-\xi)p_1,p_2\big)=\frac{1}{4}\,
\frac{1}{2(1-\xi)s}\,B_{\gmtgm}\,.
\label{matgmgm}
\eeq
Equation~(\ref{W1W2res}) implies that both the $1/q^4$ and the 
$1/q^2$ appear non-trivially in eq.~(\ref{mat331exp}). By using 
the expression of $q^2$ given in eq.~(\ref{q2y}), the former term
is dealt with by means of the identity of eq.~(\ref{ybe2}). As far
as the $1/q^2$ term is concerned, another distribution identity
is necessary, namely:
\beq
\frac{1}{1-\frac{y}{1+\rho}}=
-\log\frac{\rho}{2}\,\,\delta(1-y)
+\pdistr{1-y}{+}+\ord(\rho)\,.
\label{ybe3}
\eeq
After employing eqs.~(\ref{ybe2}) and~(\ref{ybe3}) in eq.~(\ref{mat331exp}),
one can safely set $m=0$ in the latter. The resulting cross section of
eq.~(\ref{bsig331}) is the sum of two terms: one proportional to
$\delta(1-y)$, and one proportional to $1/(1-y)_+$. It is easy to
see that the latter is nothing but the real-emission contribution 
of the $(3,1)$ FKS sector to the massless-electron cross section.
The former term must be explicitly computed, which is straightforward
owing to the simplifications induced by $\delta(1-y)$. In particular,
such a $\delta$ will allow us to use eq.~(\ref{W1W2res}) with all
of the $\ord(q^2)$ terms identically equal to zero; in other words, the 
quantity on the l.h.s.~of eq.~(\ref{matgmgm}) will naturally emerge.
We also have:
\beqn
\delta(1-y)\,d\tilde{\phi}_2&=&\delta(1-y)\,d\phi_2\big((1-\xi)s\big)\,,
\label{p2vsp2}
\\
\delta(1-y)\,F&=&\delta(1-y)\,\xi+\ordmos\,,
\\
\delta(1-y)\,\Sfun_{31}&=&\delta(1-y)\,,
\eeqn
where the $d\phi_2$ on the r.h.s.~of eq.~(\ref{p2vsp2}) is the two-body
massless phase space with incoming energy squared equal to $(1-\xi)s$.
By putting all this together, some trivial algebra leads us to the
final result:
\beq
d\bsig_{\emtemg,31}^{(3)}-d\hsig_{\emtemg,31}^{(3)}=
\aemotpi\,{\cal Q}_{\gamma e}\star d\hsig_{\gmtgm}^{[0]}
+\ordmos\,,
\label{bsig331res}
\eeq
where
\beq
{\cal Q}_{\gamma e}(1-\xi)=-\ee^2\left[\frac{1+\xi^2}{1-\xi}
\left(2\log\frac{1-\xi}{\xi}+\log\mtos\right)+\frac{2\xi}{1-\xi}\right]\,.
\label{Qres}
\eeq

\subsubsection{Final result\label{sec:PDF2res}}
We can now use eqs.~(\ref{CmC}), (\ref{SmS}), (\ref{VmV}), (\ref{degdiff}),
(\ref{bsig352}), (\ref{bsig354}), and~(\ref{bsig331res}) in 
eq.~(\ref{eq2}). We obtain:
\beq
\PDF{\gamma}{e}^{[1]}(z)=-{\cal K}_{\gamma e}(z)
+{\cal Q}_{\gamma e}(z)\,.
\label{PDFKQ}
\eeq
Therefore:
\beq
\PDF{\gamma}{e}^{[1]}(z,\mu^2)=\ee^2\,\frac{1+(1-z)^2}{z}\left(
\log\frac{\mu^2}{m^2}-2\log z-1\right) +\ee^2 K_{\gamma e}(z)\,.
\label{Ggesol2}
\eeq
By comparing this result with that of eq.~(\ref{G1sol2}), we see that:
\beq
\PDF{\gamma}{e}^{[1]}(z)-\ee^2 K_{\gamma e}(z)=
\PDF{e}{e}^{[1]}(1-z)-\ee^2 K_{ee}(1-z)\,,
\;\;\;\;\;\;\;\;
z<1\,.
\eeq
We thus find that the $z\to 1-z$ symmetry property characteristic of the
unsubtracted Altarelli-Parisi kernels $P_{ee}$ and $P_{\gamma e}$ also
holds for the unsubtracted QED PDFs at $\ord(\aem)$, up to scheme-change 
terms. This is nicely consistent with the idea of collinear factorisation. 
It would be tempting to conclude that it is also an indication that:
\beq
K_{\gamma e}(z)=K_{ee}(1-z)\,,
\;\;\;\;\;\;\;\;
z<1\,.
\label{KvsK}
\eeq
While eq.~(\ref{KvsK}) constitutes a valid choice (in particular, it is
trivially true in $\MSb$), the current computation cannot possibly suggest
that it is the {\em only} suitable choice.

We finally point out that the procedure followed above implies that 
the quantity:
\beq
\aemotpi\,{\cal Q}_{\gamma e}(z)=\ee^2\,\aemotpi\,\left[
\frac{1+(1-z)^2}{z}\log\frac{(1-z)^2 s}{z^2 m^2}-\frac{2(1-z)}{z}
\right]
\label{QasWW}
\eeq
is the kernel that collects all universal (i.e.~independent of the 
process and of the subtraction scheme) purely collinear terms of
the $e(1)\to\gamma(z)e(1-z)$ splitting. As such, it must be equivalent
to the Weizsaecker-Williams (WW) function. This is indeed the case:
eq.~(\ref{QasWW}) coincides e.g.~with eq.~(27) of ref.~\cite{Frixione:1993yw},
provided that in the latter: {\em a)} one identifies $E^2\theta_c^2$ with
$s$ (which can be understood from the physical viewpoint by noting that
$E^2\theta_c^2$ is the ``large'' scale of the problem); {\em b)} one
neglects all terms which are suppressed by powers of $m^2$ (some of these
have been kept in ref.~\cite{Frixione:1993yw}, while they are strictly
discarded here). Equation~(\ref{PDFKQ}) thus clarifies the relationship
between the WW function and $\PDF{\gamma}{e}^{[1]}(z)$, which do not
coincide (as is clear in general from the fact that the PDF depends
on an arbitrary mass scale ($\mu$) and is defined in an arbitrary scheme).

\section{Initial conditions for PDFs through collinear 
factorisation\label{sec:PDFcoll}}
The results obtained in sects.~\ref{sec:Gee} and~\ref{sec:Gge}
have, among other things, three key properties, which we now
enumerate. One of these will provide a way to compute the PDFs
in an alternative and simpler way w.r.t.~what was done so far
(namely through collinear-factorisation properties),
thus allowing a quick determination of the {\em photon} PDFs,
$\PDF{e}{\gamma}$ and $\PDF{\gamma}{\gamma}$.

~

\noindent
$\bullet$ {\em Consistency with evolution equations}. Electron and
photon PDFs must obey the DGLAP equations~(\ref{APeq2}). By employing 
eqs.~(\ref{Gexp}) and~(\ref{G0sol}), in the case of the electron PDFs 
eq.~(\ref{APeq2}) reads:
\beq
\frac{\partial\PDF{i}{e}^{[1]}}{\partial\log\mu^2}=
P_{ie}^{[0]}\,,
\label{APeq3}
\eeq
where we have added (w.r.t.~to the notation used in the rest of this
paper) an index $[0]$ to the lowest-order splitting kernel, in order
to avoid confusions. Equation~(\ref{APeq3}) is manifestly fulfilled by 
$\PDF{e}{e}^{[1]}$ of eq.~(\ref{G1sol2}) and by $\PDF{\gamma}{e}^{[1]}$
of eq.~(\ref{Ggesol2}) -- note that the scheme-change terms must be, by 
construction, independent of $\mu$. In the case of the photon PDFs, 
eq.~(\ref{APeq2}) leads to:
\beq
\frac{\partial\PDF{i}{\gamma}^{[1]}}{\partial\log\mu^2}=
P_{i\gamma}^{[0]}\,.
\label{APeq4}
\eeq

~

\noindent
$\bullet$ {\em Momentum and charge conservation}. In pure QED, the condition 
of momentum conservation in the branchings of a fermion $f^\pm$ with
electric charge $\pm 1$ reads, in terms of its PDFs (by ignoring
other fermion families):
\beq
1=\int_0^1 dz\,z\Big(\PDF{f^\pm}{f^\pm}(z)+
\PDF{\gamma}{f^\pm}(z)+\PDF{f^\mp}{f^\pm}(z)\Big)\,.
\label{emomc}
\eeq
By taking $f$ to be an electron or a positron and by using eq.~(\ref{G0sol}), 
at the first non-trivial order in $\aem$ eq.~(\ref{emomc}) becomes:
\beq
0=\int_0^1 dz\,z\Big(\PDF{e}{e}^{[1]}(z)+
\PDF{\gamma}{e}^{[1]}(z)\Big)\,.
\label{emomc1}
\eeq
With the explicit results of eqs.~(\ref{G1sol2}) and~(\ref{Ggesol2})
we can verify that the terms not related to the change of scheme fulfill
eq.~(\ref{emomc1}), and therefore that the latter can be turned into
a condition that the scheme-change terms must fulfill as well:
\beq
\int_0^1 dz\,z\Big(K_{ee}(z)+K_{\gamma e}(z)\Big)=0\,.
\label{momcKe}
\eeq
The analogue of eq.~(\ref{emomc}) for the branchings of a photon reads:
\beq
1=\int_0^1 dz\,z\Big(\PDF{\gamma}{\gamma}(z)+
\PDF{f^+}{\gamma}(z)+\PDF{f^-}{\gamma}(z)\Big)\,.
\label{gamomc}
\eeq
By taking eq.~(\ref{G0sol}) into account, and by considering only
the lightest lepton family as representative of charged fermions,
at the first non-trivial order in $\aem$ eq.~(\ref{gamomc}) becomes:
\beq
0=\int_0^1 dz\,z\Big(\PDF{\gamma}{\gamma}^{[1]}(z)+
2\PDF{e}{\gamma}^{[1]}(z)\Big)\,.
\label{gamomc1}
\eeq
Finally, with the same assumptions as for eq.~(\ref{emomc}), the
charge-conservation condition for $f^\pm$ reads:
\beq
1=\int_0^1 dz\,\Big(\PDF{f^\pm}{f^\pm}(z)-\PDF{f^\mp}{f^\pm}(z)\Big)\,.
\label{echgc}
\eeq
For the electron at $\ord(\aem)$ and again thanks to eq.~(\ref{G0sol}), 
this implies:
\beq
0=\int_0^1 dz\,\PDF{e}{e}^{[1]}(z)\,,
\label{echgc1}
\eeq
which can be seen immediately to be fulfilled (possibly bar for the 
scheme-change term) by eq.~(\ref{G1sol2}). Therefore, in analogy to
eq.~(\ref{emomc1}), eq.~(\ref{echgc1}) can be used to impose a constraint
of the $\PDF{e}{e}$ scheme-change term:
\beq
\int_0^1 dz\,K_{ee}(z)=0\,.
\label{chgcKe}
\eeq

~

\noindent
$\bullet$ {\em What determines the PDFs}. As far as the contributions
to $\PDF{i}{e}^{[1]}(z)$ are concerned, the computations of 
sects.~\ref{sec:Gee} and~\ref{sec:Gge} show that they originate 
from the following three different sources.
\begin{enumerate}
\item For $z<1$, the difference between the real-emission massive-electron
cross section, and its massless-electron counterpart. Crucially, such a 
difference leads to logarithmically-enhanced and/or constant terms in 
the electron mass only in the kinematics region dominated by the collinear 
splittings of the incoming electron.
\item For $z<1$, the contribution to the $(n+1)$-body degenerate cross 
section due to the incoming electron leg.
\item For $z=1$, the soft-subtraction terms of the two contributions
above, plus the differences between the massive- and massless-electron
cross sections for all of the Born-like cross sections.
\end{enumerate}
This suggests an alternative procedure for the determination of 
the PDFs. Let us first deal with the case of $\PDF{e}{e}^{[1]}$
to be definite. Consider the $(n+1)$-body process:
\beq
e+x_2\;\longrightarrow x_3+\ldots x_{n+2}+\gamma\,,
\label{npoproc}
\eeq
with $x_i$ being particles/partons whose nature is unimportant here, 
and its $n$-body counterpart:
\beq
e+x_2\;\longrightarrow x_3+\ldots x_{n+2}\,.
\label{nproc}
\eeq
The master equation to be solved for $\PDF{e}{e}^{[1]}(z)$ 
for $z<1$ is:
\beq
d\bsig^{(n+1)}-d\hsig^{(n+1)}-d\hbsig^{(n+1)}=
\aemotpi\,\PDF{e}{e}^{[1]}\star d\hsig^{(n)}\,.
\label{mastercll}
\eeq
With abuse of notation, we have written the real-emission and
degenerate cross sections in eq.~(\ref{mastercll}) as if they were
the exact ones, but we actually mean to use simplified forms
in keeping with items 1 and 2 above. As far as the real-emissions
contribution is concerned, we shall employ:
\beqn
d\bsig^{(n+1)}&=&\Sfun_{n+3,1}\bampsqnpot d\phi_{n+1}\,,
\label{recll}
\\
\bampsqnpot&=&\frac{2e^2}{(k_1-k_{n+3})^2-m^2}\,
P_{e^\star e}^{<}\left(z\right)\,
\ampsqnt(zk_1)\,.
\label{reMEcll}
\eeqn
We have labelled the outgoing photon with index $n+3$;
thus, the $\Sfun$ function $\Sfun_{n+3,1}$ selects kinematic
configurations where the photon is collinear to the incoming 
electron. Furthermore, the real-emission matrix element of
eq.~(\ref{reMEcll}), relevant to the process of eq.~(\ref{npoproc}),
is taken equal to its collinear limits, computed according to the
procedure outlined in appendix~\ref{sec:QCsplit} (see in particular
eq.~(\ref{uMEcollFRS})). The $n$-body matrix element on the r.h.s.~of
eq.~(\ref{reMEcll}) is the one relevant to the process of eq.~(\ref{nproc}).
Finally, the absence of soft subtractions in eq.~(\ref{recll}) is due 
to the fact that the solution obtained from eq.~(\ref{mastercll}) is
valid only for $z<1$. As far as the degenerate cross section is
concerned, following item 2 we need to consider only the contribution
due to emissions from leg 1. Thus, from eq.~(\ref{hbsigpl}) we obtain:
\beqn
d\hbsig^{(n+1)}&=&\aemotpi\,{\cal K}_{ee}\star d\hsig^{(n)}\,,
\label{degcll}
\\
d\hsig^{(n)}&=&\ampsqnt d\phi_n\,,
\label{hsign}
\eeqn
and ${\cal K}_{ee}$ given in eq.~(\ref{Kdef}) (without the plus prescriptions).

Given these cross sections, the idea is the following. Since the 
leading behaviour for $m\to 0$ of the real-emission cross section
results from integrating over collinear photon-electron configurations,
the relevant kinematics quantities are explicitly given in the
prefactor on the r.h.s.~of eq.~(\ref{reMEcll}). The integration
itself is then performed, trivially, by means of identities such 
as those of eqs.~(\ref{ybe1}), (\ref{ybe2}), and~(\ref{ybe3}).
These do not allow one to obtain $d\bsig^{(n+1)}$, but rather
the difference \mbox{$d\bsig^{(n+1)}-d\hsig^{(n+1)}$}, which however 
is the only thing that matters given eq.~(\ref{mastercll}). In the
process, we expect $d\hsig^{(n)}$ to factorise.

Having done this, the solution at $z=1$ is obtained by using either the
momentum- or the charge-conservation condition. The consistency with 
evolution equations could be used as well, but it turns out not to be 
necessary. We point out that momentum conservation requires
one to solve first for both $\PDF{e}{e}^{[1]}$ and 
$\PDF{\gamma}{e}^{[1]}$ for $z<1$. This is feasible, since it 
should be clear that the procedure outlined above can in fact be 
applied to any kind of splittings (including those in which it
is an initial-state photon that splits), with only trivial 
modifications to eqs.~(\ref{mastercll}), (\ref{reMEcll}),
and~(\ref{degcll}).

~

\noindent
$\bullet$ {\em Relationships with QCD results}. 
The previous item renders it clear that there are strict similarities
between the PDFs which are computed here, and the so-called parton-to-parton
(p2p henceforth) PDFs which emerge in the context of factorisation theorems 
in QCD; the same remark applies to fragmentation functions (FFs) as well.
One must keep in mind that while the electron and photon QED PDFs and FFs 
are directly connected with physical cross sections, their p2p QCD 
counterparts enter unphysical factorisation formulae, whose connections 
with observable quantities always imply the introduction of long-distance 
objects (the proper hadronic PDFs and FFs\footnote{\label{ft:bq} This 
statement is true also in the case of the $b$-quark (taken as representative 
of any massive quark that hadronises) FF. However, this {\em is} a special 
case: the $b$-quark to $B$-hadron FF can formally be set equal to 
$\delta(1-z)$ without causing the corresponding cross sections to diverge. 
In other words, the $b$-quark can be regarded as an object which can be 
tagged in isolation.}).
With this observation in mind, and provided that one relies on QCD 
factorisation formulae written in the same way as eq.~(\ref{master0}),
one can obtain the QED PDFs and FFs from their p2p QCD analogues by
taking the abelian limit of the latter; furthermore, the p2p PDFs and FFs
must have been computed with one massive flavour (which in the abelian
limit will play the role of the electron in QED); the contributions
of the massless flavours must be discarded.

The previous conditions imply that, with the exception of final-state
$b$ quarks (for the reasons explained in footnote~\ref{ft:bq}), the QCD
results we are interested in are those generally called ``variable flavour
number scheme'' approaches, in which one studies the cross sections where
one massive flavour (typically, the $b$ quark) undergoes a transition
from a kinematical regime where its mass is not negligible to another
regime where it can be treated as if it were massless. As far as the
initial-state case (i.e.~the PDFs) is concerned, the first NLO results
can be found in ref.~\cite{Aivazis:1993pi} (see e.g.~ref.~\cite{Olness:1997yc}
for their explicit applications to cross section computations). We must
note that these results do not include the case of the $b$-to-$b$ PDF 
(i.e.~the analogue of $\PDF{e}{e}$ computed here), since this would 
require the assumption of an intrinsic $b$-quark hadronic component, 
which is typically not made. Such a quantity has been computed in the
context of DIS in ref.~\cite{Kretzer:1998ju}. Conversely, the final-state
case (i.e.~the FFs) has been considered in ref.~\cite{Mele:1990cw}
($b$-to-$b$, on which we shall further comment in sect.~\ref{sec:DeeFF})
and in ref.~\cite{Cacciari:2005ry} (the other flavours). We also point out
that the QED $e$-to-$\gamma$ FF has several analogies with the (massive)
quark-to-$\gamma$ FF, computed at the NLO in QCD in ref.~\cite{Glover:1993xc}
(we shall briefly return on this point in sect.~\ref{sec:DegFF}).

We conclude this discussion with two observations. Firstly, the QCD
results are generally derived in the $\MSb$ scheme\footnote{See sect.~8
of ref.~\cite{Cacciari:1998it} for an exception to this statement;
that paper builds upon the results of ref.~\cite{Mele:1990cw}.}.
In QED, it is convenient to be able to exploit the flexibility
associated with changing the subtraction scheme -- we shall show
this explicitly in ref.~\cite{BCCFS} and subsequent publications.
Secondly, having established at the NLO the close relationships between 
the QED PDFs and FFs as computed here and their p2p QCD counterparts, 
it appears safe to assume that these will hold true also at orders higher 
than NLO. Thus, one could exploit the (N)NLO computations of 
refs.~\cite{Buza:1995ie,Buza:1996wv,Blumlein:2018jfm} (initial state) and 
refs.~\cite{Melnikov:2004bm,Mitov:2004du} (final state).

\subsection{Kinematics\label{sec:kincll}}
We parametrise the momenta of the partons involved in the
splitting of interest as follows:
\beqn
k_1&=&\frac{\sqrt{s}}{2}(1,0,0,\beta_1)
\label{k1cll}
\\
k_{n+3}&=&\frac{\sqrt{s}}{2}\,\xi\left(1,\vet\,\sqrt{1-y^2}\,\beta_{n+3},
y\,\beta_{n+3}\right)\,,
\label{knpocll}
\eeqn
with $k_i$ the momentum of parton $i$ (as e.g.~in eq.~(\ref{npoproc})),
$s=(k_1+k_2)^2$ and:
\beq
\beta_1=\sqrt{1-\frac{4m_1^2}{s}}\,,
\;\;\;\;\;\;\;\;
\beta_{n+3}=\sqrt{1-\frac{4m_{n+3}^2}{s\,\xi^2}}\,.
\eeq
As usual, we identify $\xi=1-z$, $z$ being that which appears
in eq.~(\ref{reMEcll}).
The masses $m_1$ and $m_{n+3}$ can be either equal to the electron
mass $m$ or to zero, depending on the type of branching being studied.
We point out the following fact. The expression of the momentum of the 
incoming parton in eq.~(\ref{k1cll}) is the same as that in eq.~(\ref{hadcm}),
but differs from that in eq.~(\ref{pb1}). The difference in the latter
two forms is due to the fact that the FKS cross sections are written
in the incoming partons c.m.~frame, hence the parametrisation of the
momentum of parton 1 depends (also) on the mass of parton 2. By employing
eq.~(\ref{k1cll}), we have assumed that the mass of parton 2 is equal
to $m_1$. This might appear odd, since parton 2 never appears in the 
procedure advocated in this section for the extraction of the PDFs.
In fact, such choices do not have an impact on the final results,
and are made with the sole purpose of simplifying the computation.
This is because the PDF we seek to calculate is a purely
collinear quantity, and hence is invariant under longitudinal boosts.
Therefore, computations carried out in any two frames connected by
longitudinal boosts will necessarily lead to the same PDFs.
An {\em a posteriori} evidence in support of this argument is the 
fact that we shall obtain here the same result for $\PDF{\gamma}{e}^{[1]}$
as in eq.~(\ref{Ggesol2}), in spite of the difference between 
eq.~(\ref{k1cll}) and eq.~(\ref{pb1}).

The $(n+1)$-body phase-space is:
\beqn
d\phi_{n+1}(k_1,\ldots k_{n+3})&=&
d\tilde{\phi}_n(k_1,\ldots k_{n+3})\,d\phi_1(k_{n+3})\,,
\label{phinpo}
\\
d\tilde{\phi}_n(k_1,\ldots  k_{n+3})&=&
(2\pi)^4\delta\left(k_1+k_2-\sum_{i=3}^{n+3}k_i\right)
\prod_{i=3}^{n+2}\frac{d^3 k_i}{(2\pi)^3 2k_i^0}\,,
\phantom{aaa}
\label{tphin}
\\
d\phi_1(k_{n+3})&=&\frac{d^3 k_{n+3}}{(2\pi)^3 2k_{n+3}^0}=
\frac{1}{2(2\pi)^3}\,\frac{s\beta_{n+3}^3}{4}\,\xi d\xi\,dy\,d\varphi\,.
\label{phi1}
\eeqn
For future use, we note the property:
\beq
d\tilde{\phi}_n(k_1,\ldots  k_{n+3})\,\delta(1-y)
\;\stackrel{m_1\to 0\,,m_{n+3}\to 0}{\longrightarrow}\;
d\phi_n\big((1-\xi)k_1,\ldots  k_{n+2}\big)\,\delta(1-y)\,.
\label{phinlim}
\eeq
The leftmost quantity on the r.h.s.~of eq.~(\ref{phinlim}) is the actual 
$n$-body phase-space with incoming massless legs and incoming energy 
squared equal to \mbox{$(1-\xi)s$}.

\subsection{Determination of $\PDF{e}{e}$\label{sec:Geecll}}
In this case, the relevant branching is $e\to e^\star\gamma$,
whence $m_1=m$ and $m_{n+3}=0$. With the QED version of eq.~(\ref{Pqstq}),
eq.~(\ref{recll}) becomes:
\beqn
d\bsig^{(n+1)}&=&\Sfun_{n+3,1}e^2\ee^2\left[
\frac{1}{k_1\mydot k_{n+3}}\,\frac{1+(1-\xi)^2}{\xi}-
\frac{(1-\xi)m^2}{\big(k_1\mydot k_{n+3}\big)^2}\right]
\nonumber\\*&\times&
\ampsqnt\big((1-\xi)k_1\big)\,
\frac{s}{8(2\pi)^3}\,\xi d\xi dy d\varphi d\tilde{\phi}_n\,.
\label{recllee}
\eeqn
With eqs.~(\ref{k1cll}) and~(\ref{knpocll}):
\beq
k_1\mydot k_{n+3}=\frac{s}{4}\,\xi\left(1-\beta_1 y\right)\equiv
\frac{s}{4}\,\xi\left(1-\frac{y}{1+\rho}\right)\,,
\label{kkee}
\eeq
where as was done in eq.~(\ref{rovsbe}) we have defined:
\beq
\rho=\frac{1-\beta_1}{\beta_1}=2\,\frac{m^2}{s}
+\ord\left(\!\frac{m^4}{s^2}\!\right)\,.
\eeq
One can then replace eq.~(\ref{kkee}) into eq.~(\ref{recllee}),
and employ the identities in eqs.~(\ref{ybe3}) and~(\ref{ybe2})
for the first and second term in the square brackets, respectively,
which also allows one to set equal to zero all mass terms that are not
explicitly written in eq.~(\ref{recllee}) and in $\rho$. By doing so,
a number of things follow. Firstly, the plus-distribution contribution
of eq.~(\ref{ybe3}) results in the collinearly-subtracted massless-electron
cross section $d\hsig^{(n+1)}$. All of the remaining contributions 
are then proportional to $\delta(1-y)$, and one can exploit:
\beq
\Sfun_{n+3,1}\delta(1-y)=\delta(1-y)\,,
\eeq
and eq.~(\ref{phinlim}). From this and from the matrix element on 
the r.h.s.~of eq.~(\ref{recllee}), the reduced cross section of
eq.~(\ref{hsign}) (evaluated at the reduced c.m.~energy $(1-\xi)s$)
emerges naturally. By performing the trivial $y$ and $\varphi$
integrations we finally obtain:
\beqn
d\bsig^{(n+1)}&=&d\hsig^{(n+1)}+\aemotpi\,\ee^2\left[
\frac{1+(1-\xi)^2}{\xi}\log\frac{s}{m^2}-2\,\frac{1-\xi}{\xi}\right]
\nonumber\\*&&\phantom{d\hsig^{(n+1)}+\aemotpi\,\ee^2}\times
d\hsig^{(n)}\big((1-\xi)k_1\big)\,d\xi\,.
\label{recllee2}
\eeqn
We can now replace this result into our master equation~(\ref{mastercll}).
By using there the explicit expression of the degenerate cross section, 
by identifying $\xi$ with $1-z$, and by solving for $\PDF{e}{e}^{[1]}(z)$,
we obtain again eq.~(\ref{G1sol2}), with the exception of the plus 
prescription (since we have worked here with $z<1$). However, the 
contribution at $z=1$ can be readily obtained by exploiting the 
charge-conservation condition of eq.~(\ref{echgc1}).

As a final technical aside,
we point out that, in order to obtain the third term in the round brackets
of eq.~(\ref{G1sol2}) with the procedure we have followed here, the 
factor $1-\xi$ in the numerator of the second term in the square brackets
of eq.~(\ref{recllee}) is crucial. In turn, this is a direct consequence of 
the crossing we have employed in app.~\ref{sec:QCsplit} to determine the
massive spacelike splitting kernels.

\subsection{Determination of $\PDF{\gamma}{e}$\label{sec:Ggecll}}
In this case, the relevant branching is $e\to \gamma^\star e$,
whence $m_1=m$ and $m_{n+3}=m$. With the QED version of eq.~(\ref{Pgstq}),
eq.~(\ref{recll}) becomes:
\beqn
d\bsig^{(n+1)}&=&\Sfun_{n+3,1}e^2\ee^2\left[
\frac{1}{k_1\mydot k_{n+3}-m^2}\,\frac{1+\xi^2}{1-\xi}-
\frac{(1-\xi)m^2}{\big(k_1\mydot k_{n+3}-m^2\big)^2}\right]
\nonumber\\*&\times&
\ampsqnt\big((1-\xi)k_1\big)\,
\frac{s}{8(2\pi)^3}\,\beta_{n+3}^3\,\xi d\xi dy d\varphi d\tilde{\phi}_n\,.
\label{recllge}
\eeqn
The analogue of eq.~(\ref{kkee}) reads:
\beq
k_1\mydot k_{n+3}-m^2=\frac{s}{4}\left(\xi-\frac{4m^2}{s}\right)
\left(1-\frac{y}{1+\rho}\right)\,,
\label{kkge}
\eeq
having defined
\beq
\rho=\frac{\xi(1-\beta_1\beta_{n+3})-4m^2/s}{\xi\beta_1\beta_{n+3}}=
2\,\frac{(1-\xi)^2}{\xi^2}\,\frac{m^2}{s}
+\ord\left(\!\frac{m^4}{s^2}\!\right)\,.
\eeq
We can now perform the same algebraic operations already carried out
in sect.~\ref{sec:Geecll}, to obtain what follows:
\beq
d\bsig^{(n+1)}=d\hsig^{(n+1)}+
\aemotpi\,{\cal Q}_{\gamma e}\star d\hsig^{(n)}\,,
\eeq
namely, the very same result as in eq.~(\ref{bsig331res}). This implies
that we also obtain again the final result for $\PDF{\gamma}{e}^{[1]}$
already reported in eq.~(\ref{Ggesol2}).

\subsection{Determination of $\PDF{e}{\gamma}$\label{sec:Gegcll}}
In this case, the relevant branching is $\gamma\to e^\star e$,
whence $m_1=0$ and $m_{n+3}=m$. With the QED version of eq.~(\ref{Pqstg}),
eq.~(\ref{recll}) becomes:
\beqn
d\bsig^{(n+1)}&=&\Sfun_{n+3,1}e^2\ee^2\left[
\frac{(1-\xi)^2+\xi^2}{k_1\mydot k_{n+3}}+
\frac{(1-\xi)m^2}{\big(k_1\mydot k_{n+3}\big)^2}\right]
\nonumber\\*&\times&
\ampsqnt\big((1-\xi)k_1\big)\,
\frac{s}{8(2\pi)^3}\,\beta_{n+3}^3\,\xi d\xi dy d\varphi d\tilde{\phi}_n\,.
\label{reclleg}
\eeqn
The analogue of eq.~(\ref{kkee}) reads:
\beq
k_1\mydot k_{n+3}=\frac{s}{4}\,\xi\left(1-\beta_{n+3} y\right)\equiv
\frac{s}{4}\,\xi\left(1-\frac{y}{1+\rho}\right)\,,
\label{kkeg}
\eeq
with:
\beq
\rho=\frac{1-\beta_{n+3}}{\beta_{n+3}}=\frac{2}{\xi^2}\,\frac{m^2}{s}
+\ord\left(\!\frac{m^4}{s^2}\!\right)\,.
\eeq
A simple algebra leads one to:
\beqn
d\bsig^{(n+1)}&=&d\hsig^{(n+1)}+\aemotpi\,\ee^2\left[
\left(\xi^2+(1-\xi)^2\right)\log\frac{s\xi^2}{m^2}
+2\xi(1-\xi)\right]
\nonumber\\*&&\phantom{d\hsig^{(n+1)}+\aemotpi\,\ee^2}\times
d\hsig^{(n)}\big((1-\xi)k_1\big)\,d\xi\,.
\label{reclleg2}
\eeqn
The relevant $(n+1)$-body degenerate cross section is:
\beqn
d\hbsig^{(n+1)}&=&\aemotpi\,{\cal K}_{e\gamma}\star d\hsig^{(n)}\,,
\\
{\cal K}_{e\gamma}(z)&=&P_{e\gamma}^{<}(z)
\left(\log\frac{s}{\mu^2}+2\log(1-z)\right)
-P_{e\gamma}^{\prime<}(z)-\ee^2 K_{e\gamma}(z)\,.\phantom{aaa}
\eeqn
By using:
\beqn
P_{e\gamma}^{<}(z)&=&\ee^2\left(z^2+(1-z)^2\right)\,,
\\
P_{e\gamma}^{\prime<}(z)&=&-\ee^2 z(1-z)\,,
\eeqn
we finally obtain:
\beq
\PDF{e}{\gamma}^{[1]}(z,\mu^2)=\ee^2\left(z^2+(1-z)^2\right)
\log\frac{\mu^2}{m^2}+\ee^2 K_{e\gamma}(z)\,.
\label{Gegsol2}
\eeq
This result fulfills eq.~(\ref{APeq4}), as expected. Note that it
is also valid at $z=1$, since the branching $\gamma\to e^\star e$ cannot
induce soft singularities.

\subsection{Determination of $\PDF{\gamma}{\gamma}$\label{sec:Gggcll}}
The absence of a $\gamma\to\gamma X$ splitting at $\ord(\aem)$
implies that all of the cross sections on the l.h.s.~of 
eq.~(\ref{mastercll}) are identically equal to zero. This implies
that the photon-to-photon PDF must have the following form:
\beq
\PDF{\gamma}{\gamma}^{[1]}(z,\mu^2)=
\left(A\log\frac{\mu^2}{m^2}+B\right)\delta(1-z)\,.
\eeq
Neglecting the scheme-change term in eq.~(\ref{Gegsol2}) for the time
being, eq.~(\ref{gamomc1}) implies:
\beq
A\log\frac{\mu^2}{m^2}+B+2\ee^2\log\frac{\mu^2}{m^2}
\int_0^1 dz\,z\left(z^2+(1-z)^2\right)=0\,.
\eeq
By solving for $A$ and $B$ one obtains:
\beqn
A&=&-\frac{2}{3}\ee^2\,,
\\
B&=&0\,.
\eeqn
By introducing an arbitrary change-of-scheme function,
the photon PDFs then reads:
\beq
\PDF{\gamma}{\gamma}^{[1]}(z,\mu^2)=
-\frac{2}{3}\ee^2\log\frac{\mu^2}{m^2}\,\delta(1-z)
+\ee^2 K_{\gamma\gamma}(z)\,,
\label{Gggsol2}
\eeq
where, analogously to eq.~(\ref{momcKe}):
\beq
\int_0^1 dz\,z\Big(K_{e\gamma}(z)+K_{\gamma\gamma}(z)\Big)=0\,.
\label{momcKg}
\eeq
The result of eq.~(\ref{Gggsol2}), which can be immediately extended to
the case of several massive charged fermions, fulfills 
eq.~(\ref{APeq4}).

\section{Initial conditions for fragmentation functions\label{sec:FF}}
What has been done in sect.~\ref{sec:PDFcoll} can also be applied to the 
case in which one considers final-state, rather than initial-state,
branchings. There, it is one of the outgoing partons whose momentum
becomes constrained by the object that plays the same role as the one 
played by the PDFs so far -- namely, the fragmentation function (FF).
The FKS formalism has been extended to include FFs in 
ref.~\cite{Frederix:2018nkq}; the formulae are given in that paper for
QCD, but it is immediate to covert them into their QED counterparts, 
which is what we are interested in here, by means of eqs.~(\ref{QCDtoQED1}) 
and~(\ref{QCDtoQED2}).

We shall write the cross section for the production of a particle
$H$ as follows:
\beq
d\sigma_H=\sum_p d\hsig_p\star\Fragm{H}{p}\,,
\label{FFxsec}
\eeq
whose explicit definition can be found in 
ref.~\cite{Frederix:2018nkq}\footnote{In that paper, $\Fragm{H}{p}$ had
been denoted by $D^{(p)}_H$. The current notation is used in order to be
as consistent as possible with the case of the PDFs.}.
We shall employ eq.~(\ref{FFxsec}) up to the first non-trivial
order in $\aem$ with:
\beq
p,\,H\;\in\;\{e^+,e^-,\gamma\}\,.
\eeq
The notation of eq.~(\ref{FFxsec}) understands that:
\beq
\Fragm{H}{p}\,:\;\;\;\;\;\;\;\;
p\,\stackrel{\rm fragments}{\longrightarrow}\,H\,.
\eeq
In QED, the FFs admit a perturbative expansion:
\beq
\Fragm{H}{p}(z)=
\Fragm{H}{p}^{[0]}(z)+\aemotpi\,\Fragm{H}{p}^{[1]}(z)+\ord(\aem^2)\,,
\eeq
where, analogously to eq.~(\ref{G0sol}), we must have:
\beq
\Fragm{H}{p}^{[0]}(z,\mu^2)=\delta_{Hp}\,\delta(1-z)\,.
\label{Dzero}
\eeq
The basic idea behind eq.~(\ref{FFxsec}) is the following: compute
the particle cross section on the l.h.s.~with massive electrons,
and keep the dominant terms as $m\to 0$; compute the subtracted 
partonic cross sections on the r.h.s.~with massless electrons;
and solve for $\Fragm{H}{p}(z)$. By performing this procedure 
in the context of complete-process computations one is conceptually
on the very same footing as in sects.~\ref{sec:Gee} and~\ref{sec:Gge}.
However, we have shown in sect.~\ref{sec:PDFcoll} how such lengthy
computations can be bypassed by exploiting the fact that the PDFs
are entirely determined by collinear contributions. Since this
property holds for FFs as well, we shall use a similar technique
also in the present case. Our master equation is thus the analogue
of eq.~(\ref{mastercll}), and reads as follows:
\beq
d\bsig^{(n+1)}-d\hsig^{(n+1)}-d\hbsig^{(n+1)}=
\aemotpi\,d\hsig^{(n)}\star \Fragm{H}{p}^{[1]}\,.
\label{masterFF}
\eeq
Note that, as was already done in the case of the PDFs, the fragmenting
parton $p$ is fixed here, and not summed over as in eq.~(\ref{FFxsec}),
which is meaningful since we are working at the level of
individual branchings. As in the case of the PDFs, 
eq.~(\ref{masterFF}) is meant to be used for $z<1$.
Although we have employed the same notations for the quantities on 
the l.h.s.~of eq.~(\ref{masterFF}) as in eq.~(\ref{mastercll}), their
meanings are different. In particular, we need to consider the
$(i,j)$ FKS sector, where we identify the momentum of $j$ with
that of the particle $H$ (and obviously the flavour of $j$ is the
same as $H$, $a_j=H$), and the real matrix element is replaced
by its collinear limit:
\beqn
d\bsig^{(n+1)}&=&\Sfunij\,\bampsqnpot d\phi_{n+1}\,,
\label{reFF}
\\
\bampsqnpot&=&\frac{2e^2}{(k_i+k_j)^2-m_p^2}\,
P_{Hp^\star}^{<}\left(z\right)\,
\ampsqnt(k_p)\,.
\label{reMEFF}
\eeqn
Note the complete analogy between these equations and eqs.~(\ref{recll})
and~(\ref{reMEcll}). As before, we expect that from the 
explicit evaluation of eq.~(\ref{reFF}) the $\Sfunij$ contribution to
the massless electron cross section, namely the second term on the
l.h.s.~of eq.~(\ref{masterFF}), will emerge. As far as the third term
there is concerned, this degenerate $(n+1)$-body cross section in the 
presence of a FF can be read from eq.~(4.85) of ref.~\cite{Frederix:2018nkq}.
We write it in a similar fashion as eq.~(\ref{degcll}):
\beq
d\hbsig^{(n+1)}=\aemotpi\,d\hsig^{(n)}\star {\cal K}_{Hp}^{\rm FIN}\,,
\label{degFF}
\eeq
where:
\beqn
&&{\cal K}_{ab}^{\rm FIN}(z)=(1-z)\,
\Bigg\{-\pdistr{1-z}{+}P_{ab}^{\prime<}(z)-
\frac{K_{ab}^{\rm FF}(z)}{1-z}
\label{Kfin}
\\*&&\phantom{\;}
+\left[\pdistr{1-z}{+}\left(\log\frac{s}{\mu^2}+
2\log\frac{2zE_p}{\sqrt{s}}\right)+
2\lppdistr{1-z}{+}\right]P_{ab}^<(z)\Bigg\}\,.
\nonumber
\eeqn
In ref.~\cite{Frederix:2018nkq} the analogy between eqs.~(\ref{Kfin})
and~(\ref{Kdef}) has been commented upon at length, and the interested
reader is referred to that paper for more details on this. Similarly to 
sect.~\ref{sec:PDFcoll}, given that eq.~(\ref{masterFF}) can be used 
only for $z<1$, the plus prescriptions of eq.~(\ref{Kfin}) will become
ordinary functions, thus allowing one to simplify the \mbox{$(1-z)$}
prefactor.
Finally, the $n$-body cross section on the r.h.s.~of eq.~(\ref{masterFF})
has the same form as in eq.~(\ref{hsign}); it is understood that the 
parton $p$ is present in the final state.

\subsection{Kinematics\label{sec:kinFF}}
We parametrise the momenta of the FKS parton and its sister by
generalising what was done in sect.~4.4 of ref.~\cite{Frixione:1995ms}:
\beqn
k_i&=&\frac{\sqrt{s}}{2}\,\xi_i\,(1,\beta_i\hk_i)\,,
\;\;\;\;\;\;\;\;\,
\hk_i=\hp_i\,R\equiv\left(\vet_i\,\sqrt{1-y_i^2},y_i\right)\,,
\label{kipar}
\\
k_j&=&\frac{\sqrt{s}}{2}\,\xi_j\,(1,\beta_j\hk_j)\,,
\;\;\;\;\;\;\;\;
\hk_j=\hp_j\,R\,,
\label{kjpar}
\eeqn
with $R$ a suitable three-dimensional rotation matrix, and:
\beq
\beta_i=\sqrt{1-\frac{4m_i^2}{s\,\xi_i^2}}\,,
\;\;\;\;\;\;\;\;
\beta_j=\sqrt{1-\frac{4m_j^2}{s\,\xi_j^2}}\,,
\eeq
with:
\beq
\hp_i=(0,0,1)\,,
\;\;\;\;\;\;\;\;
\hp_j=\left(\vet_j\,\sqrt{1-y_j^2},y_j\right)\,.
\eeq
Thus:
\beq
k_i\mydot k_j=\frac{s}{4}\,\xi_i\xi_j(1-\beta_i\beta_j y_j)\,.
\label{kikj}
\eeq
We exploit this parametrisation by writing the $(n+1)$-body phase
space as follows:
\beqn
d\phi_{n+1}(k_1,\ldots k_{n+3})&=&
d\tilde{\phi}_{n-1}(k_1,\ldots k_{n+3})\,d\phi_1(k_i)\,d\phi_1(k_j)\,,
\label{phinpoout}
\\
d\tilde{\phi}_{n-1}(k_1,\ldots  k_{n+3})&=&
(2\pi)^4\delta\left(k_1+k_2-\sum_{k=3}^{n+3}k_k\right)
\prod_{k\ne i,j}\frac{d^3 k_k}{(2\pi)^3 2k_k^0}\,,
\phantom{aaa}
\label{tphinout}
\eeqn
where the $1$-body phase space is given in eq.~(\ref{phi1}). These 
expressions will be manipulated by using the change of variables which
is customary in FKS:
\beq
(\xi_i,\xi_j)\;\longrightarrow\;(\xi_p,z)\,,
\eeq
where
\beq
\xi_i=(1-z)\xi_p\,,
\;\;\;\;
\xi_j=z\xi_p
\;\;\;\;\;\;\;\;
\Longrightarrow
\;\;\;\;\;\;\;\;
d\xi_i d\xi_j = \xi_p d\xi_p dz\,.
\label{xixjzxip}
\eeq
By construction, $\xi_p$ is the rescaled energy of parton $p$ in
the incoming parton c.m.~frame:
\beq
E_p=\frac{\sqrt{s}}{2}\,\xi_p\,.
\label{Epxip}
\eeq
The parametrisation of the momenta adopted here is such that,
in the collinear limit $y_j\to 1$, the three-momentum of the mother 
particle
\beq
\vec{k}_p=\vec{k}_i+\vec{k}_j
\eeq
is parallel to $\hk_i$ of eq.~(\ref{kipar}), and its direction can
then be associated with $y_i$ and $\vet_i$. This implies that one
can recast the r.h.s.~of eq.~(\ref{phinpoout}) as follows:
\beqn
&&d\tilde{\phi}_{n-1}(k_1,\ldots k_{n+3})\,d\phi_1(k_i)\,d\phi_1(k_j)\,
\delta(1-y_j)\;
\underset{m_p\to 0}{\overset{m_i\to 0\,,m_j\to 0}{\longrightarrow}}\;
\label{phoutlim}
\\*&&\phantom{aaaa}
\frac{z(1-z)\xi_p^2 s}{8(2\pi)^3}\,
d\phi_n\big(k_1,\ldots\cancel{k}_i,\cancel{k}_j,k_p\ldots  k_{n+3}\big)\,
\delta(1-y_j)\,dz d\varphi_j dy_j\,,
\nonumber
\eeqn
having used eq.~(\ref{xixjzxip}). As was the case in eq.~(\ref{phinlim}),
the actual $n$-body phase space (with a massless mother particle) appears
in the r.h.s.~of eq.~(\ref{phoutlim}). When multiplying it by the massless
$n$-body matrix element on the r.h.s.~of eq.~(\ref{reMEFF}), one obtains the 
$n$-body cross section that is expected to factorise in all of the terms
of the master equation~(\ref{masterFF}).

\subsection{Determination of $\Fragm{e}{e}$\label{sec:DeeFF}}
In this case, the relevant branching is $e^\star\to e\gamma$,
whence $m_i=0$, $m_j=m$, and $m_p=m$. With the QED version of 
eq.~(\ref{Pqqst}), eq.~(\ref{reFF}) becomes:
\beqn
d\bsig^{(n+1)}&=&\Sfunij\,e^2\ee^2\left[
\frac{1}{k_i\mydot k_j}\,\frac{1+z^2}{1-z}-
\frac{m^2}{\big(k_i\mydot k_j\big)^2}\right]\ampsqnt(k_p)
\nonumber\\*&\times&
\left(\frac{s}{8(2\pi)^3}\right)^2\beta_j^3\,\xi_i\xi_j\xi_p
d\xi_p dz dy_i d\varphi_i dy_j d\varphi_j d\tilde{\phi}_{n-1}\,.
\label{reFFee}
\eeqn
Using eq.~(\ref{kikj}) we have:
\beq
k_i\mydot k_j=\frac{s}{4}\,\xi_i\xi_j\left(1-\frac{y_j}{1+\rho}\right)\,,
\label{kikjee}
\eeq
having defined:
\beq
\rho=\frac{1-\beta_j}{\beta_j}=2\,\frac{m^2}{s\xi_j^2}
+\ord\left(\!\frac{m^4}{s^2}\!\right)\,.
\eeq
One can then replace eq.~(\ref{kikjee}) into eq.~(\ref{reFFee}), exploit 
eq.~(\ref{xixjzxip}), and employ the identities in eqs.~(\ref{ybe3}) 
and~(\ref{ybe2}) (with the formal replacement $y\to y_j$ there)
for the first and second term in the square brackets, respectively,
which also allows one to set equal to zero all mass terms which are not
explicitly written in eq.~(\ref{reFFee}) and in $\rho$. The next
steps are identical to those we have already gone through in
sects.~\ref{sec:Geecll}--\ref{sec:Gegcll}, with only trivial differences.
In particular, we have again the natural emergence of the 
collinearly-subtracted massless-electron cross section $d\hsig^{(n+1)}$,
induced by the plus-distribution contribution of eq.~(\ref{ybe3}).
Since the remaining terms are all proportional to $\delta(1-y_j)$,
the fundamental property of the $\Sfunij$ function, namely
\beq
\Sfunij\delta(1-y_j)=\delta(1-y_j)
\eeq
can be used, which in turn (together with the masslessness condition)
allows one to employ eq.~(\ref{phoutlim}). After performing the trivial
$y_j$ and $\varphi_j$ integrations, we are left with:
\beqn
d\bsig^{(n+1)}&=&d\hsig^{(n+1)}+\aemotpi\,\ee^2\left[
\frac{1+z^2}{1-z}\left(
\log\frac{s}{m^2}+2\log(z\xi_p)\right)
-\frac{2z}{1-z}\right]
\nonumber\\*&&\phantom{d\hsig^{(n+1)}+\aemotpi\,\ee^2}\times
d\hsig^{(n)}\,dz\,.
\label{reFFee2}
\eeqn
The result of eq.~(\ref{reFFee2}) is identical to that of
eq.~(\ref{recllee2}), bar for the presence of the second logarithmic 
term. However, that term is cancelled by its analogue in the $(n+1)$-body
degenerate contribution -- see eq.~(\ref{Kfin}), where eq.~(\ref{Epxip}) must 
be taken into account. A simple algebra then leads us to\footnote{Strictly
speaking, only for $z<1$. However, a charge-conservation condition
such as that of eq.~(\ref{echgc1}) can be imposed on the  FF as well,
whence eq.~(\ref{Deesol2}).}:
\beq
\Fragm{e}{e}^{[1]}(z,\mu^2)=\left.\PDF{e}{e}^{[1]}(z,\mu^2)
\right|_{K_{ee}\longrightarrow K_{ee}^{\rm FF}}\,,
\label{Deesol2}
\eeq
with $\PDF{e}{e}^{[1]}$ given in eq.~(\ref{G1sol2}). Apart from the
(possible) difference between the two scheme-change functions associated
with PDFs and FFs, it should be stressed that we have indicated with
the same symbol $\mu$ on the two sides of eq.~(\ref{Deesol2}) two
quantities with different physical meanings: the factorisation scales
relevant to initial- and final-state splittings in general need not be 
equal to each other.

Equation~(\ref{Deesol2}) is interesting, because it emerges from a
combination of cross sections whose ingredients, in spite of being very
similar, are different from each other; in particular, the differences 
stem from the massive Altarelli-Parisi kernels (spacelike vs timelike),
and from the forms of the collinear kernels of eqs.~(\ref{Kdef})
and~(\ref{Kfin}). Having said that, it is important not to overstate
``the PDF is equal to the FF'' conclusion, since both the PDFs and
the FFs are ultimately unphysical objects, that in part depend on 
which contributions are associated with short-distance cross sections.
In our computations, such an arbitrariness is conveniently parametrised 
by means of the scheme-change functions, which enter both the PDFs/FFs
and the $(n+1)$-body degenerate cross sections.

Equation~(\ref{Deesol2}) also implies that, apart from the formal 
replacement $C_{\sss\rm F}\to\ee^2$ and except for the scheme-change 
term, $\Fragm{e}{e}^{[1]}$ is identical to the initial condition for 
the $b$-quark fragmentation function obtained in ref.~\cite{Mele:1990cw}
(in that paper, the change of scheme had been ignored). If one considers
the procedure of sect.~\ref{sec:Gee}, this may appear surprising, since
the QCD cross sections relevant to $b\to bg$ branchings have non-trivial
differences w.r.t.~the QED ones relevant to $e\to e\gamma$ branchings.
However, it makes immediate sense if instead one proceeds by means of 
collinear factorisation arguments, as is done in sect.~\ref{sec:PDFcoll}
and~\ref{sec:FF}. This is a further evidence of the fact that any
process-specific information used in the intermediate steps of the
cross-section-based procedure eventually drops out when solving for
the PDFs and the fragmentation functions.

\subsection{Determination of $\Fragm{\gamma}{e}$\label{sec:DegFF}}
In this case, the relevant branching is $e^\star\to \gamma e$,
whence $m_i=m$, $m_j=0$, and $m_p=m$. With the QED version of 
eq.~(\ref{Pgqst}), eq.~(\ref{reFF}) becomes:
\beqn
d\bsig^{(n+1)}&=&\Sfunij\,e^2\ee^2\left[
\frac{1}{k_i\mydot k_j}\,\frac{1+(1-z)^2}{z}-
\frac{m^2}{\big(k_i\mydot k_j\big)^2}\right]\ampsqnt(k_p)
\nonumber\\*&\times&
\left(\frac{s}{8(2\pi)^3}\right)^2\beta_i^3\,\xi_i\xi_j\xi_p
d\xi_p dz dy_i d\varphi_i dy_j d\varphi_j d\tilde{\phi}_{n-1}\,.
\label{reFFeg}
\eeqn
Eq.~(\ref{kikjee}) can be applied to the present case as well,
but with:
\beq
\rho=\frac{1-\beta_i}{\beta_i}=2\,\frac{m^2}{s\,\xi_i^2}
+\ord\left(\!\frac{m^4}{s^2}\!\right)\,.
\eeq
We can now repeat the same procedure as in sect.~\ref{sec:DeeFF},
which leads to:
\beq
d\bsig^{(n+1)}={\rm eq.}~(\protect\ref{reFFee2})
\Big|_{z\longrightarrow 1-z}\,.
\eeq
By using the explicit form of ${\cal K}_{\gamma e}^{\rm FIN}$,
one obtains the final result:
\beq
\Fragm{\gamma}{e}^{[1]}(z,\mu^2)=\left.\PDF{\gamma}{e}^{[1]}(z,\mu^2)
\right|_{K_{\gamma e}\longrightarrow K_{\gamma e}^{\rm FF}}\,,
\label{Degsol2}
\eeq
with $\PDF{\gamma}{e}^{[1]}$ given in eq.~(\ref{Ggesol2}). 
Equations~(\ref{Degsol2}), (\ref{Deesol2}), and eq.~(\ref{emomc1})
imply:
\beq
0=\int_0^1 dz\,z\Big(\Fragm{e}{e}^{[1]}(z)+\Fragm{\gamma}{e}^{[1]}(z)\Big)\,,
\label{emomc1FF}
\eeq
provided that the FF analogue of eq.~(\ref{momcKe}) holds as well:
\beq
\int_0^1 dz\,z\Big(K_{ee}^{\rm FF}(z)+K_{\gamma e}^{\rm FF}(z)\Big)=0\,.
\label{momcKeFF}
\eeq
In spite of both being associated with momentum conservation, it should 
be clear that the physical interpretation of eqs.~(\ref{emomc1})
and~(\ref{emomc1FF}) is different. In the former, what is conserved is
the momentum of the particle, while in the latter it is the momentum
of the parton; this is because it is the particle in the former case,
and the parton in the latter case, that plays the role of the object
that undergoes a $(1\to 2)$ splitting.

We point out that, if one introduces quark masses and neglects
QCD effects, $\Fragm{\gamma}{e}$ can be used as the quark-to-photon
fragmentation function. It has thus the same meaning as what has
been denoted by \mbox{$D_{q\to\gamma}(z,\mu_{\sss F})$} in
ref.~\cite{Glover:1993xc}. This observation allows one to identify
what is called the ``perturbative contribution to the fragmentation
function'' in ref.~\cite{Glover:1993xc} with (part of) 
${\cal K}_{\gamma e}^{\rm FIN}$ --
note the presence of a term \mbox{$\log(z(1-z))$} in both eq.~(7) of 
that paper, and in eq.~(\ref{Kfin}). In turn, this implies that the 
quantity \mbox{$\Fragm{\gamma}{e}+{\cal K}_{\gamma e}^{\rm FIN}$},
i.e.~the leading $m\to 0$ behaviour of the real-emission
cross section, up to the reduced $n$-body cross section that
factorises out, is to be identified with what is denoted by
\mbox{${\cal D}_{q\to\gamma}$} in ref.~\cite{Glover:1993xc}
(and called ``effective fragmentation function'' in that paper).

\subsection{Determination of $\Fragm{e}{\gamma}$\label{sec:DgeFF}}
In this case, the relevant branching is $\gamma^\star\to ee$,
whence $m_i=m$, $m_j=m$, and $m_p=0$. With the QED version of 
eq.~(\ref{Pqgst}), eq.~(\ref{reFF}) becomes:
\beqn
d\bsig^{(n+1)}&=&\Sfunij\,e^2\ee^2\left[
\frac{z^2+(1-z)^2}{k_i\mydot k_j+m^2}+
\frac{m^2}{\big(k_i\mydot k_j+m^2\big)^2}\right]\ampsqnt(k_p)
\nonumber\\*&\times&
\left(\frac{s}{8(2\pi)^3}\right)^2\beta_i^3\beta_j^3\,\xi_i\xi_j\xi_p
d\xi_p dz dy_i d\varphi_i dy_j d\varphi_j d\tilde{\phi}_{n-1}\,.
\label{reFFge}
\eeqn
By using eq.~(\ref{kikj}) we write:
\beq
k_i\mydot k_j+m^2=\frac{s}{4}\left(\xi_i\xi_j+\frac{4m^2}{s}\right)
\left(1-\frac{y_j}{1+\rho}\right)\,,
\label{kikjge}
\eeq
with
\beq
\rho=\frac{\xi_i\xi_j(1-\beta_i\beta_j)+4m^2/s}{\xi_i\xi_j\beta_i\beta_j}=
2\,\frac{(\xi_i+\xi_j)^2}{\xi_i^2\xi_j^2}\,\frac{m^2}{s}
+\ord\left(\!\frac{m^4}{s^2}\!\right)\,.
\eeq
By proceeding as usual, we obtain:
\beqn
d\bsig^{(n+1)}&=&d\hsig^{(n+1)}+\aemotpi\,\ee^2\Big[
\left(z^2+(1-z)^2\right)\left(
\log\frac{s}{m^2}+2\log(z(1-z)\xi_p)\right)
\nonumber\\*&&\phantom{d\hsig^{(n+1)}+\aemotpi\,\ee^2}
+2z(1-z)\Big]
d\hsig^{(n)}\,dz\,.
\label{reFFge2}
\eeqn
Using the explicit form of ${\cal K}_{e\gamma}^{\rm FIN}$,
one arrives at the final result:
\beq
\Fragm{e}{\gamma}^{[1]}(z,\mu^2)=\left.\PDF{e}{\gamma}^{[1]}(z,\mu^2)
\right|_{K_{e\gamma}\longrightarrow K_{e\gamma}^{\rm FF}}\,,
\label{Dgesol2}
\eeq
with $\PDF{e}{\gamma}^{[1]}$ given in eq.~(\ref{Gegsol2}).

\subsection{Final considerations\label{sec:finFF}}
While eq.~(\ref{emomc1FF}) is a mathematical consequence of the
results obtained for the electron-to-electron and electron-to-photon
FFs, the physics argument that underpins it has a general validity and,
as such, can immediately be applied to photon splittings. In this
way, one obtains the analogue of eq.~(\ref{gamomc}). This, at $\ord(\aem)$,
reads as follows:
\beq
0=\int_0^1 dz\,z\Big(\Fragm{\gamma}{\gamma}^{[1]}(z)+
2\Fragm{e}{\gamma}^{[1]}(z)\Big)\,,
\label{gamomc1FF}
\eeq
which is the final-state counterpart of eq.~(\ref{gamomc1}) -- as has
been done previously, we have assumed to work with a single 
massive-lepton family. Therefore:
\beq
\Fragm{\gamma}{\gamma}^{[1]}(z,\mu^2)=\left.\PDF{\gamma}{\gamma}^{[1]}(z,\mu^2)
\right|_{K_{\gamma\gamma}\longrightarrow K_{\gamma\gamma}^{\rm FF}}\,,
\label{Dggsol2}
\eeq
with $\PDF{\gamma}{\gamma}^{[1]}$ given in eq.~(\ref{Gggsol2}).

The evolution equations for FFs read as follows:
\beq
\frac{\partial\Fragm{H}{a}}{\partial\log\mu^2}=
\frac{\aem}{2\pi}P_{ba}\otimes\Fragm{H}{b}\,.
\label{APeqFF}
\eeq
By working at $\ord(\aem)$ and bearing eq.~(\ref{Dzero}) in mind, 
eq.~(\ref{APeqFF}) becomes, for $H=e$ and $H=\gamma$:
\beq
\frac{\partial\Fragm{e}{a}^{[1]}}{\partial\log\mu^2}=P_{ea}^{[0]}\,,
\;\;\;\;\;\;\;\;\;\;\;\;
\frac{\partial\Fragm{\gamma}{a}^{[1]}}{\partial\log\mu^2}=P_{\gamma a}^{[0]}\,.
\label{APeqFF1}
\eeq
The results of eqs.~(\ref{Deesol2}), (\ref{Degsol2}), (\ref{Dgesol2}),
and~(\ref{Dggsol2}) satisfy eq.~(\ref{APeqFF1}).

\section{Summary and conclusions\label{sec:concl}}
In this work we have calculated, at the next-to-leading order (NLO)
in QED, the initial conditions for the electron and the photon
structure and fragmentation functions. In the case of the electron
structure functions, two different procedures have been adopted,
which have been shown to lead to the same results. The simpler,
and physically more appealing, of the two methods has then been employed
for the computation of the photon structure functions, and of all
of the fragmentation functions. These results will constitute the
starting point for evolving such functions at the NLL accuracy,
which will be the subject of a forthcoming paper~\cite{BCCFS}.
While the present calculations have been carried out by summing over 
the polarisations of the electrons and photons, the techniques that we
have used can be extended to the cases with definite polarisation states.

The main results of this paper are the $\ord(\aem)$ contributions
to the initial conditions of the structure and fragmentation functions.
As far as the structure functions are concerned, these can be found
in eqs.~(\ref{G1sol2}) and~(\ref{Ggesol2}) in the case of the electron,
and in eqs.~(\ref{Gegsol2}) and~(\ref{Gggsol2}) in the case of the photon,
while for fragmentation functions they are given in eqs.~(\ref{Deesol2})
and~(\ref{Dgesol2}) in the case of the electron, and in eqs.~(\ref{Degsol2})
and~(\ref{Dggsol2}) in the case of the photon.

The $e^\pm$ and $\gamma$ structure functions are an essential ingredient 
for the computations of cross sections at $\epem$ colliders. Conversely,
the fragmentation functions for these particles can be used at both 
lepton and hadron colliders; however, they are relevant only for 
observables defined by tagging (i.e.~measuring the momentum of) 
the corresponding object. This is actually more useful for photons
(see e.g.~ref.~\cite{Frederix:2016ost}) than for electrons -- in the 
latter case, one ends up dealing with quantities (the bare electrons)
that are never measured as such by experiments, but only reconstructed
through Monte Carlo simulations, and which are therefore more sensible as
theoretical than as phenomenological tools.

After evolution, structure and fragmentation functions are meant to
be used in factorisation formulae, where they are convoluted with
subtracted short-distance cross sections, fully calculable in perturbation
theory. The subtraction removes logarithmic terms (of the electron mass)
that are already included in the structure and fragmentation functions,
thus preventing one from double counting them. Conversely, power-suppressed
terms may or may not be included in the results which, roughly speaking,
corresponds to computing the subtracted cross sections with massive and
massless electrons, respectively. With this in mind, the additional
complications stemming from keeping the electron mass different from
zero do not appear to be justified for the sake of the power-suppressed
terms, which are expected to be extremely small. Furthermore, modern
cross section computations tend to be highly automated, and in particular
the reduction of virtual integrals is unlikely to be well-behaved if the
electron has a minuscule but non-zero mass. Although the final results
for the initial conditions are independent from these considerations,
the intermediate steps of the calculations presented here show explicitly
how massless-electron short-distance cross sections can be used in
factorisation formulae at the NLO.

\section*{Acknowledgments}
This paper constitutes the first part of an ongoing project in collaboration
with V.~Bertone, M.~Cacciari, S.~Catani, G.~Stagnitto, M.~Zaro, and X.~Zhao, 
whom I warmly thank. I am particularly indebted to S.~Catani, who patiently 
answered many different questions of mine. Conversations with M.~Bonvini, 
S.~Forte, F.~Maltoni, M.~Mangano, D.~Pagani, F.~Piccinini, G.~Ridolfi, 
H-S.~Shao, and B.~Ward, as well as the hospitality of the CERN TH division 
where part of this work has been carried out, are also gratefully acknowledged.

\appendix
\section{On observables defined by means of Dirac delta's\label{plvsdel}}
In this appendix we seek to prove eq.~(\ref{xipdelZ}). The key observation 
is that the $\delta$ function in that equation serves the  purpose of 
enforcing the definition of the $\Zp$ observable (since it originally
stems from eq.~(\ref{ddpl})). Therefore, this $\delta$ function and its
counterpart with argument $\Zm$ (due to eq.~(\ref{ddmn})) identify 
zero-width bins in the $\Zp$ and $\Zm$ differential distributions.
This implies that one must use the plus distributions that appear in 
${\cal K}_{ee}$ to obtain an event-counterevent structure, where the 
bins for the event and the counterevent are determined by the $\delta$'s 
in eq.~(\ref{ddpl}) computed at the given $\xi$ and at $\xi=0$, respectively.
The fact that such bins never coincide (except when $\Zp=1$) is due
to their being infinitely narrow. Therefore, in order to perform a
computation which is not divergent in any of the intermediate steps,
one must adopt a regularisation procedure. A possibility is that
of replacing the $\delta$ with a member of a family of its approximants. 
An easier one is to dimensionally-regularise the $\xi$ integration variable,
by means of a factor $\xi^{-2\ep}$ as suggested by eq.~(\ref{mu3res}).
In order to be definite, let us consider the l.h.s.~of eq.~(\ref{xipdelZ})
times a regular test function $g(\xi)$:
\beq
X=g(\xi)\xidistr{+}\delta\!\left(\Zp-(1-\xi)\right)d\xi\,.
\label{tmp1}
\eeq
By using the definition of the plus distribution and by regularising the 
resulting expression as advocated above, eq.~(\ref{tmp1}) becomes:
\beq
X=\frac{1}{\xi^{1+2\ep}}\,\Big[g(\xi)\delta\!\left(\Zp-(1-\xi)\right)-
g(0)\delta\!\left(\Zp-1\right)\Big]\,d\xi\,.
\label{tmp2}
\eeq
The two terms on the r.h.s.~of eq.~(\ref{tmp2}) are separately
finite. By using the $\delta$ function to get rid of the $\xi$
integration in the first term, and by performing such an integration
explicitly in the second term, we obtain:
\beq
X=\left(\frac{1}{1-\Zp}\right)^{1+2\ep}g(1-\Zp)
+\frac{1}{2\ep}\,g(0)\,\delta\!\left(\Zp-1\right)\,.
\label{tmp3}
\eeq
By using the distribution identity:
\beq
z^{-1-2\ep}=-\frac{1}{2\ep}\,\delta(z)+\pdistr{z}{+}-
2\ep\lpdistr{z}{+}+\ord(\ep^2)
\label{distrID}
\eeq
in the first term of eq.~(\ref{tmp3}), we finally obtain:
\beq
X=g(1-\Zp)\pdistr{1-\Zp}{+}+\ord(\ep)\,,
\label{tmp4}
\eeq
where the regularisation can be safely removed. As was anticipated,
this shows that by {\em formally} replacing $\xi$ with $1-\Zp$ directly
in eq.~(\ref{tmp1}) one obtains the correct result, thus proving 
eq.~(\ref{xipdelZ}). The same conclusion can be easily shown to apply 
to the log-plus distribution term of ${\cal K}_{ee}$, thus leading
to eqs.~(\ref{hbsigplres}) and~(\ref{hbsigmnres}).

The same procedure, namely the separate computation of the event
and counterevent contributions, each dimensionally-regularised
(whence the necessity of the $d$-dimensional \mbox{$(n+1)$}-body phase 
space of eq.~(\ref{mu3res}), which otherwise would not be needed in FKS),
can be applied to the massless-electron real-emission cross
section of eq.~(\ref{hsrealdZZ}). Technically, this case is more
complicated than that relevant to the degenerate cross sections, 
owing to the simultaneous presence of two non-trivial observable-defining 
$\delta$ functions. The final result is:
\beq
\frac{d\hsig_{\epem}^{(3)}}{d\Zp d\Zm}=
\ee^2\,\aemotpi\,f_0\,B(s,0)\,,
\label{hsrealdZZfin}
\eeq
where:
\beqn
f_0&=&\Bigg[\frac{\pi^2}{6}\,\delta(\Zp-1)-
\frac{1+\Zp^2}{\Zp}\lppdistr{1-\Zp}{+}
\nonumber\\*&&\phantom{\Bigg[}
-\frac{1+\Zp^2}{\Zp}\log\left(\frac{1+\Zp}{2\Zp}\right)
\pdistr{1-\Zp}{+}\Bigg]\delta(\Zm-1)
\nonumber\\*&+&
\frac{(1+\Zp\Zm)^3(\Zp^2+\Zm^2)}{\Zp(1+\Zp)\Zm(1+\Zm)(\Zp+\Zm)^2}
\pdistr{1-\Zp}{+}\pdistr{1-\Zm}{+}
\nonumber\\*&+&
\Big[\Zp\,\longleftrightarrow\,\Zm\Big]\,.
\label{f0res}
\eeqn
In order to cross-check this result, we have integrated analytically the 
r.h.s.'s of eqs.~(\ref{hsrealdZZ}) and~(\ref{hsrealdZZfin}) over $\Zp$ and 
$\Zm$ (the former integral being trivial, since one can immediately get rid 
of the $\delta$ functions), by imposing an arbitrary cut on the invariant 
mass of the $u\ub$ pair, $\puub^2\ge cs$, with \mbox{$0<c<1$}. We have 
found that two integrals coincide, for any value of the parameter $c$.

\section{Splitting kernels with massive partons\label{sec:QCsplit}}
In this appendix we consider initial- and final-state collinear
branchings that involve massive partons, and report or compute the forms 
of the relevant kernels. The results for the final-state cases can
be found in ref.~\cite{Catani:2000ef}; to the best of our knowledge, 
those for the initial-state ones are derived here for the first time.
We work in QCD, and obtain the corresponding QED results simply by
replacing colour factors with charge factors (see eqs.~(\ref{QCDtoQED1})
and~(\ref{QCDtoQED2})).

The kinematics of a final-state branching where a parton $p$ with
flavour $S(a_i,a_j)$ splits into two partons $i$ and $j$ with flavours 
$a_i$ and $a_j$
\beq
S(a_i,a_j)\;\longrightarrow\;a_j+a_i\,,
\label{FSRflav}
\eeq
is denoted as follows:
\beq
k_p(1)\;\longrightarrow\;k_j(z)+k_i(1-z)\,.
\label{FSRkin}
\eeq
The notation of eq.~(\ref{FSRflav}) is the standard FKS one. The
momenta are such that:
\beq
k_p^2=m_p^2\,,\;\;\;\;\;\;
k_i^2=m_i^2\,,\;\;\;\;\;\;
k_j^2=m_j^2\,.
\label{masses}
\eeq
In the cases we are interested in, two of the masses in eq.~(\ref{masses})
will be different from zero, with the remaining one equal to zero.
The results of ref.~\cite{Catani:2000ef} can be written as follows.
The collinear limit of an $(n+1)$-body matrix elements, that includes,
consistently with FKS conventions, the flux and average factors, is:
\beq
\bampsqnpot\;\stackrel{k_i\parallel k_j}{\longrightarrow}\;
\frac{2\gs^2}{(k_i+k_j)^2-m_p^2}\,P_{a_jS(a_i,a_j)^\star}^{<}(z)\,
\ampsqnt\,.
\label{MEcollFRS}
\eeq
In keeping with the notation used previously, the matrix element
on the l.h.s.~(r.h.s.) of eq.~(\ref{MEcollFRS}) does (does not)
include mass effects (the masses being those of partons with
flavours $S(a_i,a_j)$, $a_i$, and $a_j$).
The splitting kernel that appears in eq.~(\ref{MEcollFRS}) has been
denoted as is customary in FKS. See in particular appendix~B of
ref.~\cite{Frixione:1995ms}, where a symbol $a^\star$ indicates
that the parton with flavour $a$ is possibly off-shell, thus
clarifying immediately whether a splitting is timelike or spacelike. 
Although the standard (i.e.~massless) $\ord(\aem)$ Altarelli-Parisi kernels 
have the same forms regardless of whether they are relevant to time- or
space-like branchings\footnote{This is not true for their azimuthal
correlated counterparts: 
$Q_{ab^\star}(z)\ne Q_{a^\star b}(z)$~\cite{Frixione:1995ms}.}, 
this is not the case 
when massive partons are involved, which justifies the notation 
employed in eq.~(\ref{MEcollFRS}). From ref.~\cite{Catani:2000ef} 
we obtain:
\beqn
P_{qq^\star}(z)&=&\CF\left[\frac{1+z^2}{1-z}-
\frac{m^2}{k_i\mydot k_j}\right]\,,
\label{Pqqst}
\\
P_{gq^\star}(z)&=&\CF\left[\frac{1+(1-z)^2}{z}-
\frac{m^2}{k_i\mydot k_j}\right]\,,
\label{Pgqst}
\\
P_{qg^\star}(z)&=&\TR\left[z^2+(1-z)^2+
\frac{m^2}{k_i\mydot k_j+m^2}\right]\,,
\label{Pqgst}
\eeqn
which are relevant to the splittings $q\to qg$, $q\to gq$, and $g\to q\bq$,
respectively, and we have denoted the quark mass squared by $m^2$. 
Note that the denominator of the rightmost terms in square brackets in 
eqs.~(\ref{Pqqst})--(\ref{Pqgst}) is identical to that in the prefactor 
on the r.h.s.~of eq.~(\ref{MEcollFRS}). 
Equations~(\ref{MEcollFRS})--(\ref{Pqgst}) can immediately be applied
to the case of QED splittings involving a photon $\gamma$ and a fermion
$f$ of electric charge $e(f)$ (in units of the positron charge) and number 
of colours $N_c(f)$ by means of the formal substitutions:
\beqn
&&\gs^2\;\longrightarrow\;e^2\,,
\label{QCDtoQED1}
\\
&&\CF\;\longrightarrow\;e(f)^2\,,
\;\;\;\;
\TR\;\longrightarrow\;N_c(f)\,e(f)^2\,.
\label{QCDtoQED2}
\eeqn
In order to obtain the kernels relevant to spacelike branchings, we
follow the procedure of appendix~B of ref.~\cite{Frixione:1995ms},
i.e.~we obtain them by crossing the corresponding timelike expressions,
computed with an unphysical kinematic configuration where some of
the energies are negative. In order to be definite, let us consider
the case where the FKS parton $i$ is collinear to parton $1$ (i.e.~$j=1$).
The analogues of eqs.~(\ref{FSRflav}) and~(\ref{FSRkin}) are:
\beqn
a_1&\longrightarrow&\bar{S}(a_i,\bar{a}_1)+a_i\,,
\label{ISRflav}
\\
k_1(1)&\longrightarrow&k_p(z)+k_i(1-z)\,,
\label{ISRkin}
\eeqn
with $1$ and $p$ incoming and $i$ outgoing; charge conjugation has been
denoted by means of a bar. Conversely, their unphysical 
counterparts, where all partons are outgoing, read as 
follows\footnote{In ref.~\cite{Frixione:1995ms}
what is denoted here by $\tilde{k}$ has been denoted by $\bar{k}$; we 
presently have not used the latter form in order to avoid confusion
with massive-parton momenta.}:
\beqn
S(a_i,\bar{a}_1)&\longrightarrow&\bar{a}_1+a_i\,,
\label{uISRflav}
\\
\tilde{k}_p(1)&\longrightarrow&\tilde{k}_1(y)+k_i(1-y)\,,
\label{uISRkin}
\eeqn
with
\beq
\tilde{k}_p=-k_p\,,\;\;\;\;\;\;\;\;
\tilde{k}_i=-k_i\,.
\eeq
Therefore, in the strict collinear limit:
\beq
k_p=zk_1=-z\tilde{k}_1=-zy\tilde{k}_p=zyk_p
\;\;\;\;\;\;\Longrightarrow\;\;\;\;\;\;\;\;\;z=1/y\,.
\eeq
The rightmost equation above is eq.~(B.37) of ref.~\cite{Frixione:1995ms}.
We now apply eq.~(\ref{MEcollFRS}) to the branching of eq.~(\ref{uISRflav}),
by taking into account differences due to normalisation, flux, and sign
prefactors, that originate from crossing, in exactly the same way
as in app.~B of ref.~\cite{Frixione:1995ms}. Thence:
\beqn
&&\bampsqnpot(k_1\ldots k_i)\;\stackrel{k_1\parallel k_i}{\longrightarrow}\;
\nonumber\\*
&&\phantom{aaa\equiv}
\frac{2\gs^2 (-)^{[\sigma(a_1)+\sigma(S(a_i,\bar{a}_1))]}}
{\big(\tilde{k}_1+k_i\big)^2-m_p^2}\,
\frac{\omega(S(a_i,\bar{a}_1))}{\omega(a_1)}\,
zP_{\bar{a}_1S(a_i,\bar{a}_1)^\star}^{<}\left(\frac{1}{z}\right)\,
\ampsqnt(zk_1)
\nonumber\\*
&&\phantom{aaa}
\equiv
\frac{2\gs^2}
{(k_1-k_i)^2-m_p^2}\,
P_{\bar{S}(a_i,\bar{a}_1)^\star a_1}^{<}\left(z\right)\,
\ampsqnt(zk_1)\,.
\label{uMEcollFRS}
\eeqn
The rightmost expression in eq.~(\ref{uMEcollFRS}) implicitly defines
the spacelike massive splitting kernels $P_{a^\star b}$ we seek to determine. 
Eq.~(\ref{uMEcollFRS}) is formally identical to the non-azimuthal
part of eq.~(B.38) of ref.~\cite{Frixione:1995ms}, except for the
denominator of the prefactor, that here must take the masses into account. 
Note indeed that when all masses are equal to zero this prefactor is equal
to that of eq.~(B.38) of ref.~\cite{Frixione:1995ms}, including an overall 
minus sign that originates from using $k_1$, rather than $\tilde{k}_1$, 
in the latter equation.

We can now use eq.~(\ref{uMEcollFRS}) to obtain the sought spacelike
splitting kernels. The $q\to q^\star g$ kernel is obtained by crossing
the $q^\star\to qg$ one. The result is:
\beq
P_{q^\star q}(z)=\CF\left[\frac{1+z^2}{1-z}-
\frac{z\,m^2}{k_1\mydot k_i}\right]\,.
\label{Pqstq}
\eeq
The $q\to g^\star q$ kernel is obtained by crossing
the $g^\star\to\bq q$ one. The result is:
\beq
P_{g^\star q}(z)=\CF\left[\frac{1+(1-z)^2}{z}-
\frac{z\,m^2}{k_1\mydot k_i-m^2}\right]\,.
\label{Pgstq}
\eeq
Finally, the $g\to q^\star\bq$ kernel is obtained by crossing
the $\bq^\star\to g\bq$ one. The result is:
\beq
P_{q^\star g}(z)=\TR\left[z^2+(1-z)^2+
\frac{z\,m^2}{k_1\mydot k_i}\right]\,.
\label{Pqstg}
\eeq
As was expected, the mass-independent terms in
eqs.~(\ref{Pqstq})--(\ref{Pqstg}) are identical to their timelike
counterparts in eqs.~(\ref{Pqqst})--(\ref{Pqgst}), while this is not
the case for the mass-dependent terms.

\phantomsection
\addcontentsline{toc}{section}{References}
\bibliographystyle{JHEP}
\bibliography{eepdfic}

\end{document}